\documentclass[11pt, oneside]{article}   	
\usepackage{geometry}                		
\usepackage{booktabs}
\usepackage{graphicx}				
\usepackage{amssymb}
\usepackage{url}
\usepackage{hyperref} 
\usepackage{multirow}
\usepackage{subcaption}
\usepackage{comment}
\usepackage{xcolor}
\usepackage{titlesec}
\titleformat{\paragraph}
{\normalfont\normalsize\bfseries}{\theparagraph}{1em}{}
\titlespacing*{\paragraph}
{0pt}{3.25ex plus 1ex minus .2ex}{1.5ex plus .2ex}
\title{Midterm Status Report of the ILC Technology Network Activities}
\author{By the members of ILC Technology Network (ITN)\footnote{The authors of this document and list of the ITN members can be found in Section~\ref{ITN-members}}}
\date{28 February 2026\\{\tiny  (Revised from the first version in 24 December 2025, 
IDT-ITN-2015-1.V1)}}

\begin{document}
\maketitle
\abstract{
The ILC Technology Network (ITN) was established in 2022 by the ILC International Development Team, a subcommittee of the International Committee for Future Accelerators, to advance engineering studies toward the realisation of the International Linear Collider (ILC). While the ITN work packages focus on engineering activities for the ILC, their topics are also relevant to a broad range of accelerator applications in particle physics and beyond. These work packages are being carried out now by laboratories in Asia and Europe in close collaboration. This report summarises the current status of the ITN activities.
}
\newpage
\tableofcontents
\newpage
\section{Introduction}
Technical solutions for the International Linear Collider (ILC) were described in the Technical Design Reports (TDRs) \cite{ILC-TDR}, published in 2013 by the Global Design Effort (GDE) team. The next phase of engineering studies required for construction was outlined in the proposal for the ILC Preparatory Laboratory (Pre-lab) in 2022 \cite{Pre-lab}, prepared by the International Development Team (IDT). 

Both the GDE and the IDT were created under the 
auspices
of the International Committee for Future Accelerators (ICFA), which promotes the realization of a linear collider as a global project. In particular, the IDT was tasked with supporting the Japanese community’s proposal to host the ILC in Japan. 

Since the Japanese government judged a transition to the Pre-lab as premature in 2023, the IDT has instead been working to establish the ILC Technology Network (ITN), a worldwide collaboration of laboratories sharing resources to advance engineering studies for the ILC to be ready for the construction.

The Pre-lab proposal identified 18 work packages across three technical areas:
\begin{itemize}
\item Superconductivity
\item Electron and positron sources
\item Small spot size beam (Nano-beam)
\end{itemize}
Of these work packages, the IDT selected 15 \cite{ITN-WP} as especially critical, requiring substantial time and effort. These became the ITN work packages described in Table~\ref{WPPP-tab}.
\begin{table}[hb]
\centering
\begin{small}
\begin{tabular}{|l|c|l|}
     \hline
     Area & WPP & Name \\ \hline
     \multirow{3}{3cm}{Superconductivity} 
     & 1 &Cavity production \\
     & 2 & Cryomodule design\\
     & 3 & Crab cavity\\ \hline 
     \multirow{7}{3cm}{Sources} 
     & 4 &E-source \\
     & 6 & Undulator target\\
     & 7 & Undulator focusing\\ 
     & 8 & Electron-driven target\\ 
     & 9 & Electron--driven focusing\\
     & 10 & E-driven capture\\ 
     & 11 & Target replacement\\ \hline 
\multirow{7}{3cm}{Nano-beam} 
     & 12 & Damping Ring System design \\
     & 14 & Dmping Ring Injection/extraction\\
     & 15 & Final focus\\ 
     & 16 & Final doublet\\ 
     & 17 & Main dump\\
\hline
\end{tabular} 
\end{small}
   \caption{List of ITN work packages}\label{WPPP-tab}
\end{table}

The ITN was initiated by the IDT, with the convener of the IDT Working Group for Accelerator serving as project leader and the IDT Executive Board providing oversight as an interim arrangement. KEK has been playing a central role, supported by a dedicated budget. Through the CERN–KEK agreement, CERN acts as the European hub laboratory, facilitating participation by European institutes via agreements with CERN. Several European laboratories have already joined under this framework contributing to various work packages. In the United States, laboratories have expressed interest in contributing to the ITN and have been actively participating in discussions. However, their full involvement remains pending due to uncertainties in funding.

The ITN was foreseen to develop into an international collaboration, comparable in structure to those established for major particle physics experiments, once a sufficient number of laboratories had joined. Given the current uncertainty regarding U.S. participation, it was agreed at the ITN meeting held in August 2024 to maintain the interim organizational model for the time being. 

The following sections describe the ITN work packages, provide lists of the institutes committed to each, and summarize the progress achieved to date.
\newpage
\section{Work Packages}
\subsection{Superconducting Technology Area}
\subsubsection{Production of SRF (WPP-1)}
\paragraph{Description}

WPP-1 involves manufacturing nine-cell Superconducting Radiofrequency (SRF) cavities compliant with the 
Japanese
High Pressure Gas Safety Act in each 
participating
nation.
Common specifications across nations will be necessary for successful international industrialisation of the ILC cavities.
After applying the latest surface treatment in each location, performance evaluation is conducted through vertical measurements. Following confirmation of performance via vertical measurements, the cavities are integrated into cryogenic modules and subjected to horizontal measurements. The target performance is as shown in Table~\ref{SRF_target}.

\begin{table}[hbt]
    \centering
    \caption{TDR target of ILC performance of SRF cavities}
        \begin{tabular}{ccc}
            \toprule
                & Parameters &  Design \\
            \midrule
                Vertical test & Gradient, Q-value ($Q_0$) & 35.0 MV/m with $Q_0 \ge 0.8 \times 10^{10}$ \\ 
                Cryomodule test & Gradient, Q-value ($Q_0$) & 31.5 MV/m with $Q_0 \ge 1.0 \times 10^{10}$ \\  
            \bottomrule
        \end{tabular}
    \label{SRF_target}
\end{table}

\begin{figure}[h]
    \centering
    \includegraphics[width=1.0\columnwidth]
    {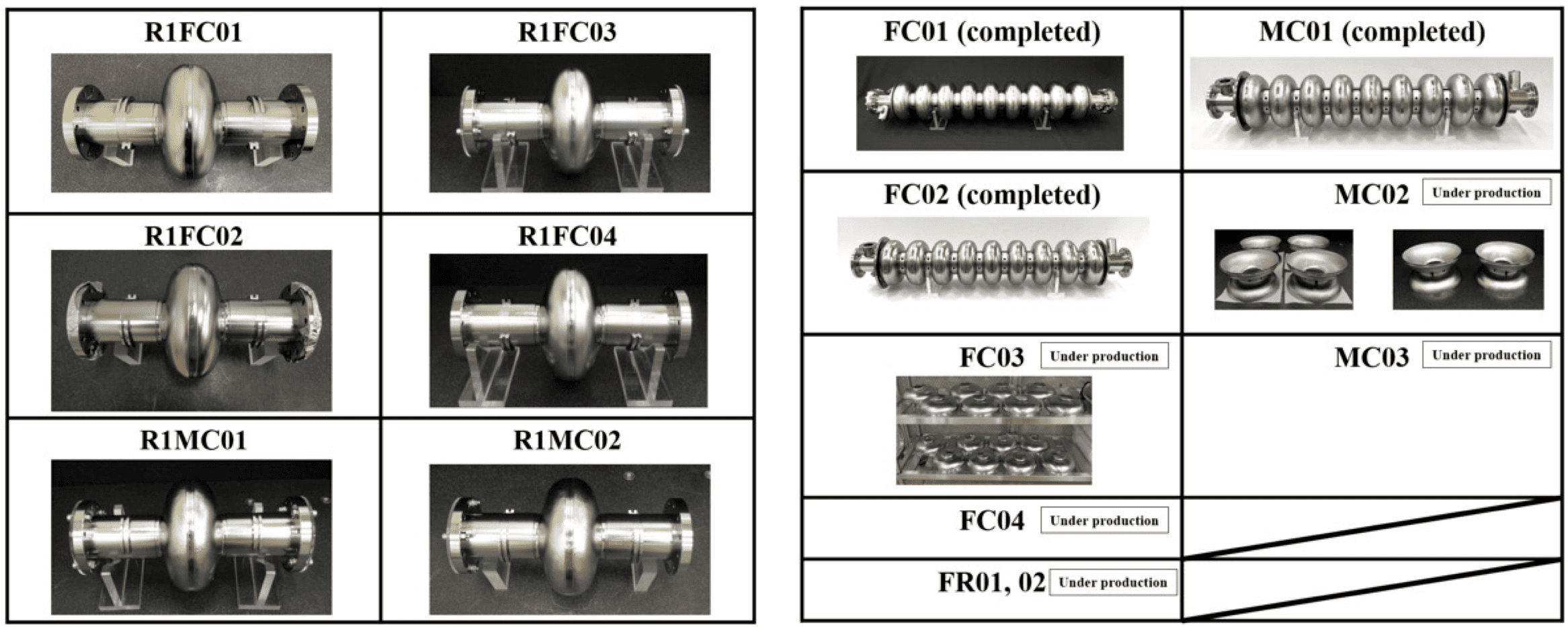}
    \caption{List of Japanese cavities}   
    \label{fig:JapanCav}
\end{figure}

\paragraph{Status in Asia}
Although cavity manufacturing has experienced some delays, it is generally progressing smoothly. To date, six single-cell cavities and three nine-cell cavities have been completed (including one nine-cell cavity made from medium grain, MG, material, a world first). This work is being carried out at KEK's cavity manufacturing facility in collaboration with domestic manufacturers. Furthermore, for the first time at KEK, orders have been placed with overseas manufacturers for fiscal year 2025, with nine cavities scheduled to be completed by the end of fiscal year 2025 (initially 12 were planned). Subsequently, surface treatment and vertical measurements will be performed. Welding of the helium tanks is scheduled for fiscal years 2026 to 2027, completing preparations for the cavity strings. All cavities are manufactured in compliance with the High-Pressure Gas Refrigeration Regulations. Figure~\ref{fig:JapanCav} shows a list of completed and currently manufactured cavities.

\begin{figure}[h]
    \centering
    \includegraphics[width=1.0\columnwidth]{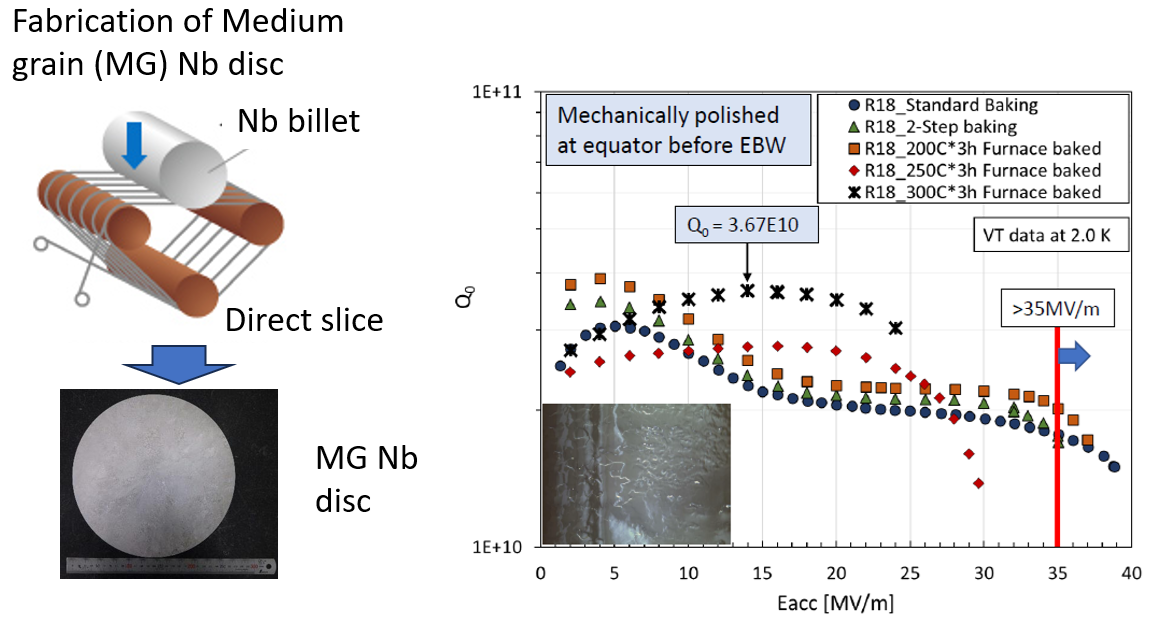} 
    \caption{Fabrication method of MG Nb disc and results of vertical test of single-cell MG cavity. ~\cite{IPAC25_sakai}}  
    \label{fig:MG.png}
\end{figure}

MG material is a new material developed in collaboration with the Jefferson Lab in the US over the past few years with cleanness and cost reduction in mind. Since its performance was confirmed in single-cell cavities as shown in Figure \ref{fig:MG.png}, it was decided to manufacture 9-cell cavities and conduct performance tests. Additionally, single-cell cavities have been manufactured in Europe and South Korea, where performance evaluations are underway.

\begin{figure}[!htb]
    \centering
    \includegraphics[width=1.0\columnwidth]{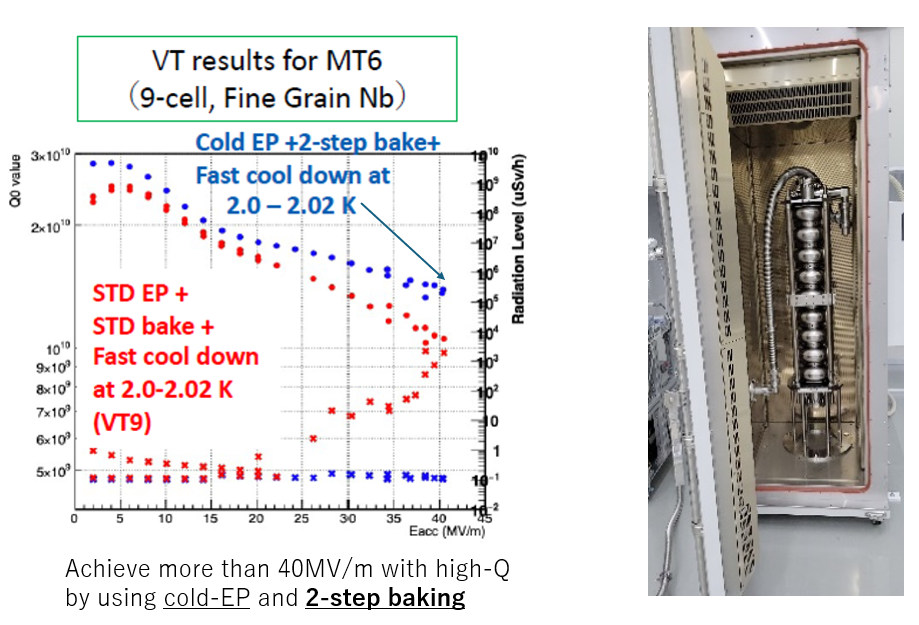}
    \caption{Results of 9-cell cavity performance by using 2 step-baking. ~\cite{IPAC25_sakai}}  
    \label{fig:2step_furnace.png}
\end{figure}

All six completed single-cell cavities underwent the latest surface treatment, 2-step baking (two-step (75/120 degC) temperature heat treatment), followed by vertical tests. Each satisfied the ILC specification (35MV/m±20\%). As expected, Q values higher than $1.0\times10^{10}\,$ were obtained. This establishes the 2-step baking process, and the same surface treatment will be applied to the nine-cell cavities as shown in Figure~\ref{fig:2step_furnace.png}.

High-pressure gas work is proceeding in compliance with refrigeration regulations. Although approximately six months behind the initial schedule, progress is steady. Approval for cavity manufacturing has now been granted, enabling production of cavities and helium tanks. Subsequent manufacturing applications will be submitted for the chimney pipe (special piping connecting the helium tank and two-phase piping, utilizing Ti-SUS dissimilar material joints) and the two-phase piping. For the chimney pipe, flange connections are anticipated.

The target performance for the 9-cell cavity in Japan is as shown in Table~\ref{SRF_newtarget}. Due to the discovery that a newly developed surface treatment method (2-step baking) enables high Q-values, only the vertical measurement performance has been revised. 

\begin{table}[hbt]
    \centering
    \caption{Target of ILC performance of newly developed SRF cavities}
        \begin{tabular}{ccc}
            \toprule
                & Parameters &  Design \\
            \midrule
                Vertical test & Gradient, Q-value ($Q_0$) & 35.0 MV/m with $Q_0 \ge \textbf{1.0} \times 10^{10}$ \\ 
                Cryomodule test & Gradient, Q-value ($Q_0$) & 31.5 MV/m with $Q_0 \ge 1.0 \times 10^{10}$ \\  
            \bottomrule
        \end{tabular}
    \label{SRF_newtarget}
\end{table}

\begin{figure}
[!htb]

    \centering

    \includegraphics[
    width=1.0\textwidth
    ]
    {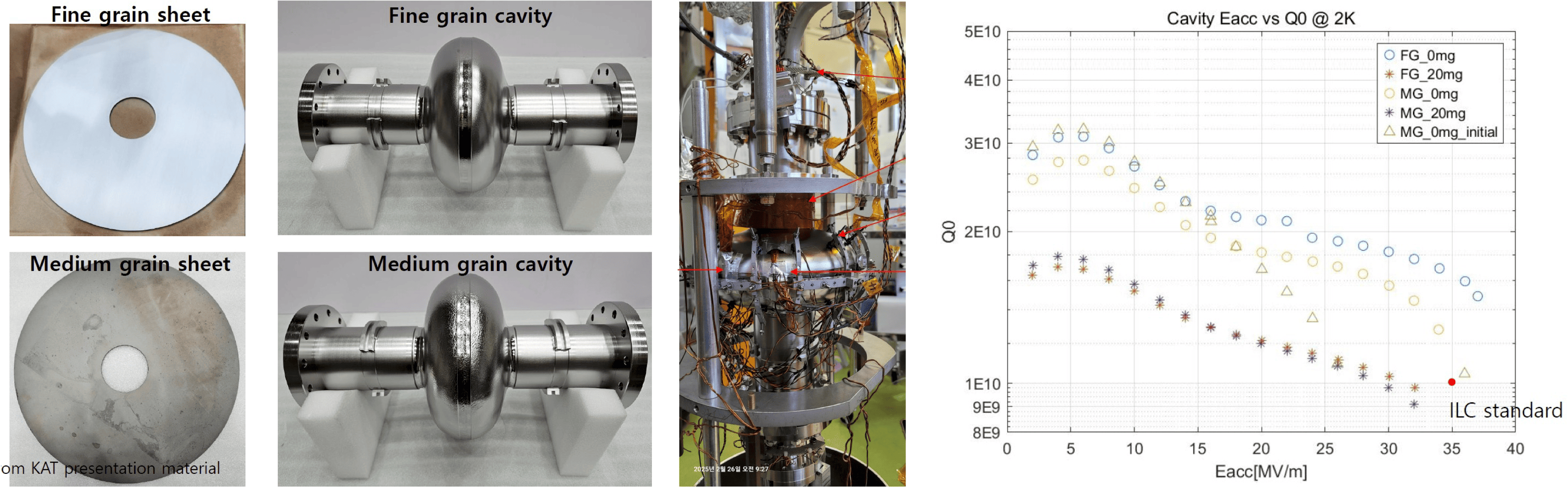}
    \caption{Peformance of Korean cavities}  
    \label{fig:KoreaCav}
\end{figure}

\subparagraph{KEK-DESY Licence Agreement}
The original designs for components such as cavities and ancillaries used in this project are based on the package constructed for the European XFEL in Europe. Therefore, a licence agreement was concluded between KEK and DESY for the creation of drawings and specifications. This agreement permits the use of E-XFEL specifications and drawings solely for ITN purposes.
\subparagraph{KEK-EU collaboration}
KEK provided niobium material to evaluate cavity performance using identical materials. This material is being used for manufacturing both single-cell and 9-cell cavities. Additionally, cavity manufacturing utilizes standardized drawings provided by KEK.
Regular meetings are held with European institutes (CERN, CEA, INFN) regarding cavity manufacturing. Standardization of specifications is being pursued for contractual purposes. For the 9-cell cavities, specifications are more detailed due to the involvement of high-pressure gas. Surface treatment specifications are expected to differ across regions due to variations in equipment, so separate specifications will be developed.
DESY, which has a proven track record in material evaluation at E-XFEL and LCLS-II, was responsible for evaluating the niobium material (Eddy Current Scanning System). DESY manufactured a single-cell cavity using MG material and also performed performance evaluations.
\subparagraph{Korea}
Regular meetings regarding cavity manufacturing are held between KEK and Korea University. KEK supplied the necessary materials and have been providing technical instruction and advice. Two single-cell cavities (FG material and MG material) were completed by a Korean manufacturer. These were then sent to KEK for surface treatment and vertical tests. The results successfully met the the updated ILC and ITN specifications ($35MV/m\pm 20\%, {1.0} \times 10^{10}$). Students from Korea University and manufacturer staff participated in the work at KEK. Figure~\ref{fig:KoreaCav} shows the two completed single-cell cavities, the vertical measurement setup, and the results.
After the performance of the single-cell cavity was demonstrated, a 9-cell cavity using copper material is currently being prototyped. Once the manufacturing technology is established, 
it is planned
to manufacture it using niobium material.

\subparagraph {Overall Timeline}
The aim is 
to complete nine 9-cell cavities in 2026, and surface treatment and vertical tests will be performed sequentially, evaluating the success rate over two vertical test (VT) cycles. The target values are an average of 35.0 MV/m and a Q-value of $1.0 \times 10^{10}$. Following the VT, helium tank welding will be performed, completing preparations for the cavity string.

\paragraph{Status in Europe}
The ITN-EU consortium has been established under the coordination of CERN, with strong contributions from CEA-Saclay and INFN-LASA. This collaboration brings together complementary expertise in materials science, cavity fabrication, and testing, ensuring that the European contribution to SRF technology development is both technically robust and aligned with international efforts \cite{ITN-Eu-TTC2025, ITN-Eu-SRF2025}.

\begin{figure}[!htb]
    \centering

    \includegraphics[width=1.0\columnwidth]{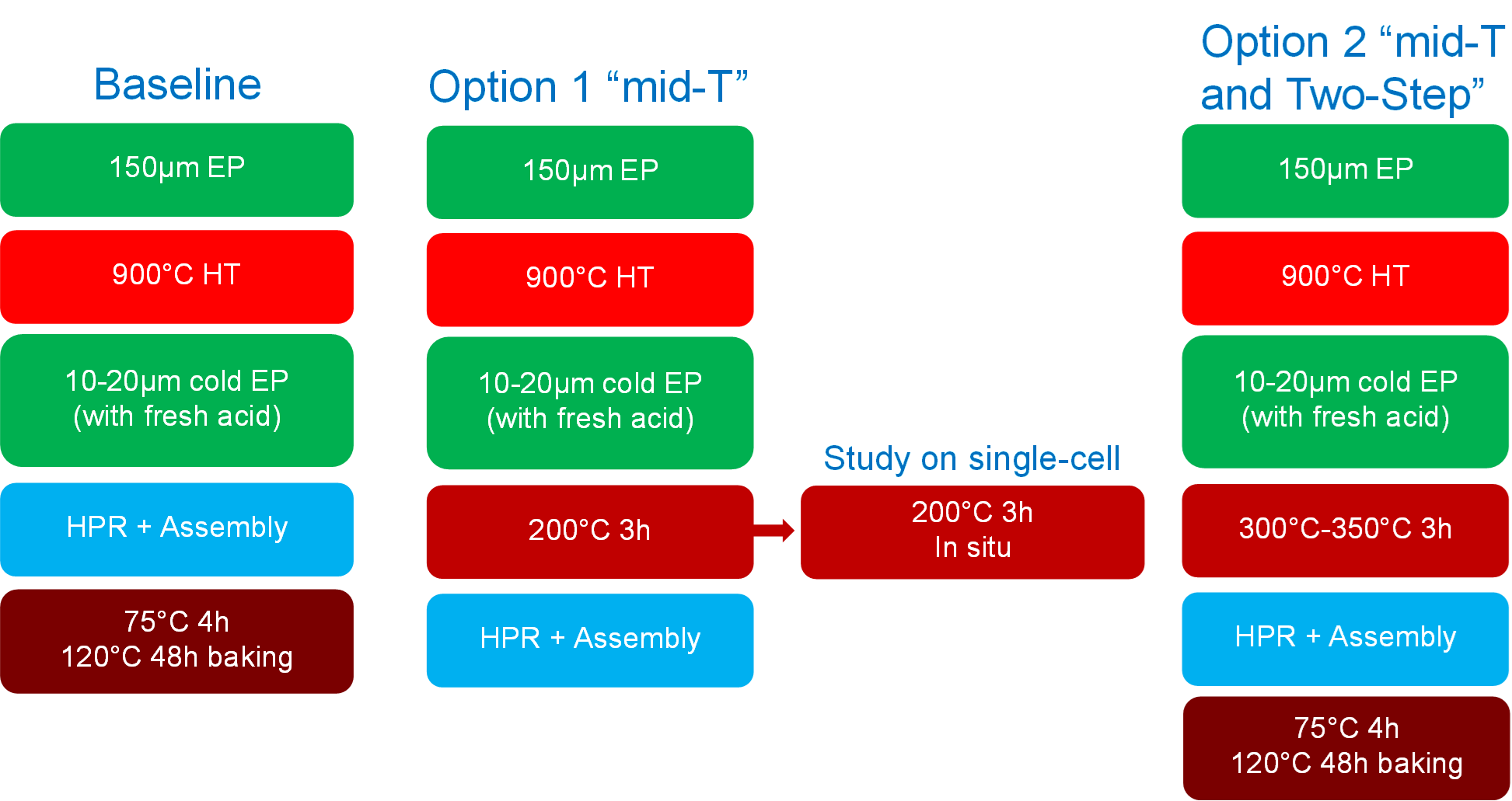}
    \caption{Surface preparation recipes for single-cell R\&D}  
    \label{fig:Flow}
\end{figure}

The material procurement for the single-cell cavities has been successfully completed, with niobium sheets and discs already delivered and qualified through Eddy Current Scanning (ECS) contributed by  DESY. This ensures that the material is free from inclusions and meets the specifications required for cavity fabrication. At present, two single-cell cavities are under manufacture, and will provide the first test bench for the preparation and surface treatment strategies as presented in Figure~\ref{fig:Flow}. 

\newpage

\subparagraph {Timeline for single-cell and nine-cell preparation}
In 2025, two single-cell cavities have been successfully manufactured, and surface preparation of these cavities  will be examined in Europe

Vertical cold tests will 
follow in 2026 to evaluate their performance and validate the fabrication process.\\
We are currently finalizing the necessary steps to award the manufacturing and surface preparation contract in ear;ly 2026. This process includes completing contractual 
requirements,
selecting qualified suppliers, and ensuring all technical specifications and quality requirements are clearly defined. Once awarded, the contract will enable the production and surface treatment of two cavities, which are scheduled to be delivered to KEK in 2026, fully prepared and ready for the jacketing process.
\subparagraph{Materials procurement and quality controls}
All materials for single-cell and 9-cell cavities have been successfully procured by KEK and delivered to CERN. 



The technical specifications for the material procurement have been prepared by KEK, taking into account the High Pressure Gas Safety (HPGS) requirements. All cavities will be produced using Nb FG (Tokyo Denkai), while one single-cell cavity will be fabricated with Nb MG (ATI) for cleanness and cost reduction studies. Moreover, spare parts for all items have been purchased for R\&D studies on surface treatments and for mechanical/welding tests: these will be used for the qualification of the EU industries necessary for the HPGS requirements in view of the harmonization of pressure vessel codes needed for the large-scale production worldwide.

\begin{figure}[!htb]
    \centering
    \includegraphics[width=1.0\columnwidth]{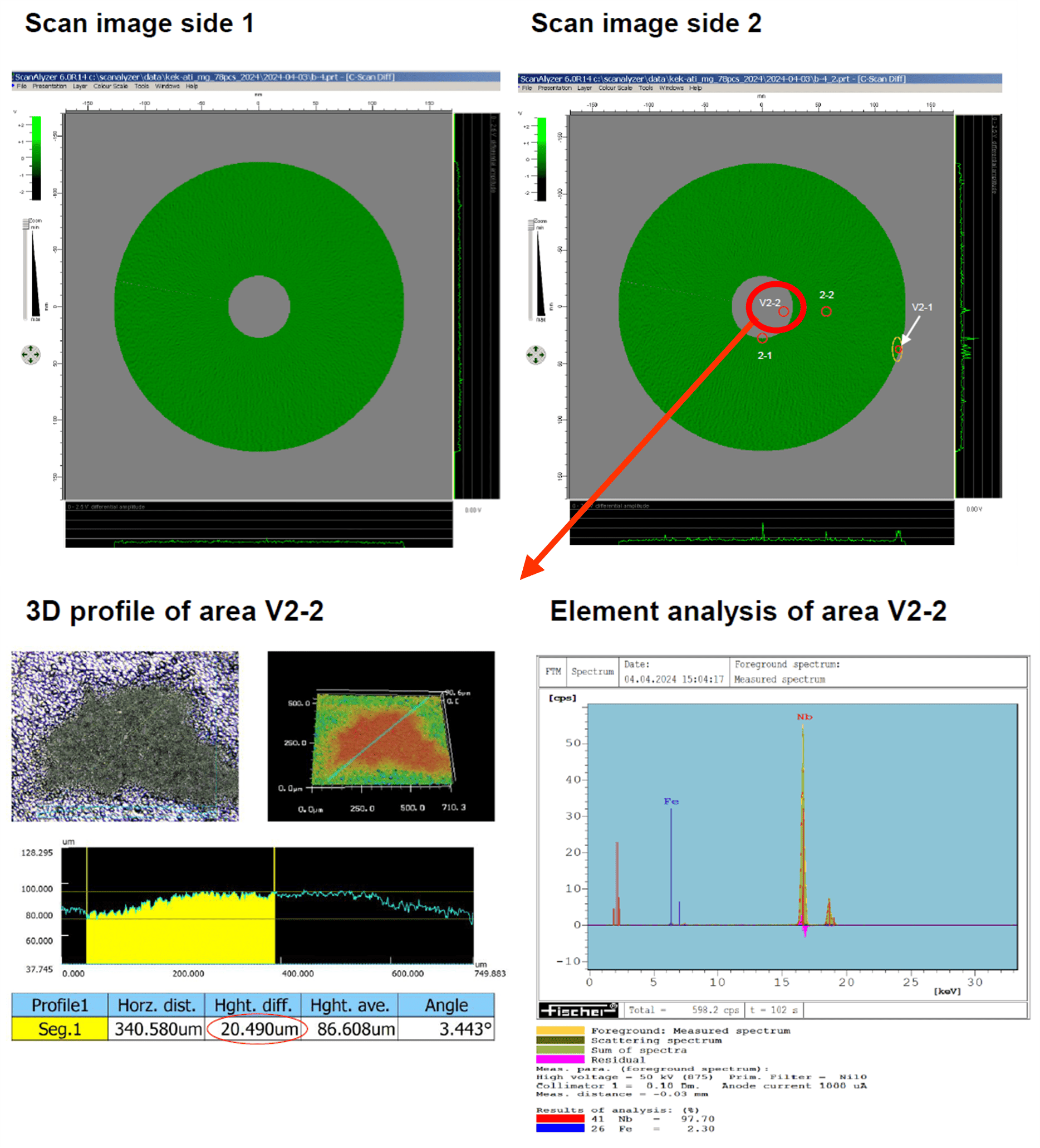}
    \caption{ECS, 3D profile and the result of the element analysis done at DESY on a MG disc.}  
    \label{fig:ECS}
\end{figure}

Given the stringent performance requirements for the 1.3 GHz 
ILC
cavities, Quality Control (QC) was performed on all FG and MG Nb sheets/discs using Eddy Current Scanning (ECS) in 2024 by DESY, as its contribution to ITN. The ECS on Nb sheets/discs resulted in full qualification in terms of specifications requirements, and without any foreign materials. Only one MG disc showed a Fe signal close to its border (end of the billet, being not a risk since it would be possibly caused by the cut out process of the billet, prior to the cell/disc fabrication). Figure~\ref{fig:ECS} shows 
ECS 
results of this MG disc and the further analysis performed due to the presence of the foreign material (Fe). 

The acceptance criteria for the niobium used in the cavities are defined as follows:
\begin{itemize}
    \item Free from non-niobium inclusions
    \item Surface roughness: $R_a \leq 1.6 \, \mu$m
    \item Maximum height: $R_t \leq 15 \, \mu$m
    \item Scratches or hollows: $R_t \leq 15 \, \mu$m
\end{itemize}

\subparagraph{Single-cell cavity activities}
Starting from 2024, activities have been devoted to the definition of the technical specifications for single-cell production, applying all QC protocols typically used for cavity production, with the aim of better preparation at  EU industries for the following 9-cell cavity production.

\begin{figure}[htb]
    \begin{subfigure}{0.5\textwidth}
    \centering
    \includegraphics[width=0.8\linewidth, angle=270]{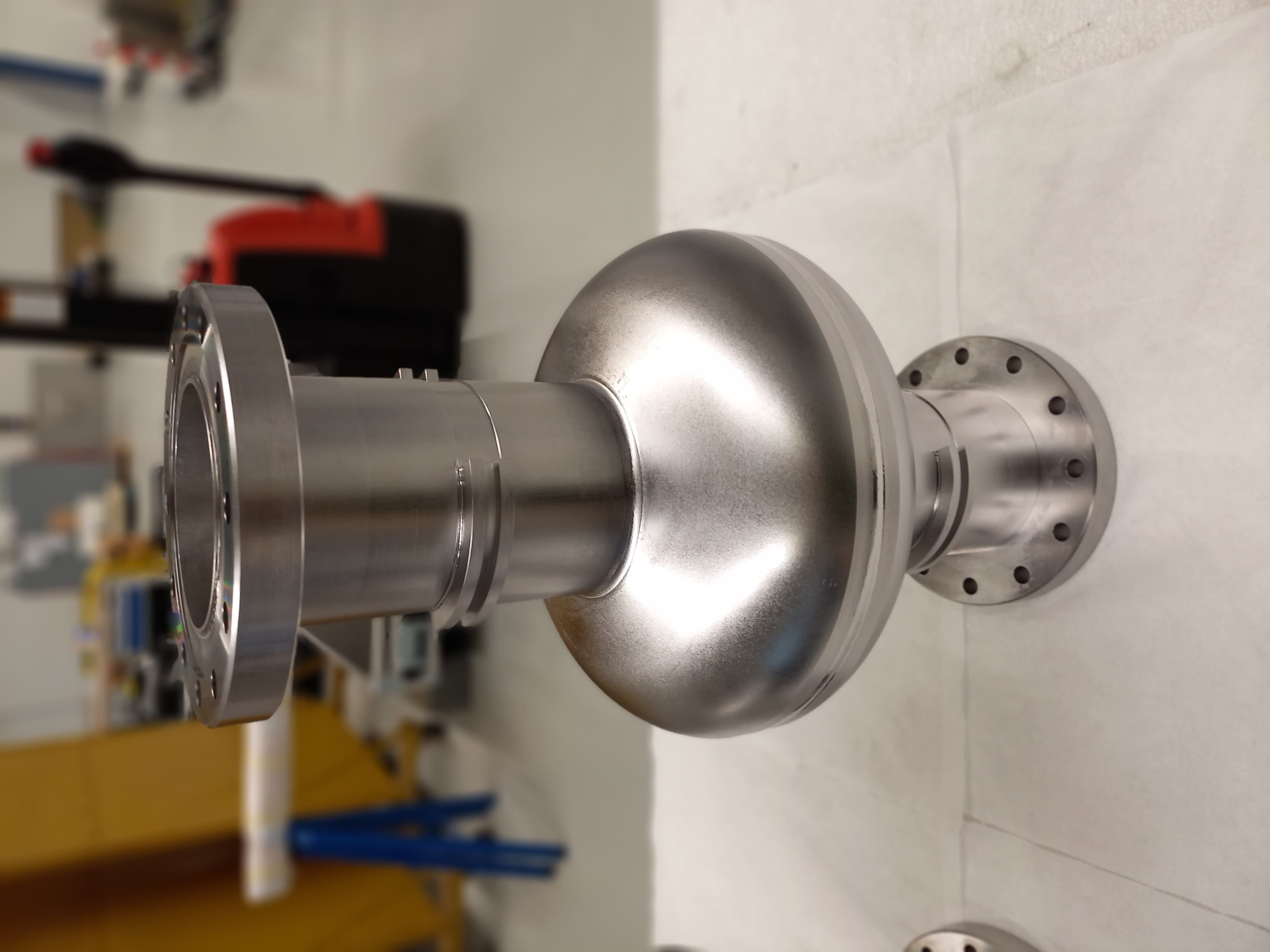}
    \caption{FG single-cell cavity.}
    \end{subfigure}
    \begin{subfigure}{0.5\textwidth}
    \centering
    \includegraphics[width=0.8\linewidth, angle=270]{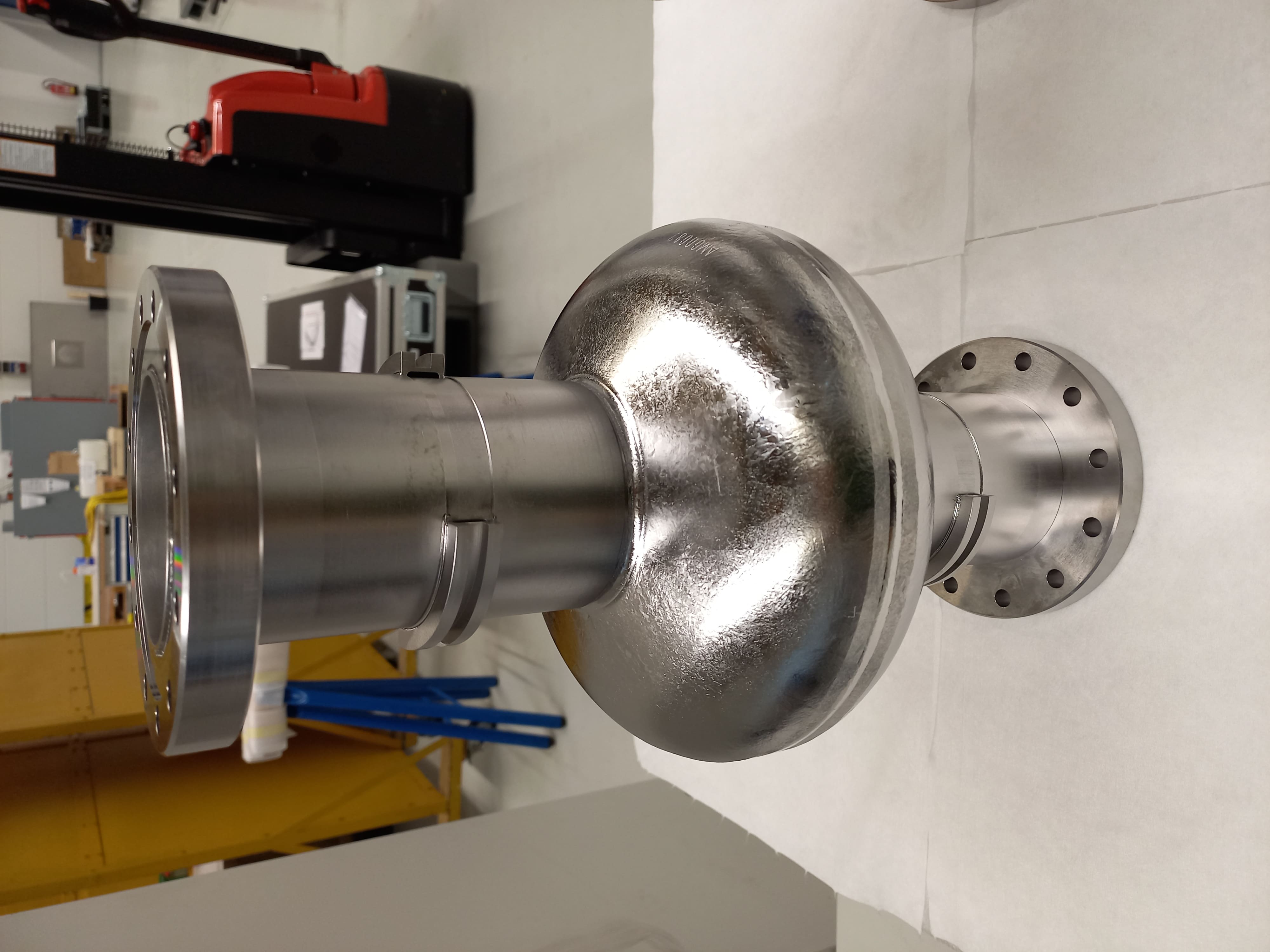}
    \caption{MG single-cell cavity.}
    \end{subfigure}
    \caption{FG and MG single-cell cavities after mechanical fabrication (RI)}
    \label{fig:SingleCells}
\end{figure}

The order was placed at RI  Research Instruments GmbH in March 2025 (order for 2 single-cells, MG and FG, and baseline surface treatments). RI received all materials at the end of April 2025 and started the mechanical production. 
The production of the two single-cell cavities 
was
completed in September 2025.
Figure~\ref{fig:SingleCells} shows FG and MG single-cell cavities after mechanical fabrication. 

The delivery of the two single-cells treated with the baseline Eu-XFEL-like process was foreseen,
and will be followed by the cold vertical tests at 2 K to check the cavity performance.
Possible performance differences 
between FG and MG Nb cavities
are to be sought.

It is important to
have a single producer
for single-cell activities 
to assure common comparison
of performance between the two Nb materials,  avoiding possible 
biases
due to different mechanical or treatment processes applied in
by second vendor.

\subparagraph{Nine-cell cavity activities}

To meet the tight ITN schedule, the immediate priority is the mechanical fabrication of four bare nine-cell cavities by EU industries, followed by the application of the optimal surface treatment recipe identified through single-cell R\&D activities. Vertical tests at 2~K will be performed and cross-checked among different ITN partner laboratories to ensure consistency and reliability.  
It is essential that both European cavity manufacturers (RI and ZRI) are actively involved in this phase to achieve full qualification in terms of HPGS, paving the way for future ILC-oriented production.

Once qualified, two bare cavities will be promptly shipped (by October~2026) to KEK for jacketing and subsequent integration into the ILC-type cryomodule. CEA-Saclay and INFN-LASA will participate on-site in the jacketing process at KEK, following up on HPGS qualification activities, string assembly operations, and the cryomodule test campaign.

The remaining two cavities will undergo jacketing at EU industrial sites, which will serve as a demonstration of their capability to deliver HPGS-qualified cavities “ready for string assembly,” as successfully achieved in projects such as Eu-XFEL and ESS. After jacketing, the cavities will be tested at 2~K to assess performance and verify the quality of the jacketing process. Although these two cavities may not be installed in the ITN cryomodule, their jacketing step
is crucial for validating the industrialization pathway (see section \ref{cryo-description}). During this phase, CEA-Saclay and INFN-LASA will provide technical support to EU industrial partners, ensuring rigorous quality control (QC) throughout the process.

The nine-cell cavity activities officially began in January~2025, with weekly meetings of a joint team of experts (ITN-EU and KEK) dedicated to defining HPGS and technical specifications. The technical specifications for the nine-cell bare cavities were prepared taking into account both HPGS requirements and the constraints of EU-based fabrication, including jacketing operations and the associated QC plan.  
In parallel, the same team identified the most effective surface treatment and annealing procedures to achieve the ILC performance targets, drawing on R\&D results obtained at KEK, FNAL, DESY, and other partner laboratories.

This collaborative effort led to the preparation of the technical specifications for the call for tender for the mechanical fabrication of two nine-cell bare cavities for KEK, awarded to RI in July~2025. Building on this success, the call for tender for the four ITN-EU cavities—currently being finalized—will adopt the same technical specifications as those used by KEK, including the requirements for surface and annealing treatments.  
Work on defining the jacketing specifications will begin immediately after the bare-cavity tender process  concluded.

\subparagraph{HPGS specifications and pressure vessel code harmonization}
The process was coordinated by KEK in close collaboration with the High Pressure Gas Safety Institute of Japan (KHK) and the Ibaraki Prefecture authorities.  
All requirements imposed by the Japanese High Pressure Gas Refrigeration Safety (HPGS) regulations have been incorporated into the corresponding technical specification. These include, among others, material traceability, wall thickness measurements, and mechanical testing of representative weld samples.

According to the current plan, the pressure test will be performed at KEK following the completion of the helium jacket welding.  
For ITN-EU activities, the possibility of conducting the pressure test at qualified EU industrial partners is being considered, following the approach successfully implemented in other large-scale projects such as Eu-XFEL, LCLS-II, ESS, and SHINE. In this case, tests will be carried out in compliance with Japanese HPGS requirements, applying a gas pressure equal to 1.5~times the maximum allowable working pressure (MAWP).  

\begin{figure}[htb]
    \centering

    \includegraphics[width=1.0\columnwidth]{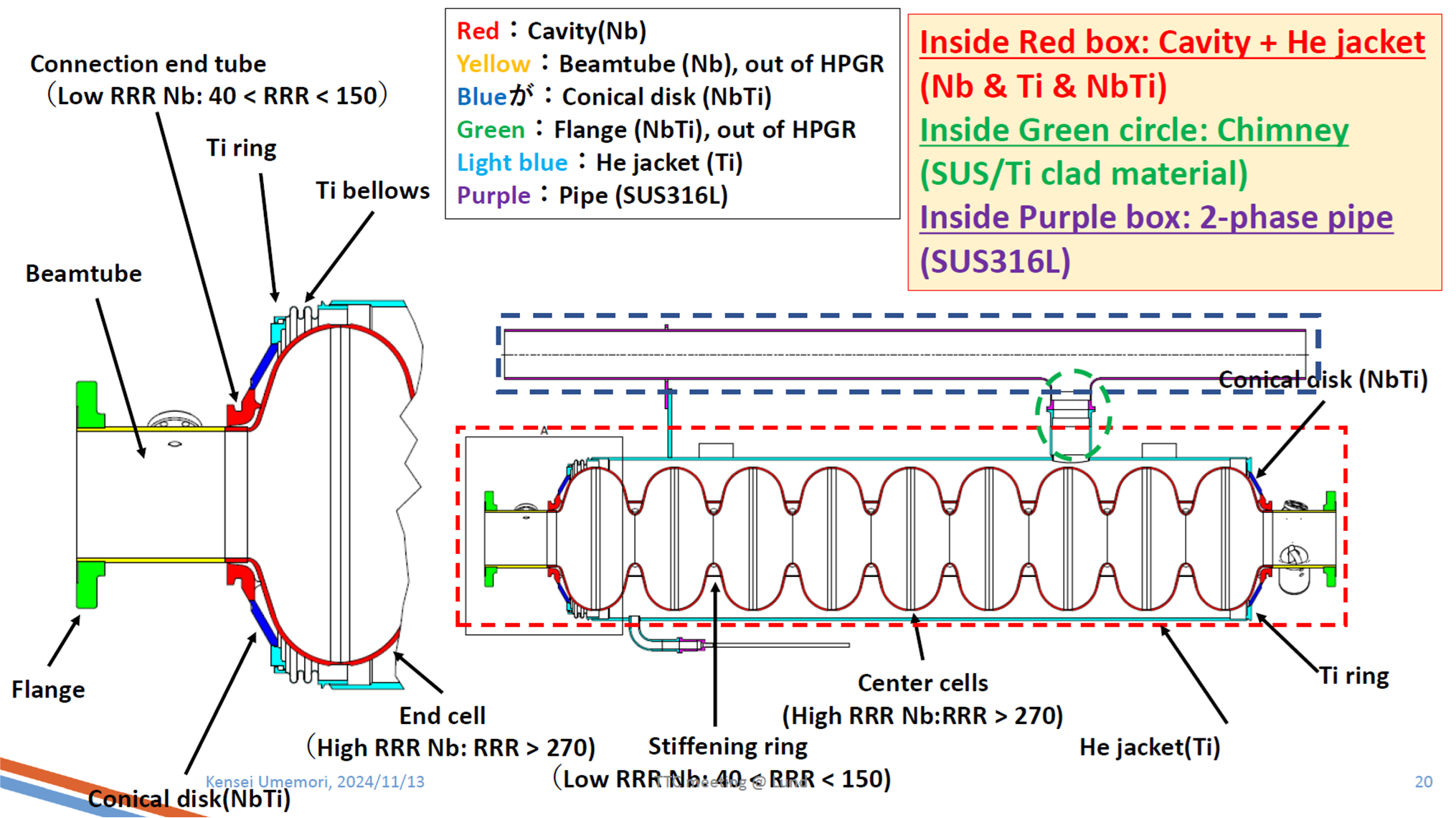}
    \caption{Sketch showing ITN cavities in the jacketing configuration.}  
    \label{fig:HPGS}
\end{figure}

Figure~\ref{fig:HPGS} illustrates a schematic representation of the jacketed nine-cell cavities, showing the materials, the bellows connection between adjacent cavities, the helium tank, and the two-phase pipe.

\subsubsection{Cryomodule Design (WPP-2)}
\paragraph{Description}\label{cryo-description}
WPP-2 involves designing the cryogenic module so called cryomodule. 

This needs to be a high-pressure gas-compatible design based on common design specifications under international cooperation.
The ILC cryomodules contain 8-9 SRF cavities and they are connected in strings  without intermediate warm sections,  as was done at the European XFEL \cite{Eu-XFEL}.  The cryomodules represent the major heat loads at liquid helium temperatures and therefore play an important role in the overall cryogenic system optimization. The design concept is to use large cryo-plants to cool kilometer-long cryo-units. Each of the 12.652 m-long cryomodules contains either nine cavities (Type A), or eight cavities and one superconducting quadrupole package, combined with horizontal and vertical dipole correctors and Beam Position Monitors (BPMs) located at the center of the cryomodule, (Type B). The cavities and quadrupole packages are assembled into the cryomodules along with their supporting structures, thermal shields and insulation, and all the associated cryogenic piping required for the coolant flow distribution along a cryogenic unit without the need for additional external cryogenic distribution lines. The ILC cryomodule design is an update of the module originally developed
for the TESLA Test Facility (TTF) in cooperation of INFN and DESY, and applied for the European XFEL project \cite{TESLA-EuXFEL-CM-Design}. For the ILC, the engineering design was then further optimized by Fermilab for the ILC-TDR \cite{ILC-TDR}. 

The design with the quadrupole package in the middle of the cryomodule allows the definition of a standard interconnection interface for all main-linac cryomodules. Fundamental-mode power couplers provide the RF power to the cavities and are connected to ports on the vacuum vessel on one side. RF cables bring the signals from the field pickup and the HOM antennas to the LLRF control system outside the cryomodule for the control of the cavity field amplitude and phase and to extract HOM power from the 2 K level.  Manually operated valves, required for the clean-room assembly, terminate the beam pipe at both module ends. 

The largest component of the transverse cross section is a 300 mm-diameter helium-gas return pipe (GRP) that allows recovery of the mass flow of He vapors at a negligible pressure drop along the cryo-strings, to preserve temperature stability. It acts as the structural backbone for supporting the string of beamline elements. The GRP is supported from the top by three composite posts with small thermal conduction from the room-temperature environment. The center post is fixed to the vacuum vessel, while the two remaining posts are laterally flexible and can slide on the flanges to allow for the longitudinal contraction/expansion of the GRP with respect to the vacuum vessel during thermal cycling.

\paragraph{Status}
In Asia, Japan participates in WPP-2, planning to undertake not only design but also manufacturing, assembly, and testing \cite{IDT-ADT-CM-Development}. Furthermore, beyond the cryomodule, design, manufacturing, assembly, and testing are also progressing for the input coupler\cite{IDT-ADT-CM-components development}, frequency tuner, magnetic shield, and superconducting magnet. Additionally, studies are underway on improving cleanroom work efficiency (e.g., introducing robotic technology) and demagnetizing cryovessels.
\subparagraph{Cryomodule}
Based on the Technical Design Report issued in 2013, as described above, a redesigned 3D model for the cryomodules has been completed 
\cite{ITN-CM-3D-model}.
ILC TDR's key modifications include the removal of the 5K radiative heat shield, the addition of a frequency tuner access port, and changes to the ports for the magnet current leads. Manufacturing is scheduled for fiscal years 2025–2026, with assembly completion planned for fiscal year 2027. The figure below, Figure~\ref{fig-cryomodule}, shows the completed 3D model.
\begin{figure}[htb]
    \centering

    \includegraphics[width=1.0\columnwidth]{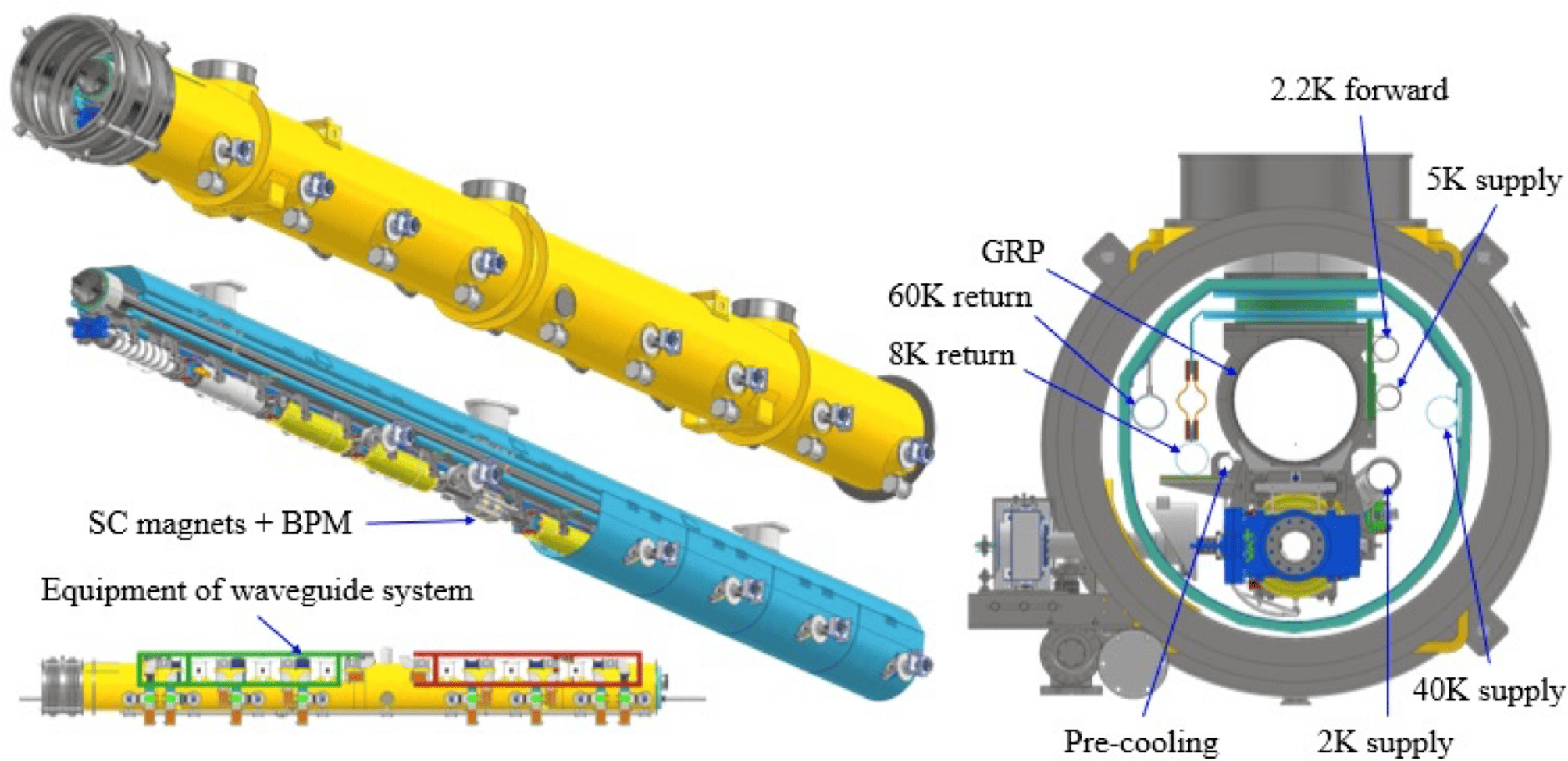}
    \caption{A complete 3D model of the cryomodule.}\label{fig-cryomodule}  
\end{figure}

\begin{figure}[htb]
    \centering
    \includegraphics[width=1.0\columnwidth]{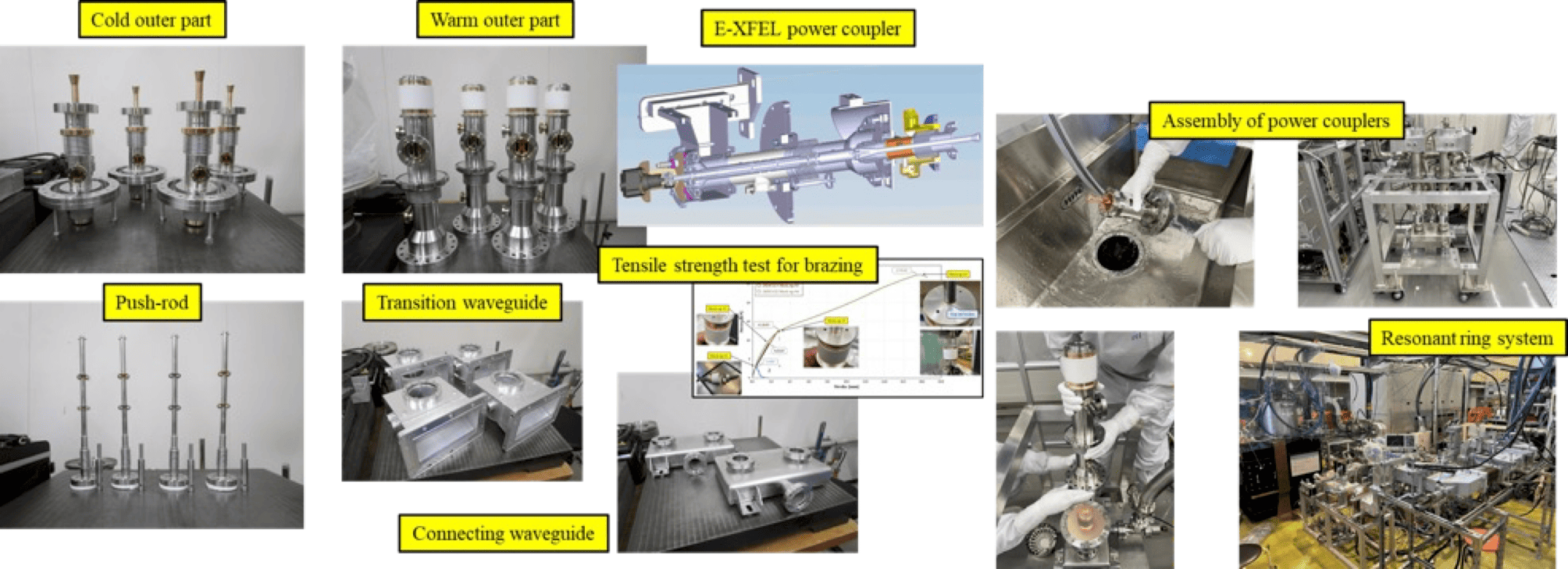}
    \caption{Four completed couplers.}\label{fig-couplers}  
\end{figure}

\subparagraph{Input Coupler} Although the input coupler design is based on the TTF and Eu-XFEL design \cite{TTF-EuXFEL-coupler}, multiple modifications were made during its design and manufacturing under US-Japan Cooperation. The prototype input power couplers have been completed and tested in 2025. KEK is under preparation of contract for real eight power couplers.They will be produced and tested in 2026-2027. After high power test, they will be installed in CM. 
Figure~\ref{fig-couplers} shows the prototype input couplers and the assembly for high-power testing.

\begin{figure}[htb]
    \centering
    \includegraphics[width=1.0\columnwidth]{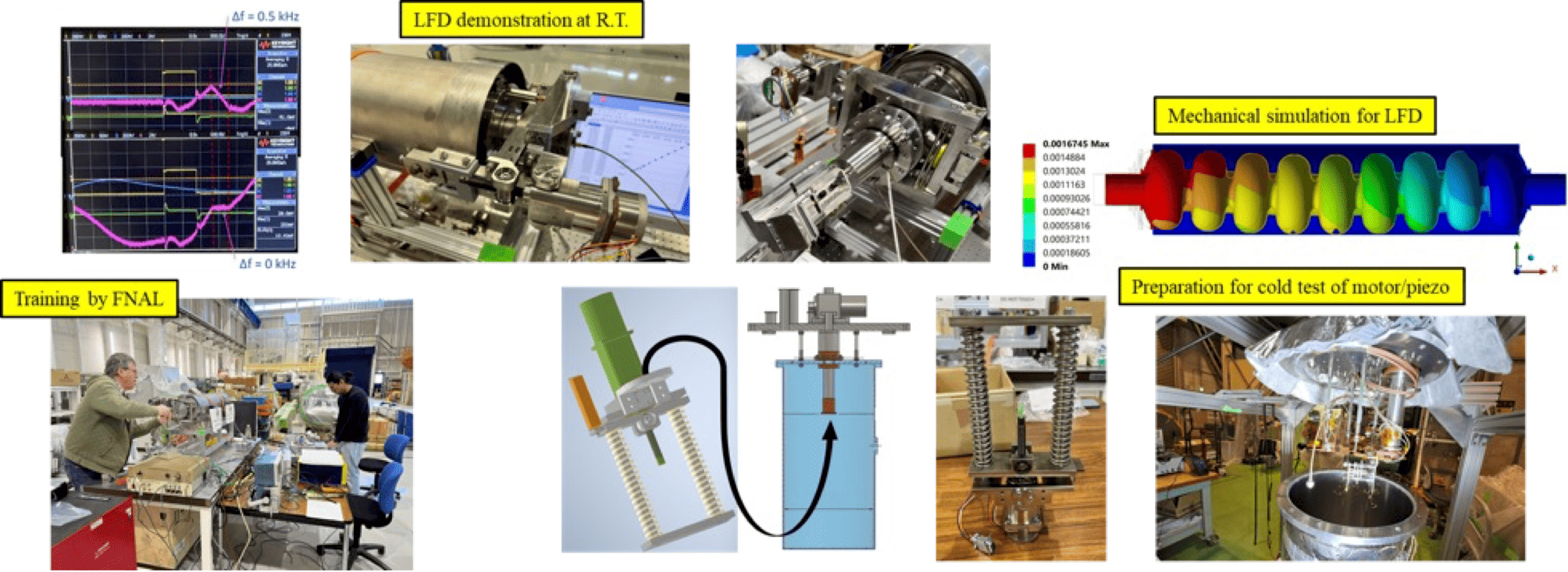}
    \caption{Simulation results for tuner drive testing.}\label{fig-tuner}  
\end{figure}

\subparagraph{Frequency Tuner}
The LCLS-II tuner will be used for the frequency tuner \cite{LCLS-II-tuner}. FNAL sent a set of cavities and tuners, and training was conducted on assembly, testing, and evaluation by FNAL staff under US-Japan Cooperation. Demonstration tests for Lorentz force-induced detuning, which is problematic during high acceleration gradients, have also been completed (at room temperature conditions). A prototype tuner has been manufactured and similarly tested. Motor and piezo selection is now complete, and cryogenic testing is underway \cite{ITN-frequency-tuner}. Figure~\ref{fig-tuner} shows simulation results for tuner drive testing and mechanical deformation due to Lorentz forces, along with the cryogenic test preparation status.

\begin{figure}[htb]
    \centering
    \includegraphics[width=1.0\columnwidth]{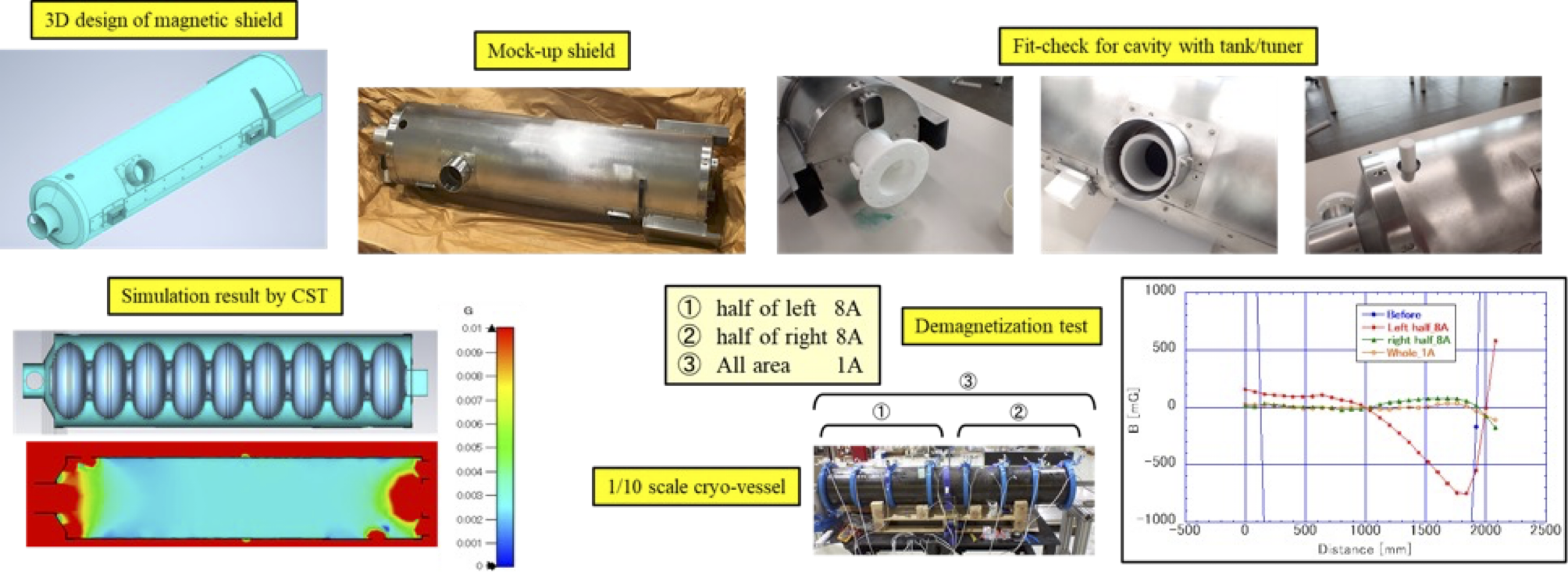}
    \caption{Prototype test of magnetic shielding.}\label{fig-shielding}  
\end{figure}

\subparagraph{Magnetic Shielding and Demagnetization}
The basic design of the magnetic shielding is complete, and a prototype has been manufactured. This prototype has undergone a fit check to verify 
there is no
interference with the tuner. Initial production of the actual unit is currently underway \cite{ITN-magnetic-shield}. Once evaluation is complete, production of eight units will commence. Figure~\ref{fig-shielding} show the fit check using the prototype shield and the demagnetization test using a mock-up vessel.

\begin{figure}[htb]
    \centering
    \includegraphics[width=1.0\columnwidth]{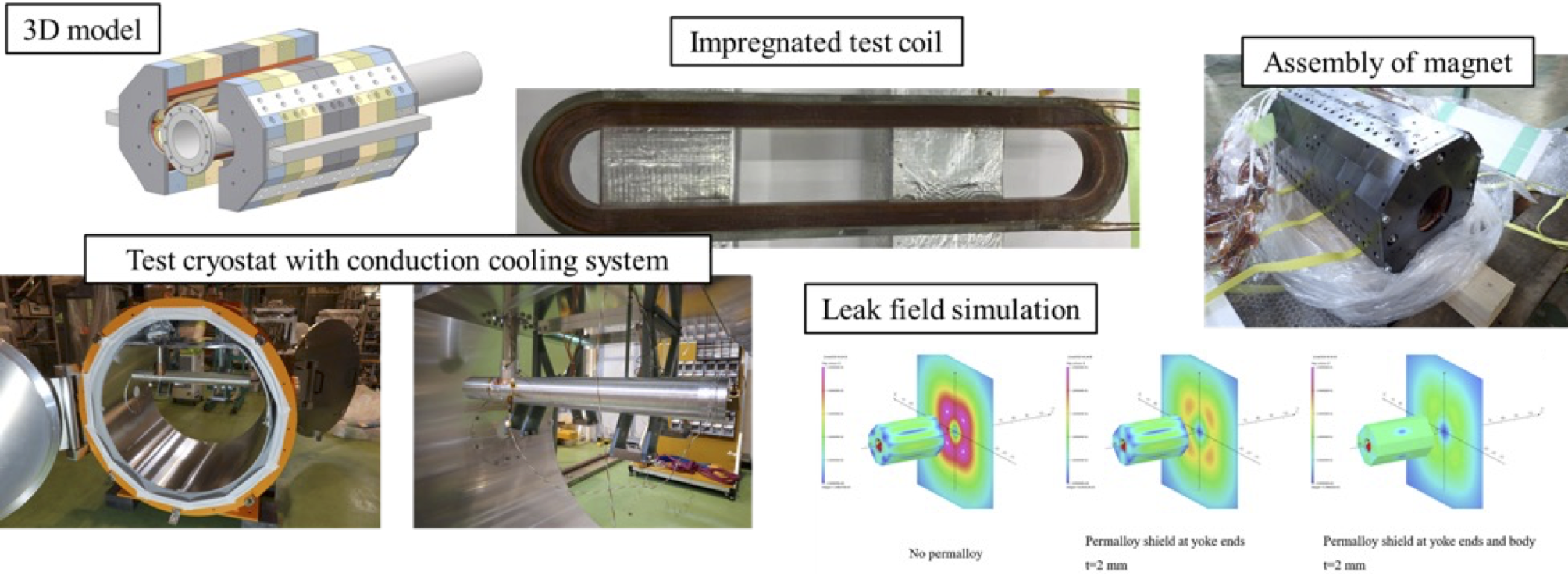}
    \caption{Prototype test of magnetic shielding.}\label{fig-superconducting}  
\end{figure}

Regarding demagnetization of the cryogenic vessel (insulation layer), tests using a mock-up vessel have been completed, and tests on an actual-size vessel are scheduled.
\subparagraph{Superconducting Magnet}
The superconducting magnet is scheduled for manufacturing and testing over a three-year period from fiscal years 2024 to 2026 \cite{ITN-superconducting-magnet} \cite{ITN-superconducting-magnet Cryostat}. The NbTi coil is expected to be completed in fiscal year 2025. Following assembly, performance evaluation will be conducted during cooling tests. The evaluation insulation layer has already been manufactured, and the conduction cooling system has been installed. Figure~\ref{fig-superconducting} shows the coil under construction, the completed test cryostat, and the simulation results for the leakage magnetic field.

Another development approach by using Nb3Sn superconductor has been complementary carried out by CIEMAT. It aims to be further  sustainable to absorb  heat deposition caused by field emission  electron flux induced by the high electric field on the SRF cavity surface.  The basic design has been completed and R\&D work for coil fabrication is in progress.  A model magnet is to be developed in the next step \cite{CIEMAT-Nb3Sn-SCQ-design, CIEMAT-Nb3Sn-SCQ-temperature-margin}.

R\&D work for the ILC Maing Linca Beam Position Monitor (BPM) is in progress at IFIC in Spain. High precision BPM with a spatial resolution of less than 1 micron-meter and a time resolution smaller than 500 nano-meters. is a critical requirement, as well as very precise   aligned with the SCQ \cite{ILC linac re-entrant caivty BPMs at IFIC},  Therefore  KEK and IFIC is closely cooperating in particular for the interface.  The current SCQ design with iron-yoke dominated quadrupole magnetic would be adequate to the reliable and stable alignment. The assembly test is expected in 2026-2027.

\subparagraph{Cleanroom Operations and Robot Implementation}
Cleanroom-related work primarily involves preparing necessary jigs for hollow string operations. Rails have been installed within the cleanroom. Next, posts for mounting the hollow strings will be manufactured and introduced. Vacuum components required for hollow strings (interconnecting bellows, clean gate valves) are also undergoing testing.

\begin{figure}[htb]
    \centering
    \includegraphics[width=1.0\columnwidth]{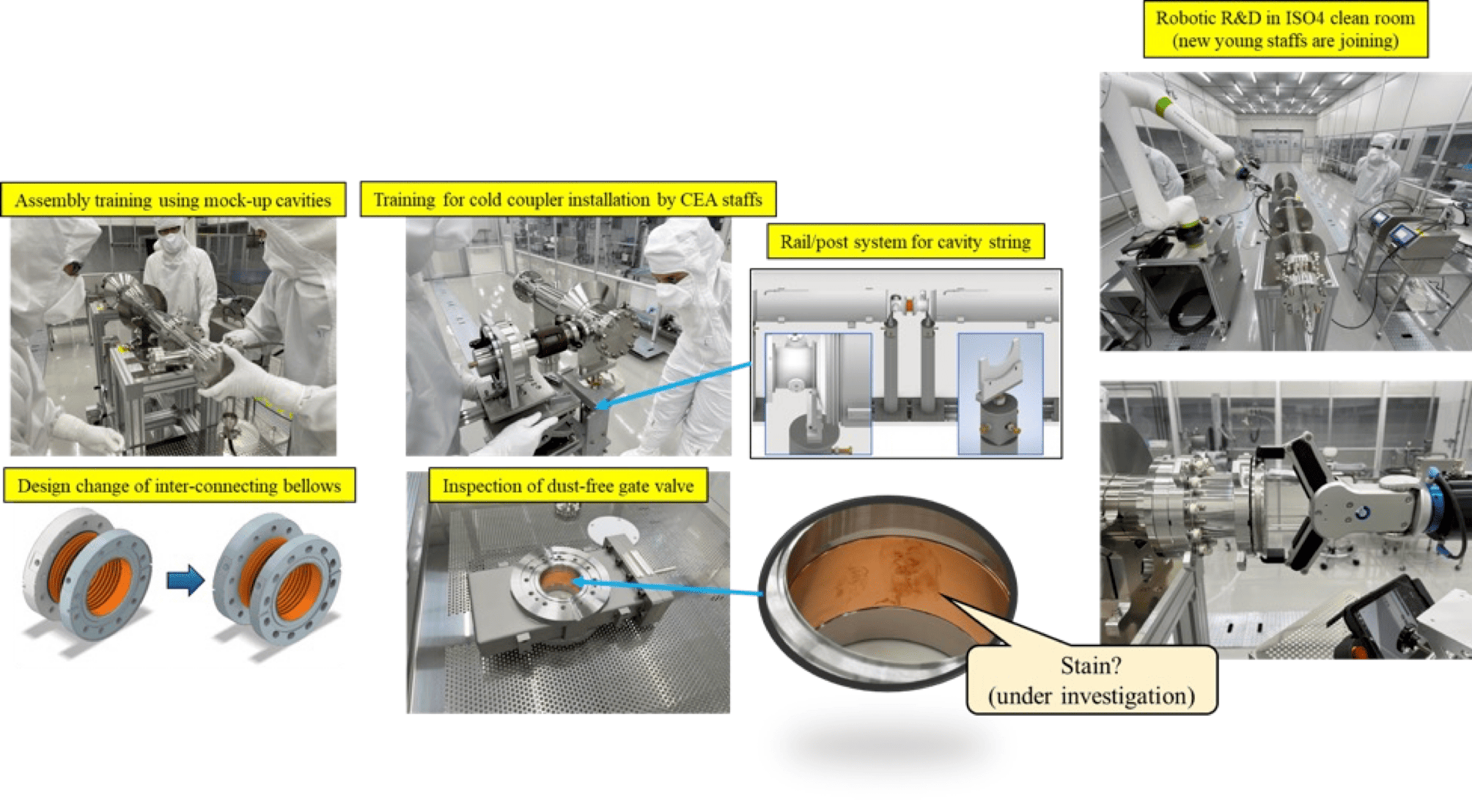}
    \caption{Clean-room and work being conducted there.}\label{fig-clean_room}  
\end{figure}

Regarding the robot, one unit has been introduced, and demonstration tests for automated cleaning and assembly have been completed. A remote control system is currently under construction, while the selection of the optimal end-effector (An end effector or tool head is the device at the end of a robotic arm, designed to interact with the environment) 
is being pursued in parallel. Following this, robot studies using cavities and demonstrations of input coupler installation are planned. Figure~\ref{fig-clean_room} shows the clean-room work and robot introduction.
\subparagraph {Timeline}
We will manufacture a cryomodule in 2026-2027. Concurrently, four ancillaries (power coupler, frequency tuner, magnetic shield, SC magnet) will be manufactured and evaluated for standalone performance. Cavity string assembly and cryomodule assembly will commence in 2027. Cooling tests scheduled after CM construction will be conducted in two stages: the first
being a low-power test, followed by a high-power test. In the high power test, the performance of cavities will be compared with those in vertical test.

\subsubsection{Crab Cavity (WPP-3)}
\paragraph{Description}
The International Linear Collider (ILC) requires crab cavity systems to counteract the 14 mrad crossing angle at the interaction point (IP), ensuring optimal luminosity. These cavities rotate particle bunches transversely to align them for head-on collisions. The ILC’s baseline energy is 250 GeV, with scalability to 1 TeV \cite{SRF23_1}. Since 2021, a global consortium under the ILC International Development Team (IDT) has developed five crab cavity designs to meet stringent RF and mechanical specifications \cite{CERN-ESU-004}, culminating in a down-selection process to identify the most promising technologies for prototyping.

The crab cavities must fit within a compact cryomodule space of 3.85 m longitudinally and 0.198 m transversely (see Figure~\ref{fig-crab-location}). 
RF parameters were defined for three operating frequencies: 1.3 GHz, 2.6 GHz, and 3.9 GHz, with key design constraints that include:
\begin{figure}[h]
 \centering
    \includegraphics[width=1.0\textwidth]{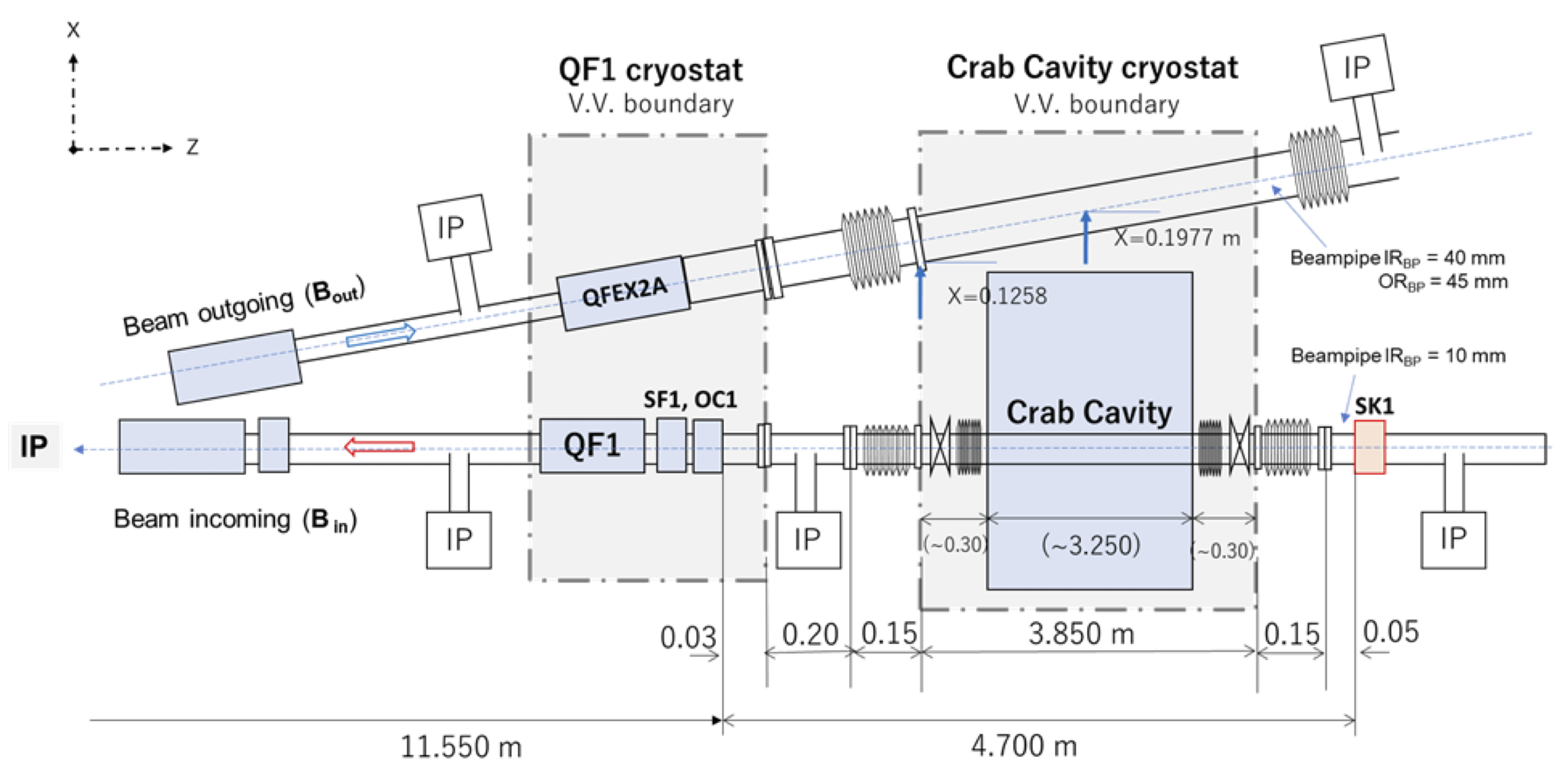}
\caption{ILC Crab Cavity IP Location}\label{fig-crab-location}
\end{figure}
\begin{itemize}
\item Peak electric field (Ep): $\rm \le 45 MV/m$
\item Peak magnetic field (Bp): $\rm \le 80 mT$
\item Beam-pipe aperture: $\rm \ge 25 mm$
\item Kick voltages: up to 7.4 MV for 1 TeV operation
\item HOM impedance thresholds and alignment tolerances
\end{itemize}

\begin{figure}[hbt]
 \centering
    \includegraphics[width=1.0\textwidth]{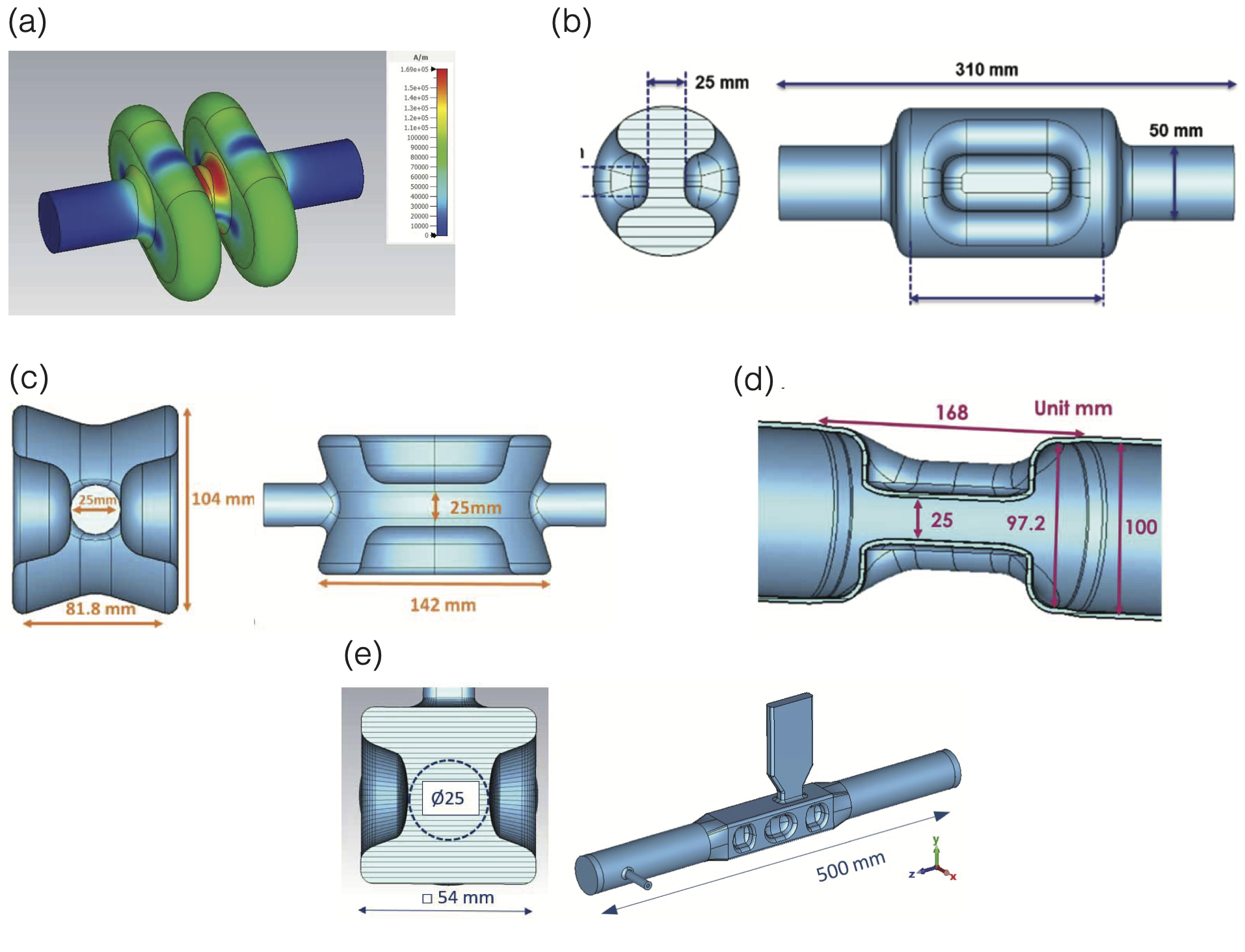}
\caption{Different crab cavity designs: (a) 2-cell, 3.9 GHz Racetrack Crab Cavity, (b) 1.3 GHz RF Dipole (RFD), (c) 1.3 GHz, DQW Crab Cavity, (d) 1.3 GHz WOW crab cavity,(e) 2.6 GHz QMiR Crab Cavity.}\label{fig-all-crab-cavities}
\end{figure}

\subparagraph{3.9 GHz Racetrack Cavity}
Developed by Lancaster University and STFC, originally based on a 9-cell elliptical design\cite{ILC-TDR}, this cavity was redesigned to mitigate issues with Same Order Modes (SOMs) and Lower Order Modes (LOMs). A racetrack geometry was adopted to improve mode separation and reduce Bp. A 2-cell version with optimized aperture and HOM damping (Figure~\ref{fig-all-crab-cavities}-a) was developed, achieving strong suppression of unwanted modes and good mechanical stability.
\subparagraph{1.3 GHz RF Dipole (RFD)}
Developed by Old Dominion University and JLab, the RFD cavity (Figure~\ref{fig-all-crab-cavities}-b) operates in a TE11-like mode. A 1-cell design was selected for its simplicity and effectiveness\cite{Linac22_1}, similar to designs developed for both HL-LHC\cite{SRF17_1}, and EIC\cite{NAPAC22_1}. It features TESLA-style HOM couplers \cite{PRAB2010_1} and meets impedance thresholds for both 125 GeV and 500 GeV operations. Mechanical analysis shows acceptable stress levels\cite{SLAC2023_1}, and fabrication will use medium-grain niobium (MG Nb) for cost efficiency and uniformity.
\subparagraph{1.3 GHz Double Quarter Wave (DQW)}
Developed by BNL and CERN, adapting from HL-LHC\cite{IPAC13_1} and EIC\cite{PRAB2021_1}  designs, the DQW cavity (Figure~\ref{fig-all-crab-cavities}-c) offers ultra-compact geometry and excellent surface field-to-kick voltage ratios. It requires 2 cavities for 125 GeV and 6 for 500 GeV. HOMs are well-separated from the fundamental mode, allowing simple damping schemes. Integration studies show feasibility within ILC space constraints.
\subparagraph{1.3 GHz Wide Open Waveguide (WOW)}
Developed by BNL, the WOW cavity (Figure~\ref{fig-all-crab-cavities}-d) traps the fundamental mode while allowing HOMs to leak through large-diameter beampipes. This decouples the damper from the cavity, simplifying fabrication\cite{SRF21_1} . A waveguide-coax damper design was proposed, and simulations confirmed impedance compliance and mechanical robustness. A cryomodule with 5 WOW cavities fits within the available space.
\subparagraph{2.6 GHz Quasi-Waveguide Multicell Resonator (QMiR)}
Developed by Fermilab, QMiR (Figure~\ref{fig-all-crab-cavities}-e) features open ends and integrated beam vacuum chambers, eliminating the need for dedicated HOM couplers. An early-stage technology prototype  demonstrated a 2.6 MV transverse kick in tests\cite{IPAC14_1} . The design was adapted for ILC with a 25 mm aperture and 2.6 GHz frequency. Simulations confirmed HOM suppression and mechanical tunability. The cryomodule design meets ILC requirements\cite{SRF23_2}.
\subparagraph{Niobium Material Considerations}
Three types of niobium are considered for ILC crab cavity fabrication\cite{SRF21_2}\cite{SRF21_3}\cite{snowmass}:
\begin{itemize}
\item Fine-Grain (FG): Uniform but costly and prone to contamination in rolling process.
\item Large-Grain (LG): Clean but mechanically inconsistent.
\item Medium-Grain (MG): Clean and uniform mechanical properties at lower cost.
\end{itemize}
MG Nb, developed collaboratively by KEK, JLab, and ATI, was deemed to be the preferred solution for its balance of performance and manufacturability.
\subparagraph{Down-Selection Outcome}
A review held at KEK in April 2023 evaluated all five designs based on compliance, maturity, HOM mitigation, RF ancillaries, mechanical stability, manufacturability and integration feasibility\cite{LCWS2023}. The two selected designs for prototyping are:
\begin{itemize}
\item 1.3 GHz RF Dipole (RFD) – Old Dominion University/JLab
\item 2.6 GHz QMiR – Fermilab
\end{itemize}

\paragraph{Status}
The ILC crab cavity development has successfully narrowed down to two promising technologies: {\bf the RF Dipole} and {\bf the Quasi-Waveguide Multicell Resonator}, ready for prototyping and validation. These designs meet the operational demands of the ILC across its energy range and offer scalable, cost-effective solutions for future implementation.
The next step is prototyping and evaluating the cavities, awaiting research institutions with expertise and  resources to participate\footnote{An agreement for prototyping of an RFD crab cavity, based on the conceptual design of ODU, is under discussion between CERN and STFC (UK).}

\newpage
\subsection{Electron and Positron Sources Area}
\subsubsection{Electron Source (WPP-4)}
\paragraph{Description}
The baseline design of the polarized electron source in the ILC TDR includes the drive laser, a 200 kV DC high voltage photo-gun, and  GaAs/GaAsP photocathodes which provide polarization $>$85\%.  
Hence, the WP4 consists of these three components. Among these the photon gun is the most urgent item.

Jefferson Lab has been developing two ILC prototype guns. They  would meet the requirements of the TDR, providing 4.8 nC bunches within a pulse duration of 1 nsec 
from a laser with diameter of 1 cm at the photocathode. 

However, experience during the past 10 years motivates further improvements to the ILC gun technical design. 
Based upon the inverted insulator geometry, improvements have been made in: 
\begin{itemize}
\item the high voltage triple point junctions, achieving higher operating voltage while maintaining maximum 
gradients $<$ 10 MV/m to prevent field emission; 
\item the cathode-anode geometry to suppress asymmetric fields within the accelerating gap to suppress 
beam deflection and aberration, and; 
\item the vacuum design to achieve extremely high vacuum and limiting ion back-bombardment, required for long photocathode quantum efficiency (QE) lifetime. 
\end{itemize}

Additionally, gun voltages $>$200 kV offer the potential for significant performance improvements. Laser pulse lengths shorter than 1 ns may relax sub-harmonic bunching requirements. Benefits also accrue from reduced ion back-bombardment QE degradation, as the ionization cross section decreases rapidly with electron beam energy.

The proposed work of WP4 over the first 2 year period includes:
\begin{itemize}
\item beam dynamics simulations of shorter than 1 ns, higher peak current bunches that define the allowable 
initial longitudinal and transverse laser pulse shapes,
\item an electrostatic design which maximizes gradient at the photocathode while limiting gradient on the 
electrode surfaces to $<$ 10 MV/m at the operating voltage of $\sim$300 kV, 
\item a triple point junction shield design to linearize the potential along the inverted insulator, 
\item a tilted biased anode design to correct for the asymmetric electrostatic field created by the insulator,
\item vacuum modeling to achieve static vacuum 
$ < 2 \times 10^{-12}$ Torr, and
\item a biased anode design to limit ion back-bombardment from entering the cathode anode gap, to extend photocathode operating lifetime.
\end{itemize}

\begin{figure}[htb]
\begin{minipage}{0.42\textwidth} 
\includegraphics[width=\textwidth]{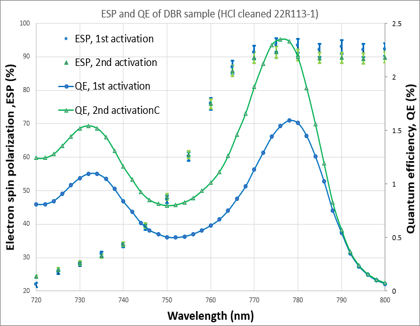}
\end{minipage}
\begin{minipage}{0.58\textwidth}
\includegraphics[width=\textwidth]{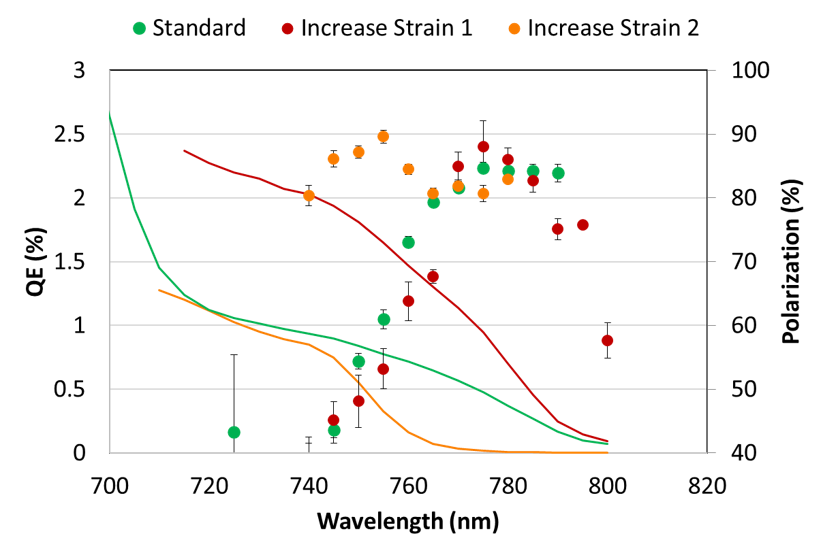}
\end{minipage}
\caption{Left: Polarization and QE from GaAs/GaAsP cathode with DBR. QE: 2.3\%, Pol: 92\% @780nm. 
Right: Polarization and QE from AlGaAs/InAlGaAs cathode (red points). 
\label{fig:photocathode}}
\end{figure}

\paragraph{Status}
There have been progresses in the area of high-voltage insulator and photocathodes.
\begin{itemize}
\item Development of commercialized insulators (KEK, Kyocera and JLAB)\cite{pes-insulator}\\
Compact inverted insulators are ideal for mating a photogun to the high voltage power supply via a cable. 
This design
reduces the electrode surface exposed to high voltage compared with cylindrical insulators. KEK received funding for Kyocera to manufacture a test insulator with low secondary electron yield and it will be high-voltage tested at JLAB.
\item Development of strained GaAs/GaAsP cathode (ODU(Old Dominion Univ.) and JLAB)\cite{pes-GaAscathode} 
A strained GaAs/GaAsP cathode was successfully grown by MOCVD (Metal Organic Chemical Vapor Deposition) with and without DBR (Diffracted Bragg Reflector). 
Polarization $>$90\% was obtained with QE $>$1\% (without DBR) and $>$2\% (with DBR).
The bunch charge reached 9 nC with an 8 mm diameter laser.
(see Figure.\ref{fig:photocathode}, left)
\item Development of strained AlGaAs/InAlGaAs cathode grown by MBE. (UCSB and JLAB)\cite{pes-AlGaAscathode}\\
InAlGaAs has potential advantage 
over GaAs/GaAsP in the availability in vendors and in the independent control of strain 
(polarization) and band gap (wavelength).
Best result showed polarization 88$\pm$4\% with QE 0.95\%.
(see Figure.\ref{fig:photocathode}, right)
\end{itemize}
\subsubsection{Undulator-scheme Positron Source (WPP-6/7)}
\paragraph{Description}
%
\subparagraph{Overview}
The Undulator-based positron scheme provides a mature, reliable source with a rather large margin w.r.t. the damping ring acceptance, fully matching the ILC requirements. The baseline scheme consists of an undulator with 
231 m 
active length and an electron drive beam of 126.5 GeV at 5 Hz
which is also capable of
10 Hz-running for high luminosity. 
\\
A specific GigaZ running mode with a repetition rate of 3.7+3.7 Hz~\cite{Moortgat-Pick:2024fcy} is foreseen. 
The undulator-based positron source allows for
the generation of polarized positrons that would be of high physics impact, leading to significant control of systematic uncertainties, better statistics and offers additional new observables only accessible with simultaneously polarized electron and positron beams~\cite{Fujii:2018mli}.

\begin{figure}[h]
    \centering
    \includegraphics[width=0.35\textwidth,height=0.32\textwidth]{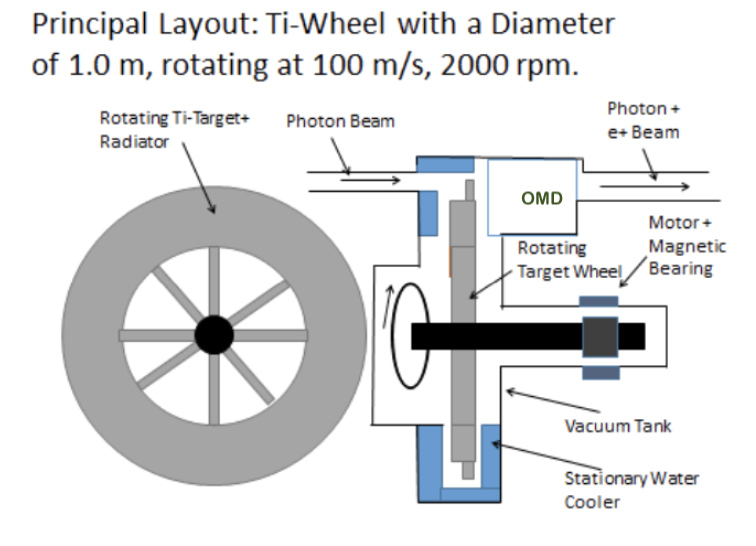}
    \hspace{1cm}
    \includegraphics[width=0.3\textwidth]{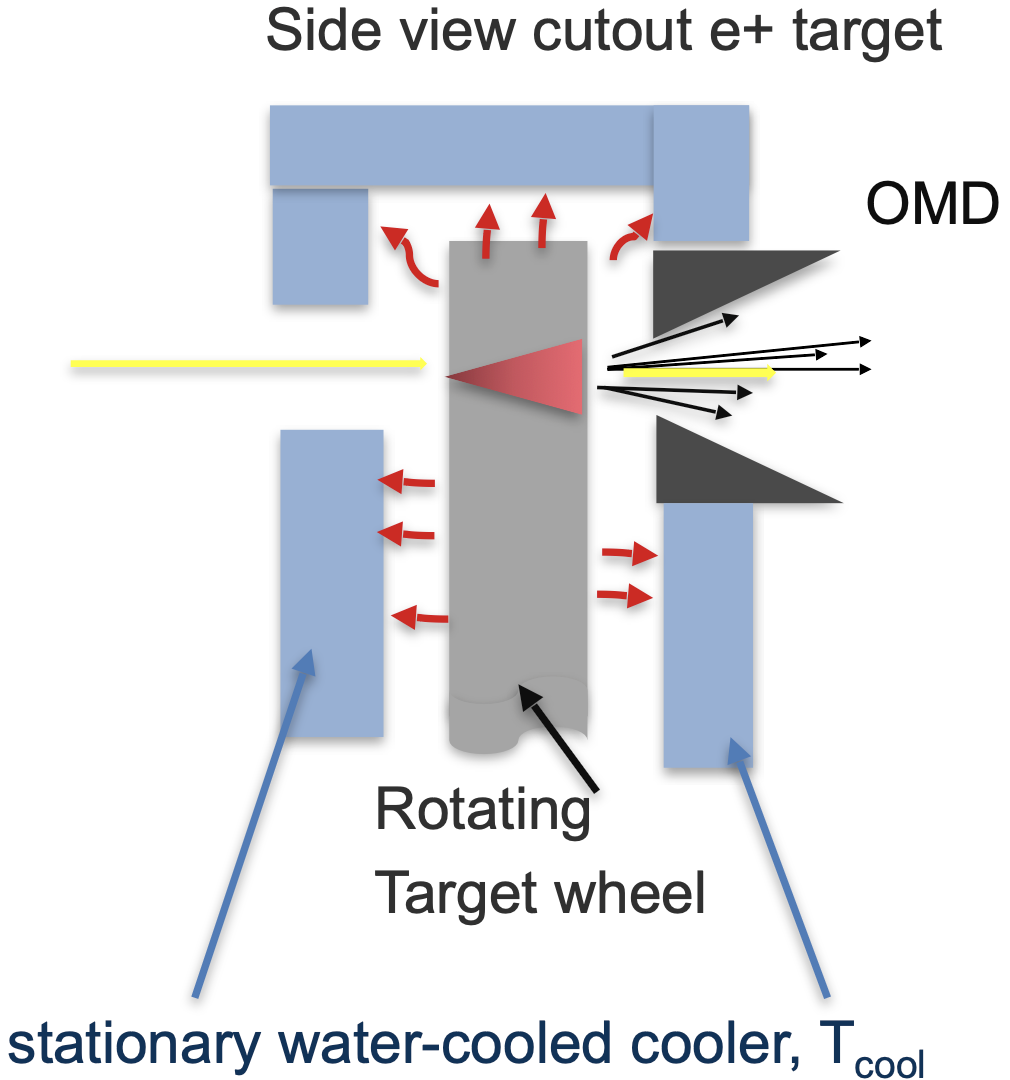}    
    \caption{Overview  of target wheel set-up: rotating in vacuum; supported by magnetic bearings; 
    embedded in a
stationary water cooler (left panel); picture detail of the target rim where the photon beam hits the target wheel (right panel).
}
    \label{und-fig-rootwheel}
\end{figure}

\begin{figure}
    \centering
    \includegraphics[width=0.9\textwidth]{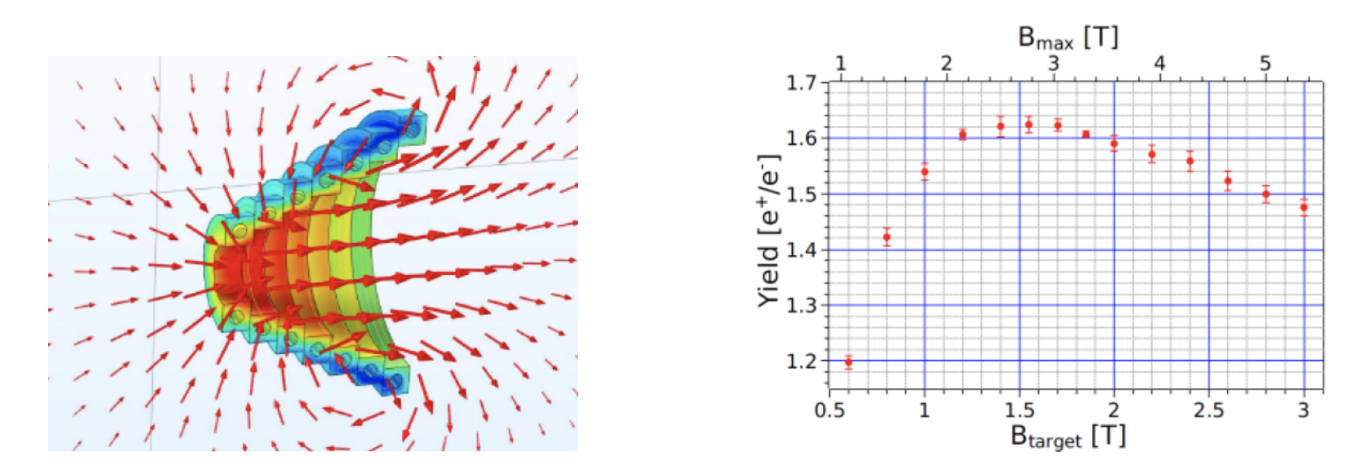}
    \caption{Pulsed solenoid fields, half-sine current pulse with 4 ms and a peak current of 50 kA
and peak field of 5.3 T.  Surface: Magnetic flux density norm (T), Arrow volume: Magnetic flux density
(spatial frame), generated with COMSOL-Multiphysics Code (left panel); Expected positron yield
depending on the field $B_{\rm target}$ at the target exit and the maximum field $B_{\rm max}$ (right panel)~\cite{Moortgat-Pick:2024gbw}.}
    \label{ps-eps23-fig5}
\end{figure}

\begin{figure}
    \centering
    \includegraphics[width=0.9\textwidth,height=0.25\textwidth]{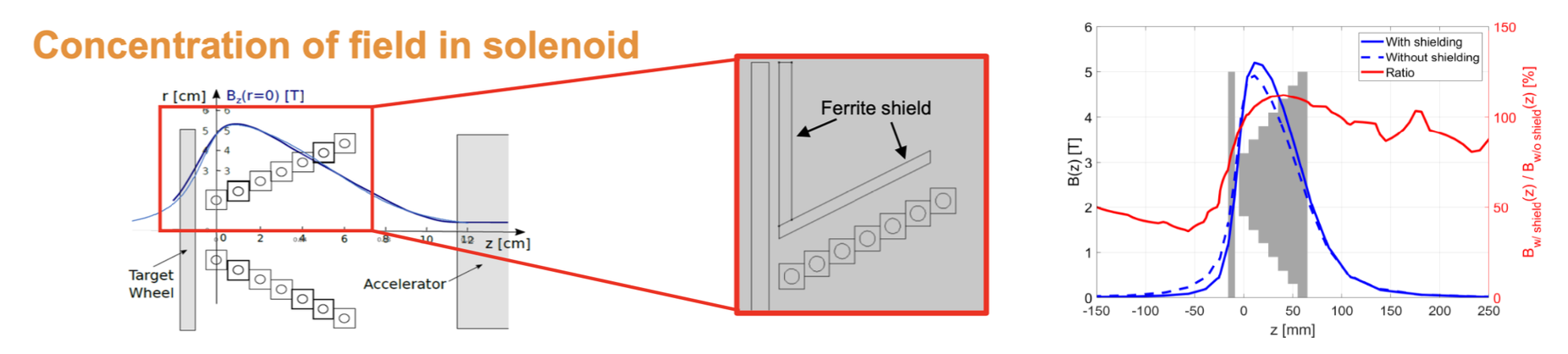}
    \caption{Ferrite shielding inside the pulsed solenoid, 2D and 3D simulations with COMSOL
including the rotating target (100 m/s) and a peak current of 45 kA (left panel). Resulting magnetic
field with and without shielding. The induced heat as well as the peak force are reduced by more than a factor
two with shielding, the peak field is slightly increased by $B_z=10\%$ (right panel)~\cite{Moortgat-Pick:2024gbw}.}
    \label{ps-eps23-fig6}
\end{figure}

The main two work packages within ITN are, workpackage WPP-6 'Rotating Target' and WPP-7 'Magnetic Focusing'.
A wheel with 1m-diameter is rotating at 2000 rpm (i.e. 100 m/s tangential speed) and is cooled via radiation 
into a stationary water-cooled cooler
(see Figure~\ref{und-fig-rootwheel}). 
Due to the long bunch train length an optic matching device 
(OMD)
matching  the flat-top pulse width (0.726 ms for 1312 bunches) is needed. 
Therefore the current flux concentrators can not be used. Instead a pulsed solenoid is foreseen (see Figure~\ref{ps-eps23-fig5})
,
generating a half-sine current pulse of about 4 ms and a peak field of 3-5~T at the target,
(see Figures ~\ref{ps-eps23-fig5} and \ref{ps-eps23-fig6}.
This design will
achieve a yield of Y$\ge 1.5 e^+/e^-$ (see Table.~\ref{tab-ps} ) ~\cite{Moortgat-Pick:2024gbw}.  

An active plasma lens OMD---generating a tapered azimuthal magnetic flux in the plasma---is discussed as a novel application that has a high potential for improving the yield further  and for collecting the highly divergent positrons.
A downscaled prototype of a tapered plasma lens is already under work at the University of Hamburg in collaboration with DESY and the results will also be finalized within the ITN period.

 The efficiency of the source, however, 
 also depends 
 strongly on the undulator design and on the target material performance. On all the above mentioned topics decisive progress has been made. 

\begin{table}[b]
    \centering
    \includegraphics[width=0.6\textwidth]{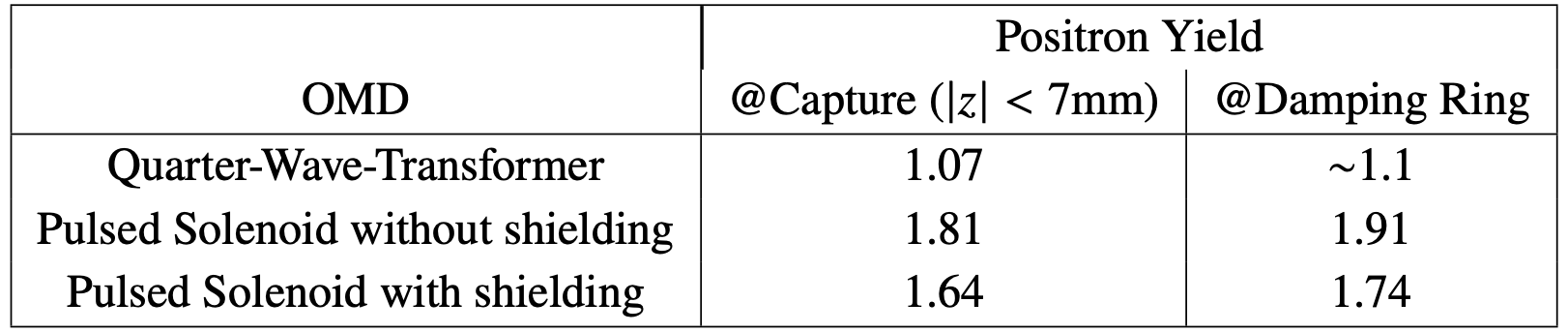}
    \caption{Expected positron yield at the undulator-based positron source for different optic matching devices, the Quarter-Wave Transformer(QWT) and the Pulsed Solenoid w/wo the ferrite shielding~\cite{Moortgat-Pick:2024gbw}.}
    \label{tab-ps}
\end{table}

\begin{figure}[ht]
    \centering
    \includegraphics[width=0.5\textwidth,height=0.2\textwidth]{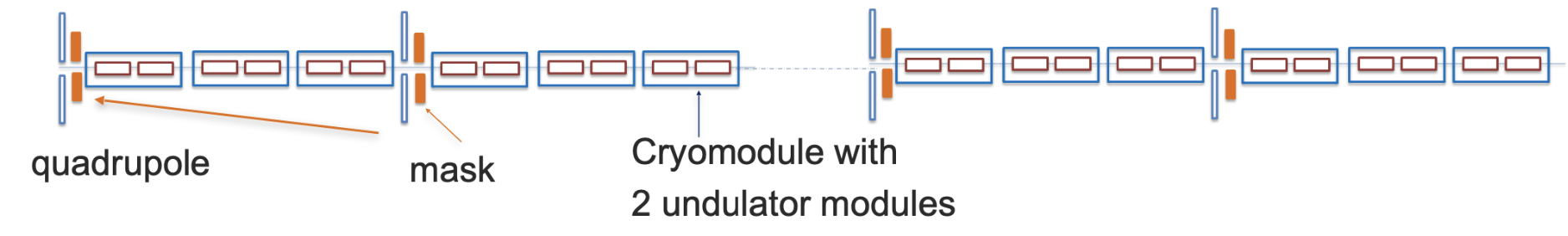}
    \caption{Design of the mask system to protect the undulator walls, keeping the incident power of photons
(including secondary particles) below the acceptable limit of 1 W/m~\cite{Moortgat-Pick:2024fcy}.}
    \label{und-fig-mask1}
\end{figure}

\begin{figure}[ht]
    \centering
    \includegraphics[width=0.8\textwidth]{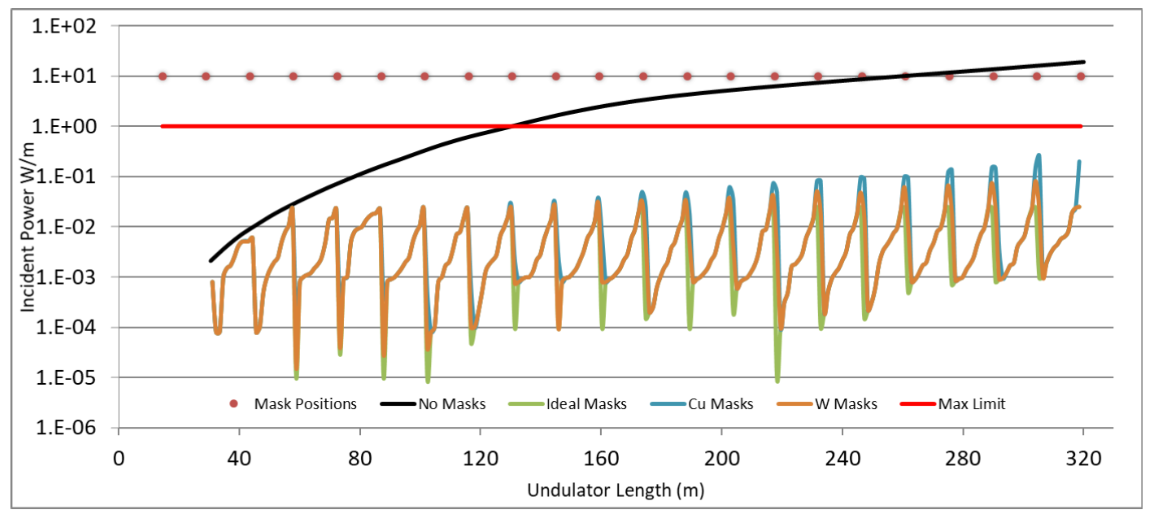}
    \caption{Impact of implementing the masks (allowed limit of 1W/m assumed(redline)) for protecting the
undulator walls for the ILC250. Different materials iron, copper, tungsten have been used~\cite{Moortgat-Pick:2024fcy}. 
}
    \label{und-fig-mask2}
\end{figure}

\subparagraph{Undulator Design}
%
There has now been about 10 years operating experience with long planar undulators at XFEL including stable operation with  a beam alignment up to 10-20 microns
for 200 m undulator length~\cite{Moortgat-Pick:2024fcy}. This is remeasured every 6 months. During beam operation the beam trajectory is controlled better than 3 microns both with slow and fast feedback systems. The beam requirements at XFEL are even more tighter than at ILC due to the FEL 
requirements on the photon beam.
The ILC undulator is helical with a K-value of about 0.92. 
Simulations 
including
a non-ideal field and misalignments have been performed.
A mask system
has been designed
 Figure~\ref{und-fig-mask1}
 to protect the undulator walls from synchrotron radiation 
absorption~\cite{Alharbi:2022okv}, as demonstrated in Figure~\ref{und-fig-mask2}.
\subparagraph{Target Material Performance}
The incident photons have an energy of about 7.5 MeV (1st harmonic) and hit every 7-8 s on the same position at the target, i.e.\ in 5000 h roughly 5$\times 10^6$ load cycles.
For the  250~GeV CMS operation, a target thickness of 7mm (0.2 radiation length) of Ti-alloy grade-5 is foreseen.
It is expected to result in a $\Delta T_{\rm max}=60$-120~K per pulse. 
Thus the maximum energy deposition  of about 60  J/g
is expected 
with an effective pulse length on the material of about 25-55 $\mu$s and effective repetition rate of 0.17 Hz causing annual displacement per atom (dpa) of about 0.3-0.5 per year~\cite{Lengler:2024znl}.

\subparagraph{WPP-6: Rotating Wheel Design}
The Titanium-target wheel with 1~m diameter is rotating at 2000 rpm and has two major components: the cooling system  via radiation, and
 a rotation system turning on magnetic bearings.
 Magnetic bearings
are standard components to support elements rotating in vacuum, allowing long time operations high rotation speed without major maintenance. The whole system sealed in an UHV-tank.
The deposited thermal power will be about 2~kW, accumulated close to the rim of the target. Therefore the heat deposition can be removed via thermal radiation from the spinning target to a stationary water-cooled Cu-cooler following 
$~T^4_{\rm radiator}-T^4_{\rm cooler}$, keeping $T_{\rm ave}<460^0$~C.

Within the ITN work package WPP-6, it is planned to 
produce
mechanical drawings and to
generate
the final engineering design for the complete wheel system. Tests of a partial model and a downscaled prototype are foreseen by University of Hamburg in collaboration with DESY and CERN.
\subparagraph{WPP-7: Magnetic Focusing System}
Pulsed Solenoid: Matching of the flat-top pulse of about 1~ms can be achieved with a Cu pulsed solenoid generating a half-sine current pulse with 
4~ms duration so that at the peak of the half-sine pulse eddy currents have died out and a stable field over the duration of the beam pulse is achieved. In order to achieve multi-Tesla fields, peak currents
of about 50 kA or above are required. During the pulse a heat load on the target is induced, generating
a force on the rotating target. Implementing a ferrite shield
inside the pulsed solenoid will only slightly reduce the peak magnetic flux but significantly reduce
the heat load (by more than a factor two) as well as the peak force on the rotating wheel. A yield of about $Y = ~1.7$ 
has been achieved in 2D- and 3D-simulations with COMSOL after implementing such a ferrite shield.

Within the ITN work package Wpp-7, fabrication and tests of the pulsed solenoid prototype are performed by University of Hamburg in collaboration with DESY. Field measurements with 1kA (pulsed and DC) and with 50kA both in a single pulse mode and finally with a pulse duration of 5ms at 5 Hz are intended to be performed in final tests at CERN and DESY.

\paragraph{Status}
The Institutes and Labs currently involved in the undulator-based positron source are CERN, DESY, University of Hamburg, Helmholtz-Zentrum Hereon Geesthacht, Johannes-Gutenberg University of Mainz, Helmholtz Institute Mainz, SKF Magnetic Bearings Canada, Université Paris-Saclay Orsay, Thomas Jefferson National Accelerator Facility, Newport New, US.

%
\subparagraph{Overview}
The 
results 
achieved to date
and 
the developments underway 
show that the undulator-based positron is feasible to
match the ILC positron source requirements. The ITN work packages 
mainly concern 
engineering issues and prototyping test work is of high demand to guarantee the timely readiness of the components. The undulator design for the ILC250 has been finalized, the target tests have shown that the targets will stand the load, see Table~\ref{und-fig-target1}:
the targets did not show any material damage even under a multiple of the annual expected PEDD.
Also the luminosity upgrade is feasible with the chosen Ti-alloy grade-5 target material.
\begin{table}
    \centering
    \includegraphics[width=0.7\textwidth]{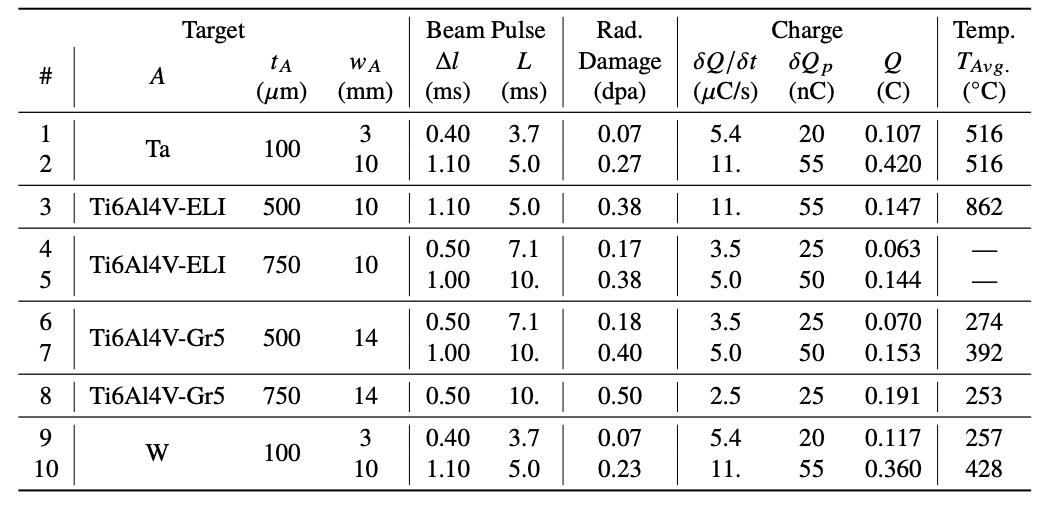}
    \caption{Material Irradiation Specifications: Positron source candidate materials of different composition
($A$), thickness ($t_A$), and width ($w_A$) have been exposed to
the electron beam: titanium alloys, pure tantalum, and pure
tungsten~\cite{Lengler:2024znl}.}
    \label{und-fig-target1}
\end{table}

Concerning the ITN work packages, a) WPP-6 Rotating Wheel and b) WPP-7 Magnetic Focusing System, the design of suitable prototype tests as well as  measurements of first (down-scaled) focusing sysytem prototypes are expected soon. Industrial contacts for magnetic bearings has been established.

\subparagraph{Engineering Design and Prototype Status of pulsed Solenoid}
Both mechanical 
and
CAD drawings have been 
produced
for the pulsed solenoid design. In order to achieve 
5 Tesla fields,
peak currents
of  50 kA or above are required.
Both 2D and 3D detailed simulations have been performed with COMSOL
w/wo the 
Titanium plate moving at the speed of 100 m/s, and a peak current of 45 kA. 
Implementing a 
ferrite shield reduces significantly the induced
heat load by more than a factor two, but only slightly reduces the peak magnetic field. 

The yield of the undulator-based source with the pulsed solenoid as OMD has been simulated
up to the damping ring (DR).
It has been shown to have a 
significant yield increase in comparison with  
the
quarter-wave-transformer
(QWT) as the OMD  (see Table~\ref{tab-ps}).

The stress in the coil of the pulsed solenoid has also been simulated.
An average power of
about 10kW is expected for a peak magnetic flux of 4.6T. The von-Mises stress in the coil amounts
to about 570 MPa and provides a conservative estimate on the expected stress. However, the soft
tensile strength of copper is about 200 MPa. The actual stress in the coil is very sensitive to the
exact shape. Therefore the mechanical design including the solenoid, the support structures and the
connectors 
require further 
iteration.
Even in case the stresses get too high, one could find solutions, for
instance, using multiple layers or reducing the peak current. 

\begin{figure}
    \centering
    \includegraphics[width=0.35\textwidth,height=0.25\textwidth]{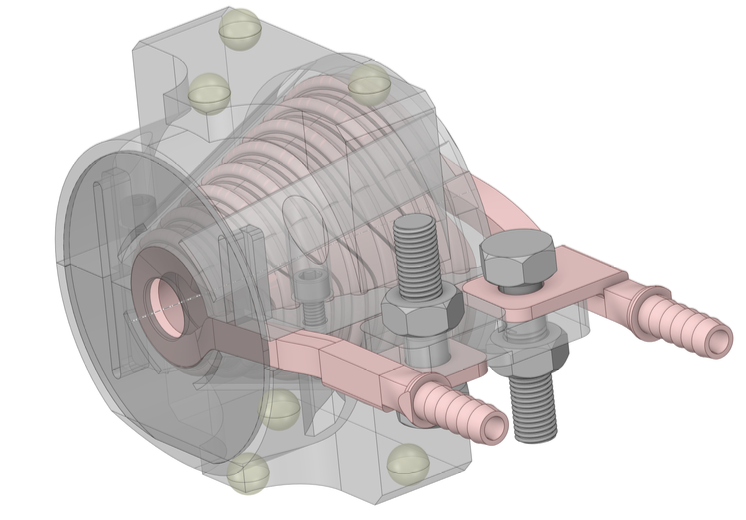} \includegraphics[width=0.35\textwidth,height=0.25\textwidth]{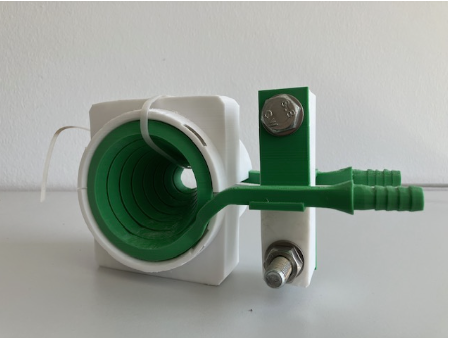}
    \caption{Design of the tapered solenoid coil with 8 windings, metal support bridges and
insulated support rods (left panel). 3d-printed test model (right panel) of the pulsed solenoid design.
    }
    \label{fig-pulsed-solenoid}
\end{figure}

The global optimization is therefore still outstanding 
and
awaita first results from the prototype
of the pulsed solenoid design. Mechanical drawings for such a prototype have been done recently,
(see  left panel Figure~\ref{fig-pulsed-solenoid} ).
It consists of a solenoid coil with tapered planar windings and interconnections.

First down-scaled prototypes was 3D-printed during
initial ITN work (see right panel  Figure~\ref{fig-pulsed-solenoid}).
A copper prototype has been ordered,  delivery is awaited and first measurements will follow.

\begin{figure}[h]
    \centering
    \includegraphics[width=0.35\textwidth,height=0.25\textwidth]{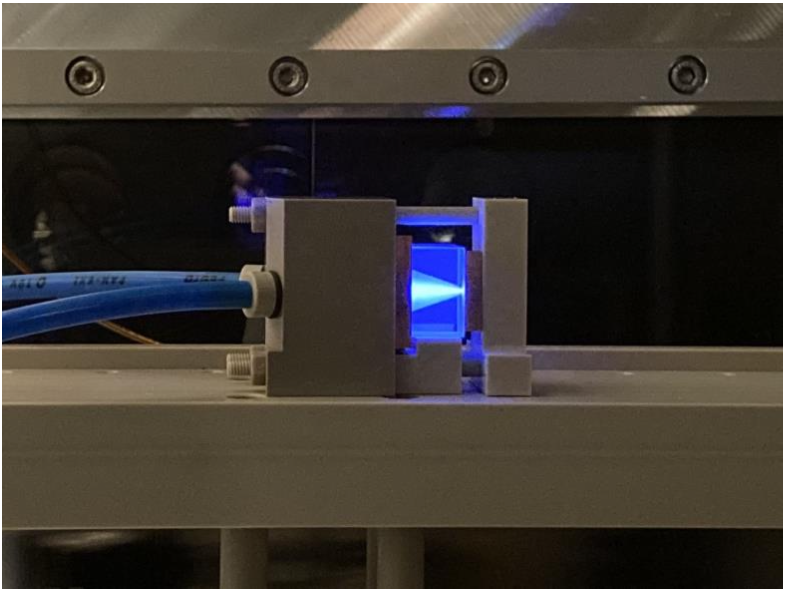} 
    \hspace{2cm}
    \includegraphics[width=0.35\textwidth,height=0.27\textwidth]{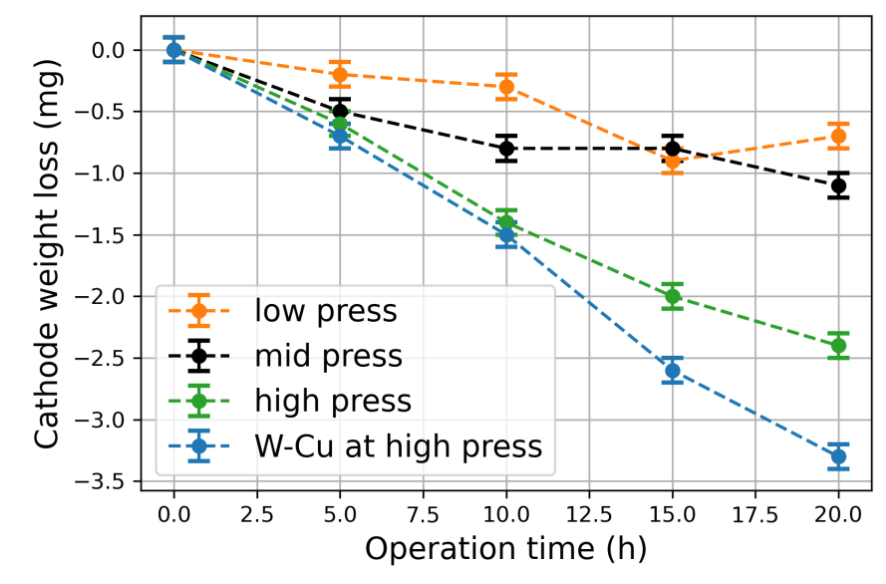}
    \caption{ Scaled-down
    prototype ILC active plasma lens during operation (left panel)~\cite{Moortgat-Pick:2025xve};
    Electrode erosion under different pressure conditions and different materials (right panel)~\cite{Moortgat-Pick:2025xve}.    }
    \label{fig-plasma-lens}
\end{figure}

An automated specialized simulation code enabling Bayesian optimization of a parameterized solenoid geometry has been developed. These geometries are simulated in COMSOL to generate magnetic field distributions, which are then used to track positrons through the initial accelerator structure. The ultimate aim is to refine the solenoid’s shape to maximize positron yield based on the prototype measurements.
\subparagraph{OMD prototype active plasma lens}
A quarter-scale
prototype of an ILC active plasma lens as optic matching device
has been simulated, constructed and
tested in operation (see left panel  Figure~\ref{fig-plasma-lens}).
The experimental tests are ongoing. Plasma discharge has been observed after around 90 minutes of operation. However, its exact source 
is still unknown.
Studies of different electrode materials, for instance W-Cu, under different conditions of pressure, etc., are ongoing ( see right panel Figure~\ref{fig-plasma-lens}~\cite{Moortgat-Pick:2025xve}.

\begin{figure}
    \centering
    \includegraphics[width=0.9\textwidth]{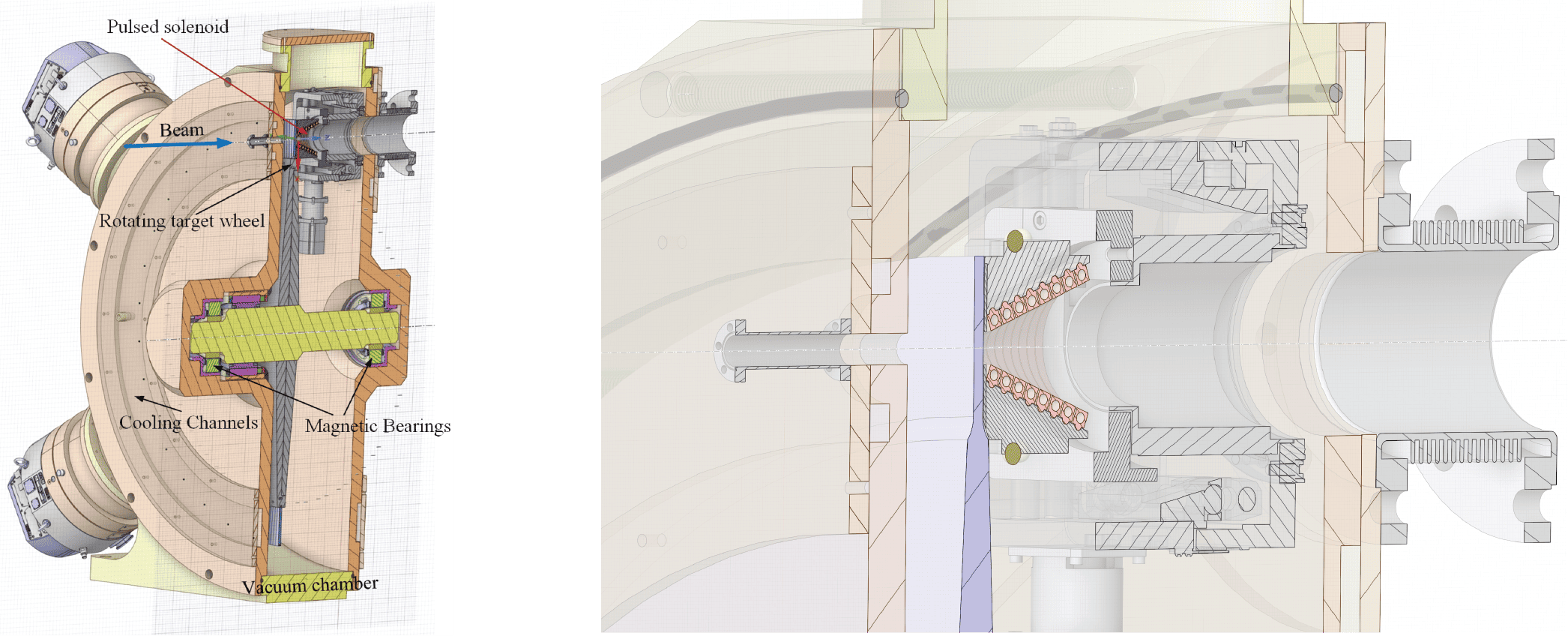}
    \caption{Design of the rotating wheel based on ANSYS simulations including magnetic bearings, pulsed solenoid and vacuum vessel (left panel); Picture detail of the pulsed solenoid device (right panel).}
    \label{fig-rotwheel-engineering}
\end{figure}

\subparagraph{Engineering Design and Magnetic Bearings of Rotating Wheel Prototype}
An ANSYS-based design for the rotating wheel including the magnetic bearings, the vacuum vessel and the OMD has been performed (see Figure~\ref{fig-rotwheel-engineering}). 

The technical specifications concerning the magnetic bearings have been worked out and the step file has been shared with industrial manufacturer.
A  Letter of Intent concerning the 
magnetic bearing spindles including a control system and matching the full ILC requirements (target weight, speed and tolerances) has been exchanged with the manufacturer within the first ITN period.

In order to pursue
validation of 
the full wheel system 
prototypes 
 commercially 
 available items will be used.
 For example,
scaled down versions of the ILC wheel can be purchased from commercial suppliers,
for testing 
radiation cooling from a heated target.
\subparagraph{Summary and Outlook}
The undulator specifications and the target material tests have been finalized. The undulator is 
feasible and will allow stable running without operational issues based on actual long-time operation experience at XFEL. The precise target analyses performed both 
via
scanning method and by synchrotron diffraction show practically no material damage even under a multiple of the annual expected PEDD. Therefore 
the luminosity upgrade option is also feasible.
A CAD-based engineering design for the full rotating wheel system has been 
completed 
and contact with the manufacturer concerning the magnetic bearings has been established. In the 1st half of 2026 it is intended to identify suitable places for prototype tests. 
Scaled-down
prototype tests are intended to be constructed in the second half of 2026. 
Mechanical as well as CAD-based engineering design for the pulsed solenoid as optic matching device (OMD) has been accomplished.
First measurements of a 
scaled-down 
prototype has been ordered and  delivery is awaited.
Further prototypes and the finalizing of the design based on the current measurements are foreseen for the first half of 2026.
No showstoppers have been identified so far concerning the undulator-based ILC positron source.
\subsubsection{Electron-driven Positron Source (WPP-8-11)}

\paragraph{Description}
\subparagraph{Overview}

Electron driven positron sources are widely used
in
previous and present accelerator facilities.
The operation principle 
uses
simple and proven technology.
However, to achieve 
the
ambitious requirement of ILC, $ 131.2\times10^{12}\, e^{+}/\mathrm{s} $ must be delivered.

Table~\ref{Tab1} summarizes
the
basic parameters of electron driven positrons sources for ILC and SuperKEKB. 
Although care should be taken when comparing the values since the bunch structure is different project by project, considering the positron source for the SuperKEKB,
the
world's most intense positron source in operation, one can understand how demanding the requirement is.

\begin{table}[hbt]
    \centering
    \caption{Comparison of Parameters on Positron Sources for SuperKEKB and ILC}
        \begin{tabular}{lcc}
            \toprule
                Project & SuperKEKB & ILC \\
            \midrule
                $e^-$ beam on target \\
                \quad energy (GeV) & 3.5 & 3 \\
                \quad bunches per pulse & 2 & 66 \\
                \quad repetition (Hz) & 50 & 100 (300) \\
                \quad beam power (kW) & 3.5 & 74 \\
                $e^+$ beam after Dumping Ring \\
                \quad $N_{e^+} /\ N_{e^-}$ & 0.4 & $>$1 \\     
                \quad bunch charge (nC) & 4 & 3.2 \\
                \quad number of positron ($10^{12}e^+$/s) & 2.5 & 131.2  \\ 
            \bottomrule
        \end{tabular}
    \label{Tab1}
\end{table}

The goal of this Work package is to demonstrate
the
feasibility of the engineering design of the electron driven positron source for the ILC by design and prototype production.

\begin{figure}[h]
    \centering
    \includegraphics[width=0.7\textwidth]{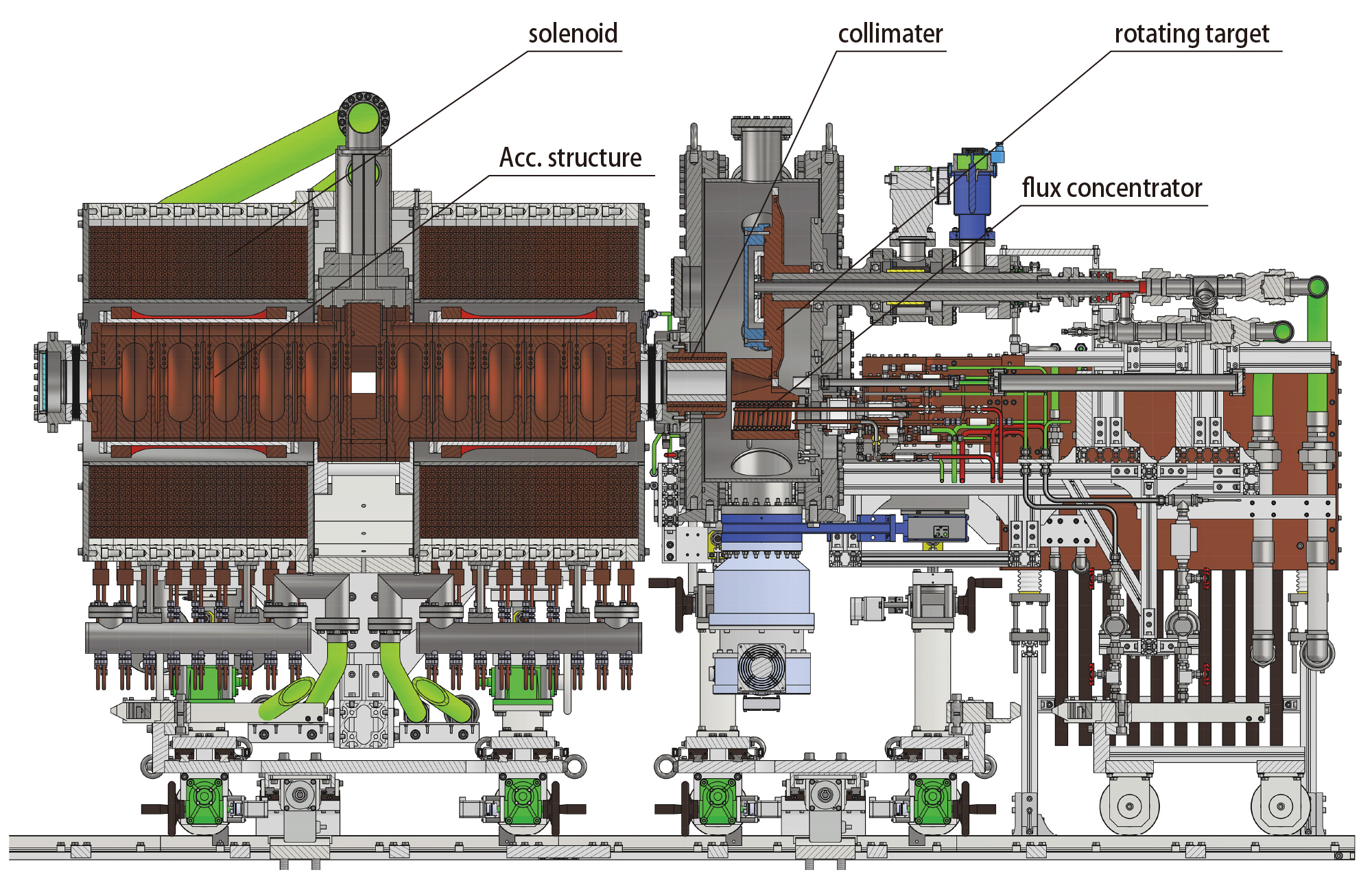}
    \caption{Cross sectional view of the electron driven positron source prototype under development in KEK iCASA.}
    \label{Fig_cross-section}
\end{figure}

Figure~\ref{Fig_cross-section} shows a cross-sectional view of the electron driven positron source currently under development.  
The positron source consists of a target, a flux concentrator (FC), solenoids, and acceleration structure for capturing and accelerating the produced positrons.
As can be seen from the figure, numerous devices are arranged in a narrow space, and even minor layout changes can have a significant impact on performance.
Therefore, in addition to the development of each component, the
overall design is important to ensure the feasibility of the entire system.  
Simulations and performance evaluations should be based on a feasible layout,
to result in a meaningful 
design.

Considering these facts, our strategy to develop a electron driven positron source is as follows.
\begin{itemize}
    \item Complete the overall design before detail designing of each of the components
    \item Adopt a relatively conservative layout to prioritize manufacturability over performance for the first prototype.
    \item Based on the first prototype design, conduct simulations to explore directions for improvement and pursue a strategy to apply these findings to subsequent prototypes.
    \item Use experiences and design of SuperKEKB as much as possible
\end{itemize}

The development goals for each components are as follows.

\subparagraph{Rotating target}
It is expected that energy and power of the electron beam impinging on the target are $\rm 3\,GeV$ and $\rm 74\,kW$ respectively.
Deposited heat is estimated as high as 17$\,$kw.
To satisfy the requirements, a water cooled rotating target compatible with UHV ($\rm 10^{-6}\,Pa$) and high radiation environment is necessary.

\subparagraph{Magnetic focusing}
A flux concentrator (FC) is a device that generates a gradually decreasing magnetic field by a pulsed current through a conductor with a tapered inner surface, which has been widely used in previous positron sources.
The positron beam generated at the target has a large angular spread.
In order to fit it into the acceptance of the downstream accelerating tubes, it is necessary to exchange the position spread with the angular spread.
The FC is a matching device which performs such a conversion.  
Since energy acceptance depends on the magnetic field, the design goal is to capture positrons over a wider energy range by using a stronger magnetic field.
Table~\ref{FC-param} shows parameters of FC for SuperKEKB and ILC.

\begin{table}[hbt]
    \centering
    \caption{Comparison of Parameters of FC for SuperKEKB and ILC}
    \begin{tabular}{lccc}
        \toprule
            & unit & SuperKEKB & ILC \\
        \midrule
            peak voltage & $\mathrm{kV}$ & 20 & 20 \\
            peak current & $\mathrm{kA}$ & 12 & 35 \\
            repetition & $\mathrm{pps}$ & 50 & 100 \\
            pulse width & $\mathrm{\mu s}$ & 6 & 11 \\
            entrance aperture & $\mathrm{mm}$ & 7 & 12 \\
            peak magnetic field & $\mathrm{T}$ & 3.5 & 5 \\
            peak power & $\mathrm{MW}$ & 240 & 700 \\
            average power & $\mathrm{kW}$ & 12 & 128 \\
            Ohmic loss on FC & $\mathrm{kW}$ & 0.8 & 8.9\\
        \bottomrule
    \label{FC-param}
    \end{tabular}
\end{table}

The
requirement of the positron capture efficiency for the ILC is 2.5 times higher than that of SuperKEKB.
Stronger magnetic fields and wider apertures are necessary.
In addition, due to bunch structure and operational requirements, the repetition rate is $100\,$pps, which is twice that of SuperKEKB.  
Combining these factors,  
the
expected heat load due to Ohmic loss is more than an order of magnitude greater than that of SuperKEKB.

\subparagraph{Capture cavity}
In order to capture and accelerate the positrons generated by the target as efficiently as possible, it is necessary to install the cavity as close to the target as possible.
On the other hand, the thermal load caused by the shower from the target increases as the cavity is moved closer to the target, and the thermal effect varies greatly depending on whether the beam is on or off.
In this design, the cavity just after the target is assumed to have an input heat of approximately 20 kW due to the shower, and the cooling structure is designed so that the resonance frequency deviation remains within the allowable range even under these conditions.
Higher acceleration gradient is desirable for quickly bunching positron beams with a large energy spread.
A
large aperture is necessary to improve capture efficiency. 
A certain group velocity is required to compensate for beam loading by multi-bunch beams. 
In addition, since it is installed in a solenoid, 
constraints 
on the external shape and support
must be met.

\subparagraph{Target replacement}
After beam operation, the target will become radioactive.  
Considering that the beam power hitting the target is approximately 20 times that of SuperKEKB, it is expected that the intensity of radioactivity will also increase by roughly the same ratio, which makes it difficult to work around the area for long time.
In addition, the positron source will be installed deep underground, severe restrictions on access from above is expected.  
Under these conditions, a mechanism capable of handling failures and regular replacements is necessary. 
In this prototype development, several critical mechanisms will be designed and evaluated.

\begin{figure}[h]
    \centering
    \includegraphics[width=0.55\textwidth]{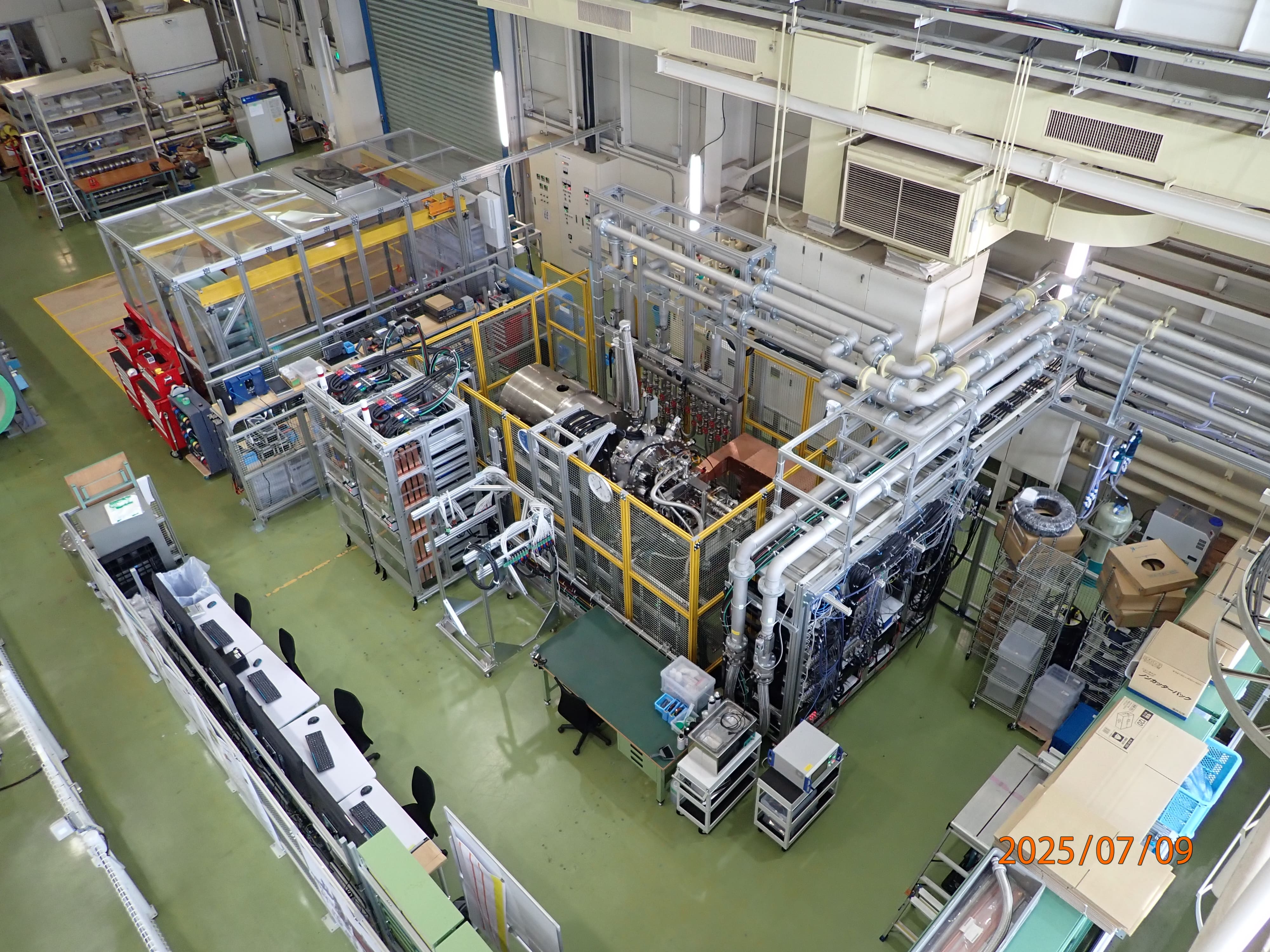}
    \caption{Overview of the experimental area in July 2025.}
    \label{test-bench-photo}
\end{figure}

\paragraph{Status}
\subparagraph{Overview}
Current prototype design and manufacturing started in 2022 and 2023 respectively.
All the components except for the 
accelerating structure in the Figure~\ref{Fig_cross-section} have been delivered and assembled.
Power supplies for the solenoids and control, monitor system except for  pulsed power supply for the flux concentrator have also installed and commissioned.
We are focusing on the development of the two missing components (accelerating structure and pulsed power supply for the FC) along with the evaluation of delivered components presently.
Figure~\ref{test-bench-photo} shows the snapshot of the prototype and test bench in July 2025.

\begin{figure}[t]
    \centering
    \includegraphics[width=0.6\textwidth]{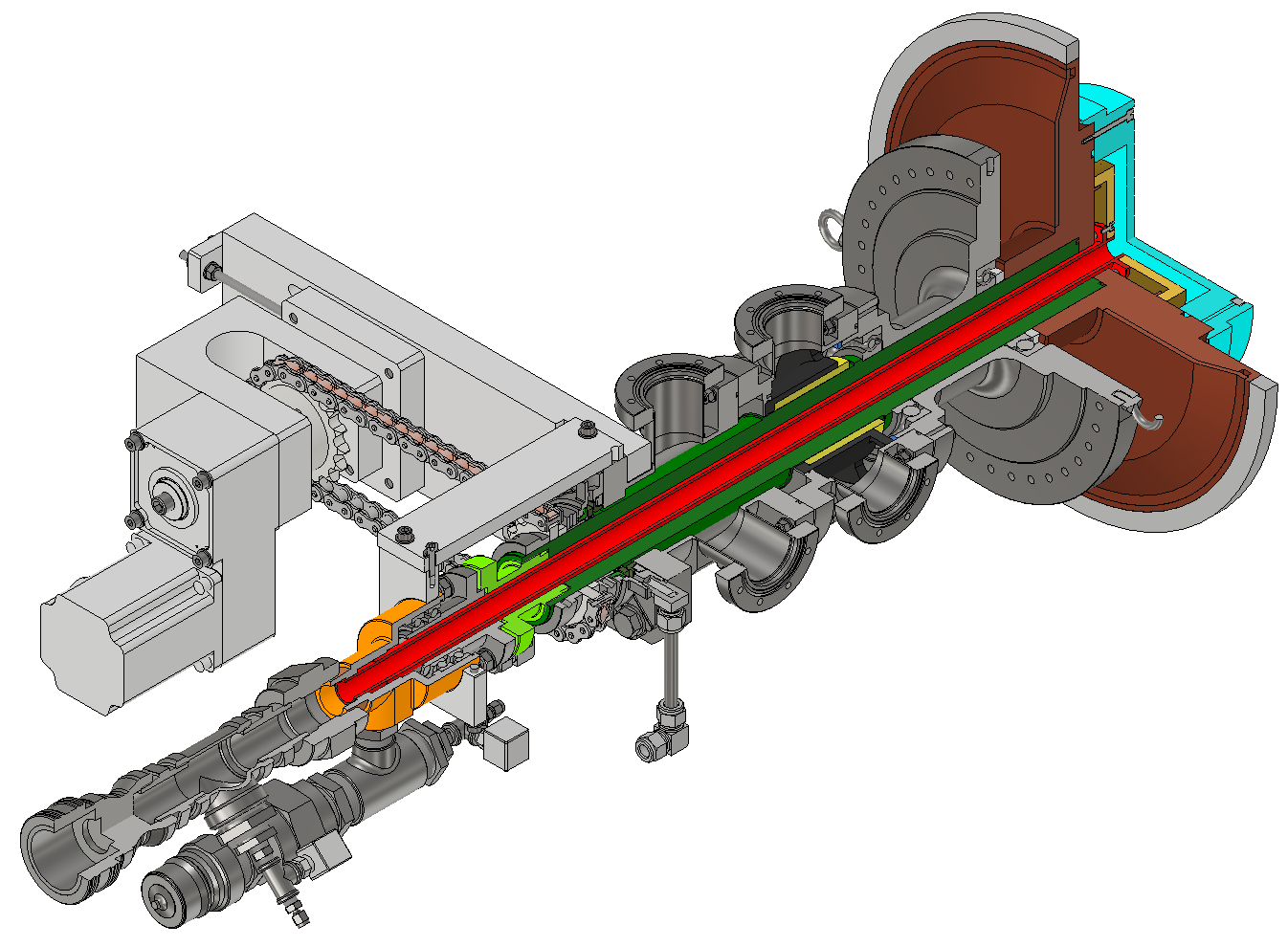}
    \caption{Cross sectional view of the rotating target.}
    \label{Fig_targte-cross}
\end{figure}

\begin{figure}[b]
    \centering
    \includegraphics[width=0.6\textwidth]{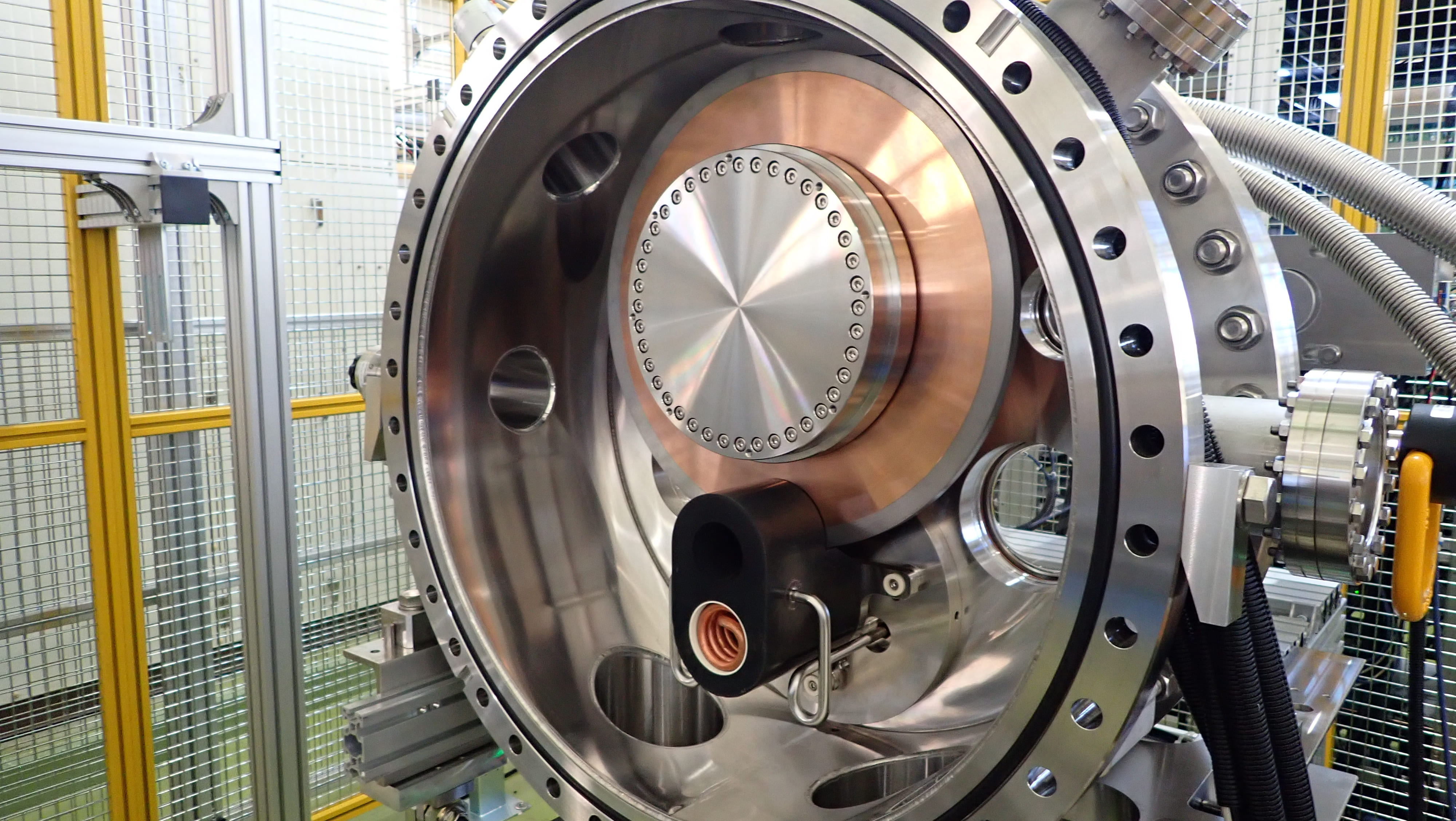}
    \caption{The rotating target assembled in the vacuum chamber with the flux concentrator.}
    \label{Fig_target-photo}
\end{figure}

\subparagraph{Rotating target}
Figures \ref{Fig_targte-cross} and \ref{Fig_target-photo} show cross sectional view of the 3D model and photo of the rotating target assembled in the vacuum chamber respectively.

Rotating mechanism consists of a two-stage differential pumping system with extremely small gaps and a SiC mechanical face seal are adopted \cite{pasj2023_3, pasj2024_3}.
A chain transmission system is used for easy adjustment of the motor position.
The motor is a brushless DC motor (400 W, 3500 rpm) integrated with a gearbox with a reduction ratio of 1:15. 
It realises rotation speed of 233 rpm at the disk axis, slightly higher 
than
the design rotational speed of 225 rpm.

The target disk consists of a tungsten ring with an outer diameter of 500 mm, an inner diameter of 460 mm and a thickness of 15.7 mm,  and a copper alloy disk with a water cooling channel to hold and cool the ring.
The tungsten ring and copper alloy disc are joined by cold pressing using liquid nitrogen.  
Thermal contact at the joint surface is an issue.
Evaluations of thermal resistance of the contact surfare are in progress through both simulation and experimentation.  

The rotation mechanism and the target disk were manufactured in 2023 and 2024 respectively.
Evaluation from
from the 
vacuum, thermal, mechanical perspectives
are in progress in 2025

The ultimate pressure was measured by cold cathode ion gauges, and the residual gas composition was analyzed by a residual gas analyzer (RGA).  
After approximately one week of evacuation, the pressure reached $1\times10^{-5} \,$Pa or less.  
The main component of the residual gas was water.
The atmospheric component, nitrogen, contributed less than one-tenth. 
No difference observed between rotating and non-rotating conditions.
It is considered that the use of rubber O-rings in some places 
to reduce cost
 and ease of testing, 
and the 
the lack of
a
baking-out process of the chamber, are 
the
main limitations
on the final pressure and residual gas composition.  
Even at the current pressure, there are no issues with actual operation.  
Furthermore, based on the atmospheric pressure, it can be said that the sealing performance is sufficient.  

\begin{figure}[t]
    \centering
    \includegraphics[width=0.6\textwidth]{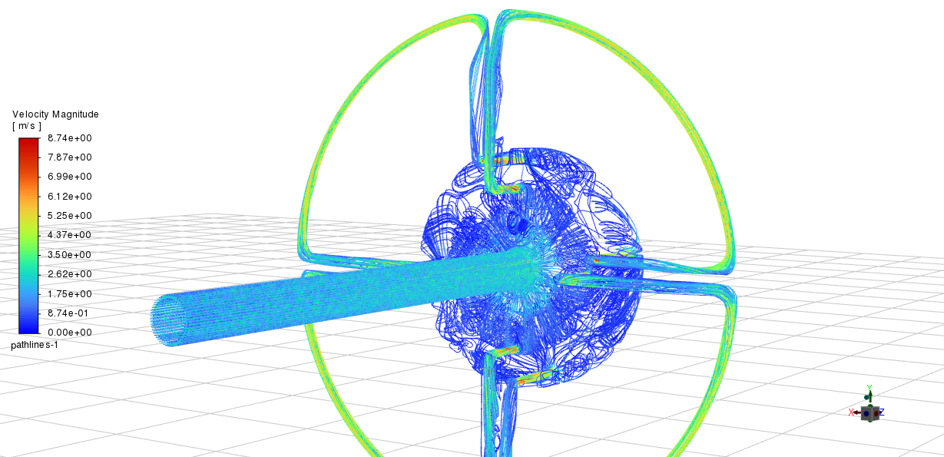}
    \caption{CFD result of water flow in the channels.}
    \label{CFD}
\end{figure}

\begin{figure}[b]
    \centering
    \includegraphics[width=0.8\textwidth]{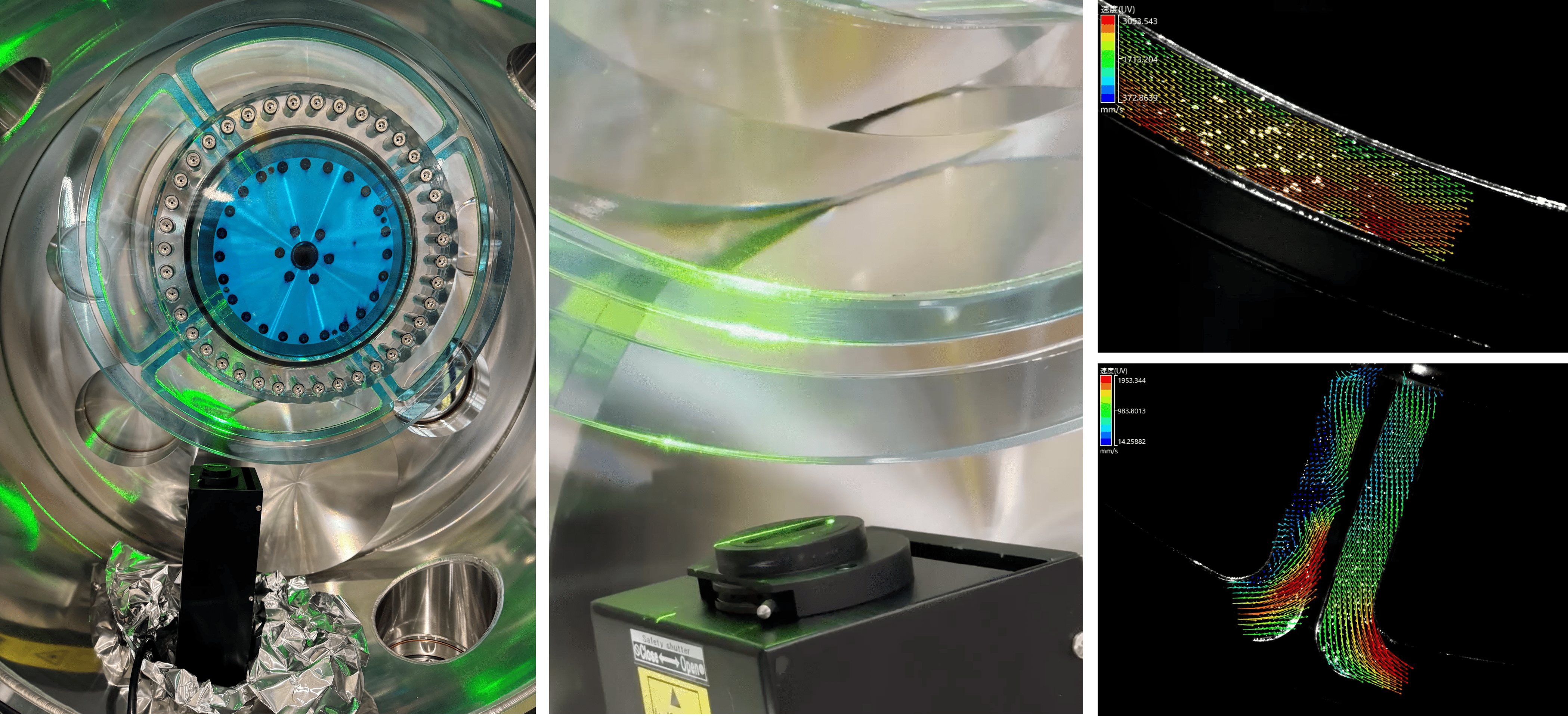}
    \caption{Setup of PIV experiment (left and center) and visualize flow speed map (right).}
    \label{PIV}
\end{figure}

Water flow speed in the channel, which is an important parameter to determine heat transfer efficiency from the target to the cooling water, was evaluated both by simulation and experiment.
Figure~\ref{CFD} shows CFD simulation result. The simulation was done by Jefferson Laboratory under collaboration of the US-Japan science and technology cooperation program in high energy physics.
Figure~\ref{PIV} shows Particle Image Velocimetry (PIV) experiment setup and visualized flow speed.
A full scale transparent target model made of Polycarbonate was used for the test.

To simulate the heat load by the beam, a copper block jigs were attached to the tungsten disc, and a cartridge heater was inserted into them for thermal tests.
The temperature distribution was measured by an infrared camera.
The heat flux and thermal resistance of the contact area were evaluated.
For a heater power of 20 kW, it was estimated that approximately 13.5 kW was transferred to the target disk based on the cooling water flow rate and temperature rise of the rotating target.
The thermal resistance of the contact area was consistent with the calculated value within an order of magnitude. 
Based on the tests conducted so far, we believe that there are no practical issues with the joint, but the expected thermal load from the beam is approximately 17 kW, and we plan to test with heat input exceeding this value in the future.

Torque measurements were performed using the output power monitor of the motor drive.
The torque was almost constant at approximately 9 N·m regardless of the rotational speed.  
The continuous rating of the motor used is approximately 12 N·m, and it was confirmed that there are no issues with continuous operation.  
The result that the torque is constant regardless of the rotational speed 
implies
there 
is
no 
rotational speed dependence of friction.

The gap between the target disk and the FC is currently designed to be $2\,$mm.  
If this distance changes, it will have a significant impact on the positron capture efficiency.
In the worst case, the target disk and FC may collide.
Since the FC is a pulse magnet, it is expected that 
a
moment load will be applied to the axis due to the electromagnetic repulsive force created by the magnetic field of the FC and the eddy current flowing on the target disk.  
Therefore, the rotational accuracy and rigidity of the axis are important factors for stable operation.  
For this reason, the disk runout was measured using a laser displacement sensor (Keyence, LK-G500).
The sensor was installed on both sides of the disk, and the results were evaluated by separating the components caused by variations in the thickness of the disk itself and the wobble of the shaft from the sum and difference of the measurement results.  
The results showed that the thickness variation was $\rm \pm 30 \mu m$ and the wobble was $\rm \pm 70~\mu m$.
The amplitude did not change when the rotation speed was changed.  
The influence of resonance due to rotation can be ignored.
The maximum amplitude of approximately $\rm \pm 100\, \mu m$ corresponds to a variation of about 5\.\% for the design distance of 2 mm.
This will not significantly affect positron capture efficiency or other factors.

The longest continuous operation tests so far was for about one month.
Sensors and software to collect various data continuously over a long period were prepared along with an archiving system using Archiver Appliance.
An additional
long-term continuous operation test was scheduled 
for
September 2025.

\subparagraph{Magnetic focusing}

\begin{figure}[b]
    \centering
\includegraphics[width=0.7\textwidth]{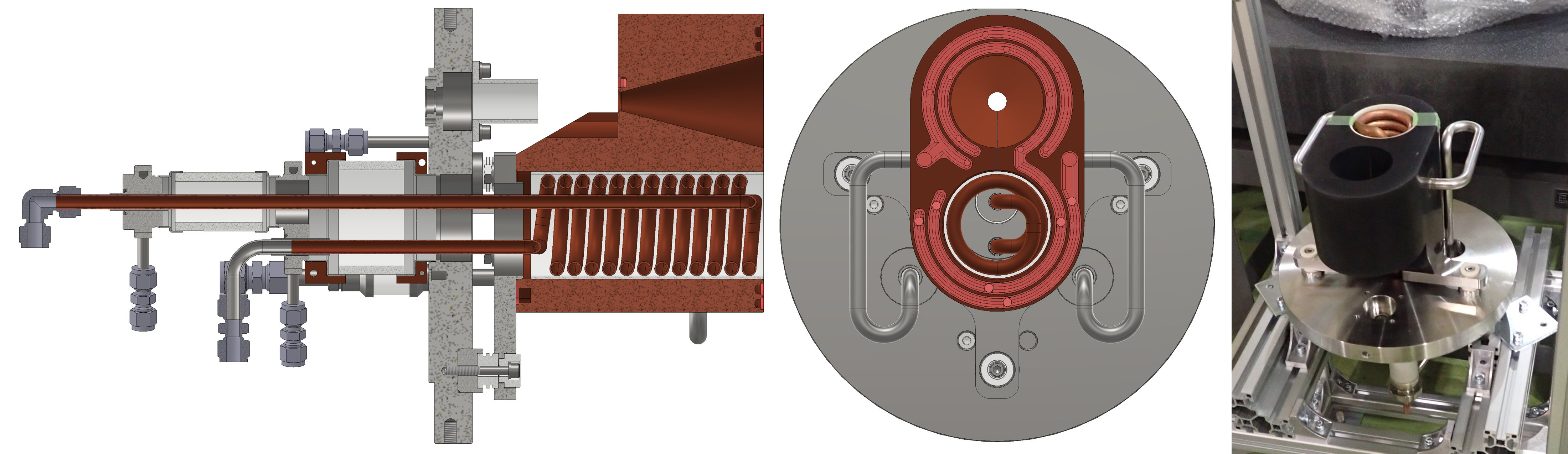}
    \caption{Cross sectional and downstream views of the FC and a photo.}
    \label{Fig_FC}
\end{figure}

To satisfy requirements, we designed a new FC by reviewing the material and improving the water-cooling structure, 
and using coupled analysis of electromagnetic field and thermal analysis with CST Studio\cite{pasj2024_1}. Figure~\ref{Fig_FC} shows cross sectional and downstream view of the FC with a photo of the assembled FC.

The pulsed power supply for the FC will be larger than that of SuperKEKB due to the increasing performance requirements.
In particular, due to the requirements of the bunch structure, although the pulse rate is 100 pulses per second, the pulse interval is 3.3 ms, which requires a shorter charging time.
In order to reduce the charging load and heat load, a power supply with an energy recovery function, which recovers the energy stored in the coil to a capacitor rather than dumping it to resistors, is under development.

Pulsed current is supplied to the FC using coaxial cables and busbars, 
similar to that at
SuperKEKB.
To prevent cable degradation due to radiation, the cables must be connected to the busbars at a certain distance from the target.
In SuperKEKB, the busbars were air-cooled, but 
here
the number of pulses per second is doubled and the peak current is approximately tripled, so the busbars which are in the limited space are water-cooled.
Selection of the cable
has been completed
and its purchase is being processed.

\begin{figure}[hbt]
    \centering
    \includegraphics[width=0.8\textwidth]{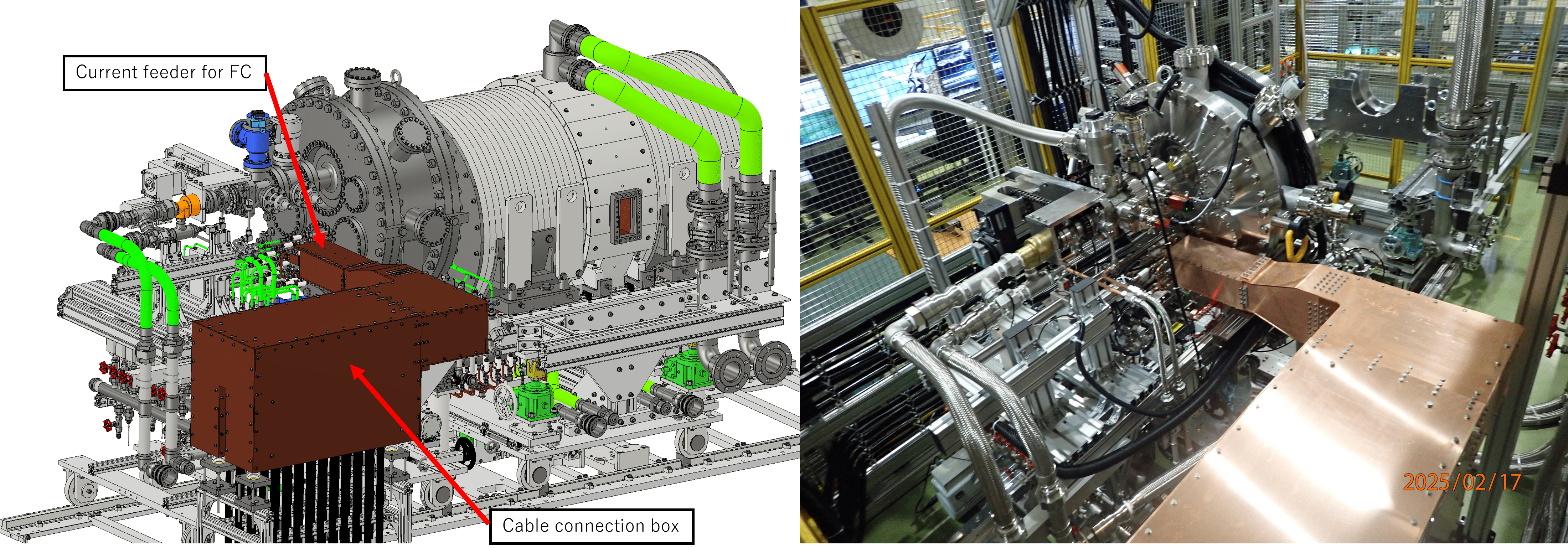}
    \caption{Cable connection box and current feeders for the FC.}
    \label{Fig_feeder-and-box}
\end{figure}

\begin{figure}[h]
    \centering
    \includegraphics[width=0.5\textwidth]{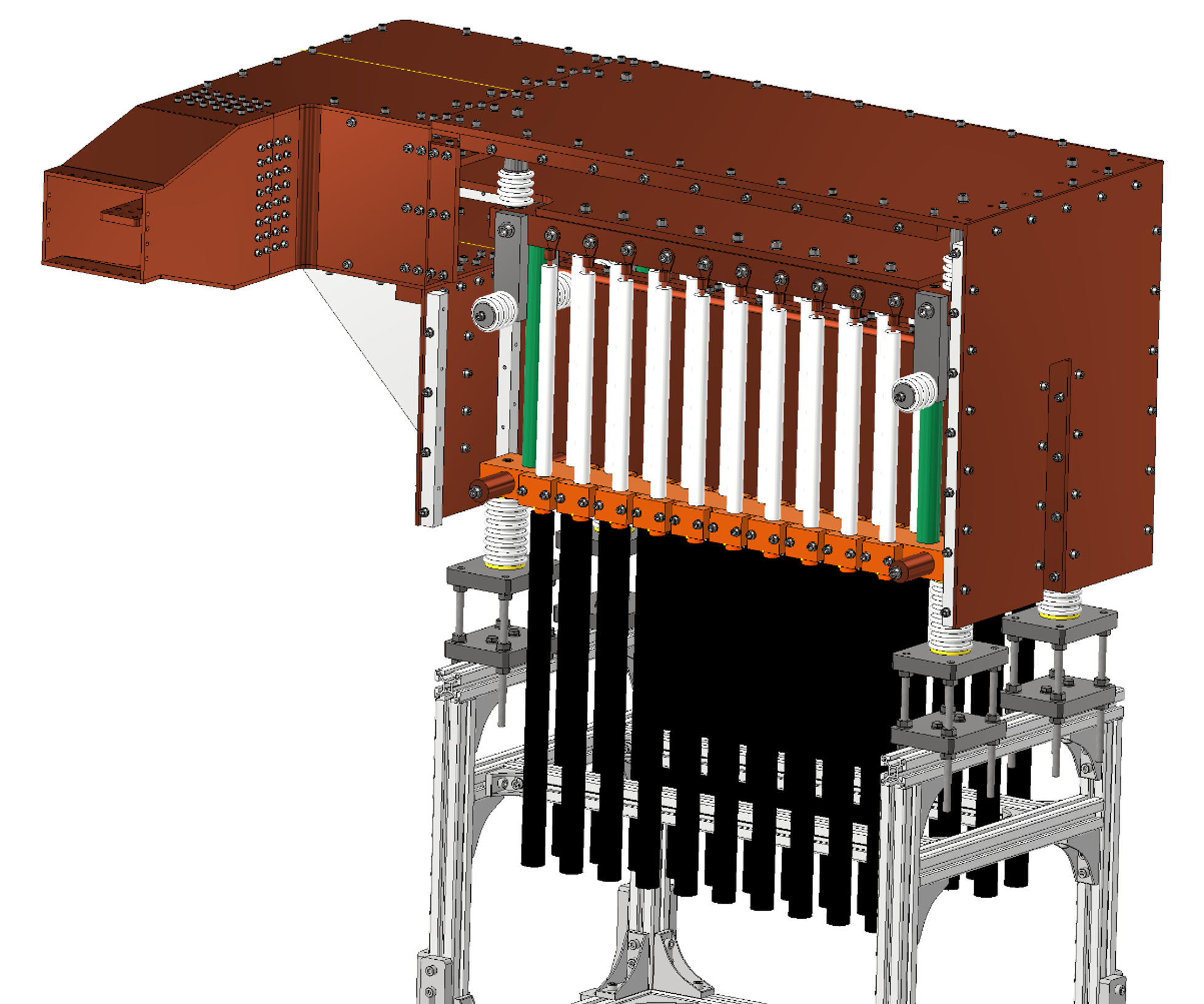}
    \caption{Cable connection box.}
    \label{Fig_connection-box}
\end{figure}

\begin{figure}[t]
    \centering
    \includegraphics[width=0.7\textwidth]{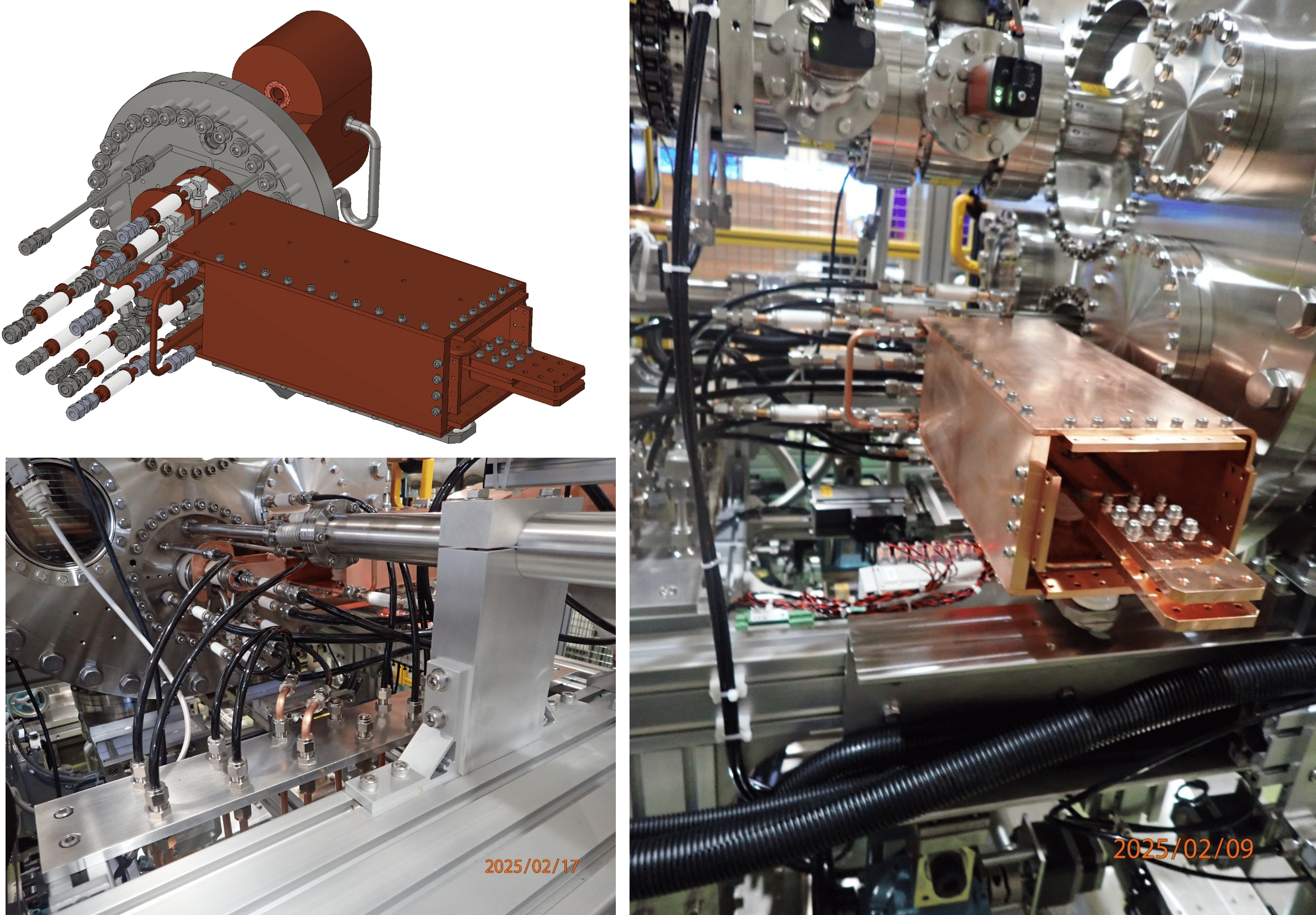}
    \caption{Current feeders for the FC.}
    \label{Fig_feeder}
\end{figure}

Busbars and the cable connection box 
have been 
designed, manufactured and installed.
Figures \ref{Fig_feeder-and-box},\ref{Fig_connection-box},\ref{Fig_feeder} shows layout and details of them.

\subparagraph{Capture cavity}

\begin{figure}[h]
    \centering
    \includegraphics[width=0.6\textwidth]{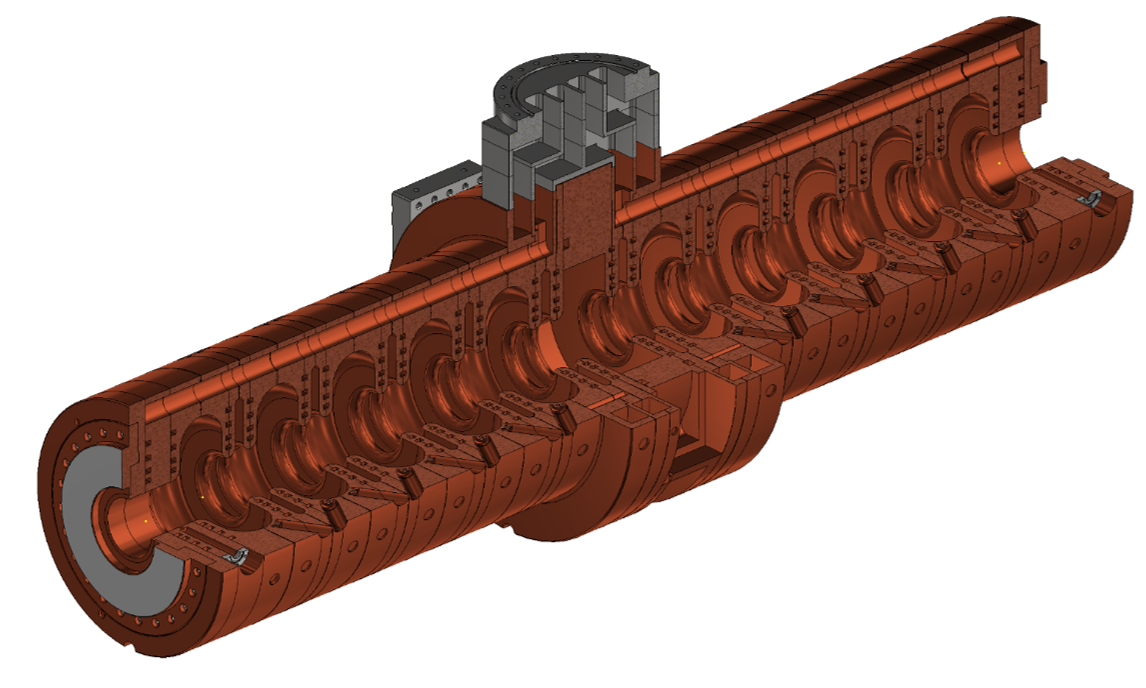}
    \caption{Cross sectional view of the large aperture L-band APS cavity with a water-cooled structure in the iris section}
    \label{cavity}
\end{figure}

A large aperture L-band APS cavity with a water-cooled structure in the iris section has been
designed \cite{pasj2023_2, pasj2024_2}. 
Figure~\ref{cavity} shows cross sectional view of the 3D model.

\begin{figure}[t]
    \centering
    \includegraphics[width=0.8\textwidth]{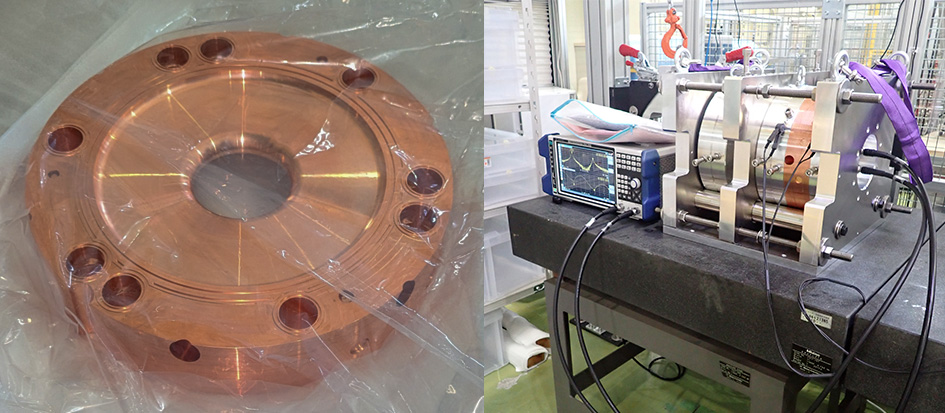}
    \caption{Prototype cell (left) and RF measurement setup (right).}
    \label{Fig_acc-cell}
\end{figure}

One regular cell was prototyped in 2024.
The resonance frequency was measured and compared with simulation to evaluate machining accuracy using this prototype.
Figure~\ref{Fig_acc-cell} shows the prototype cell (left) and RF measurement (right).

Material procurement and cutting 
were
done in 2024.
Rough machining, water channel machining and hot pressing process 
were
in progress in 2025.
Fine machining and overall bonding are expected in 2026.
At the same time, preparation of RF measurement environment and jigs for bonding are in progress.

\subparagraph{Target replacement}

\begin{figure}[h]
    \centering
    \includegraphics[width=0.6\textwidth]{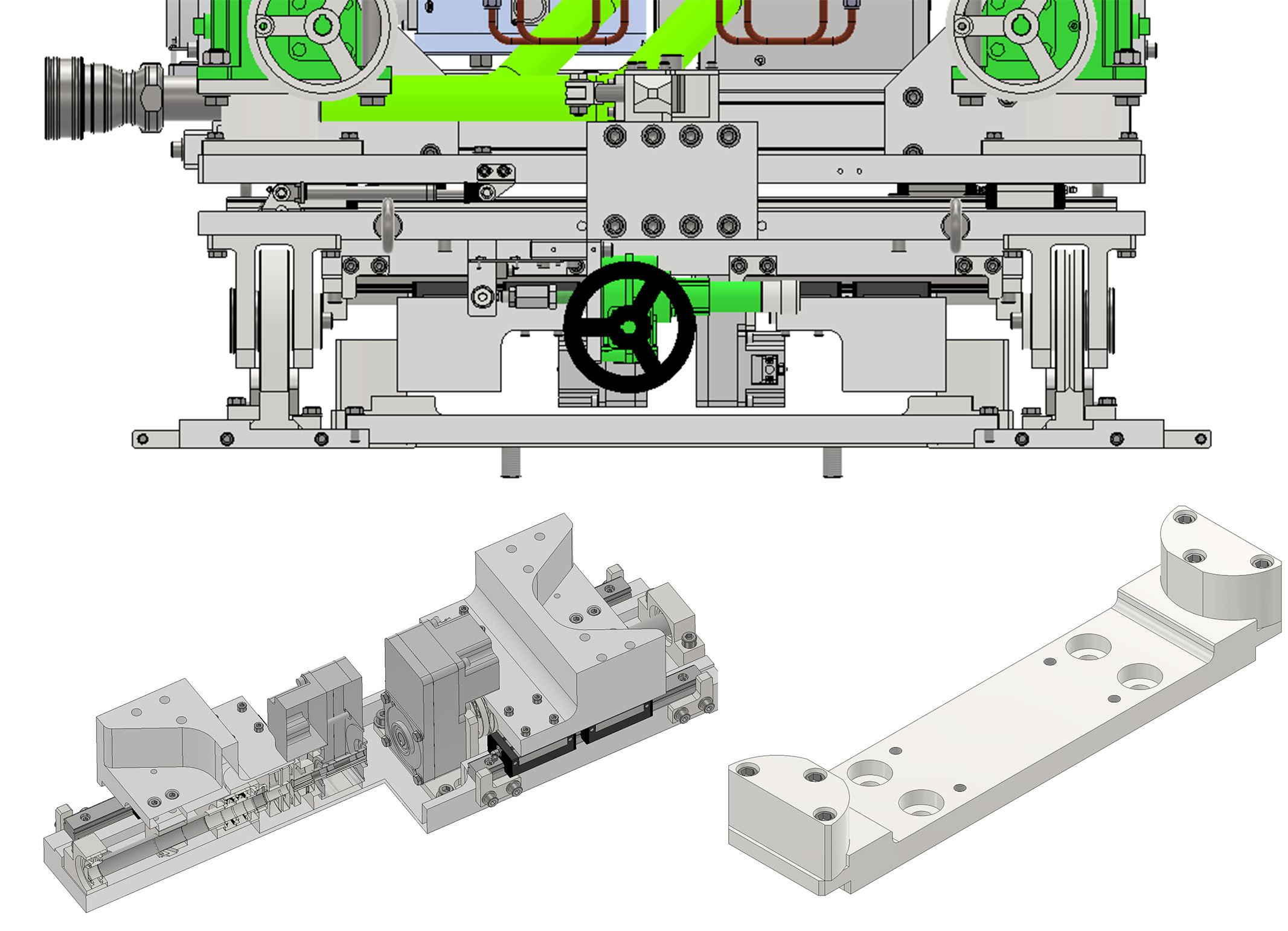}
    \caption{Unit lock mechanism. Movable module mounted on girders (left) and fixed module on  the floor (right). The movable module is shown upside down to show its mechanism.}
    \label{Fig_unit-lock}
\end{figure}

Each component is unitized and mounted on a movable support that travels along rails.
The rails are laid along the beamline direction. 
Units move along these rails and are fixed once transported to their designated positions.
As a method for fixing the units, a module shown in Figure~\ref{Fig_unit-lock} (bottom left) is installed on the unit. 
The cups of the module installed on the unit side is movable.
These cups are stretched and pressed against the cylindrical section of the module shown in Figure~\ref{Fig_unit-lock} (bottom right), which is fixed to the ground. 
During movement along the rails, the cups are fixed in a centred position to avoid collision with other ground-fixed modules.

\begin{figure}
    \centering
    \includegraphics[width=0.8\textwidth]{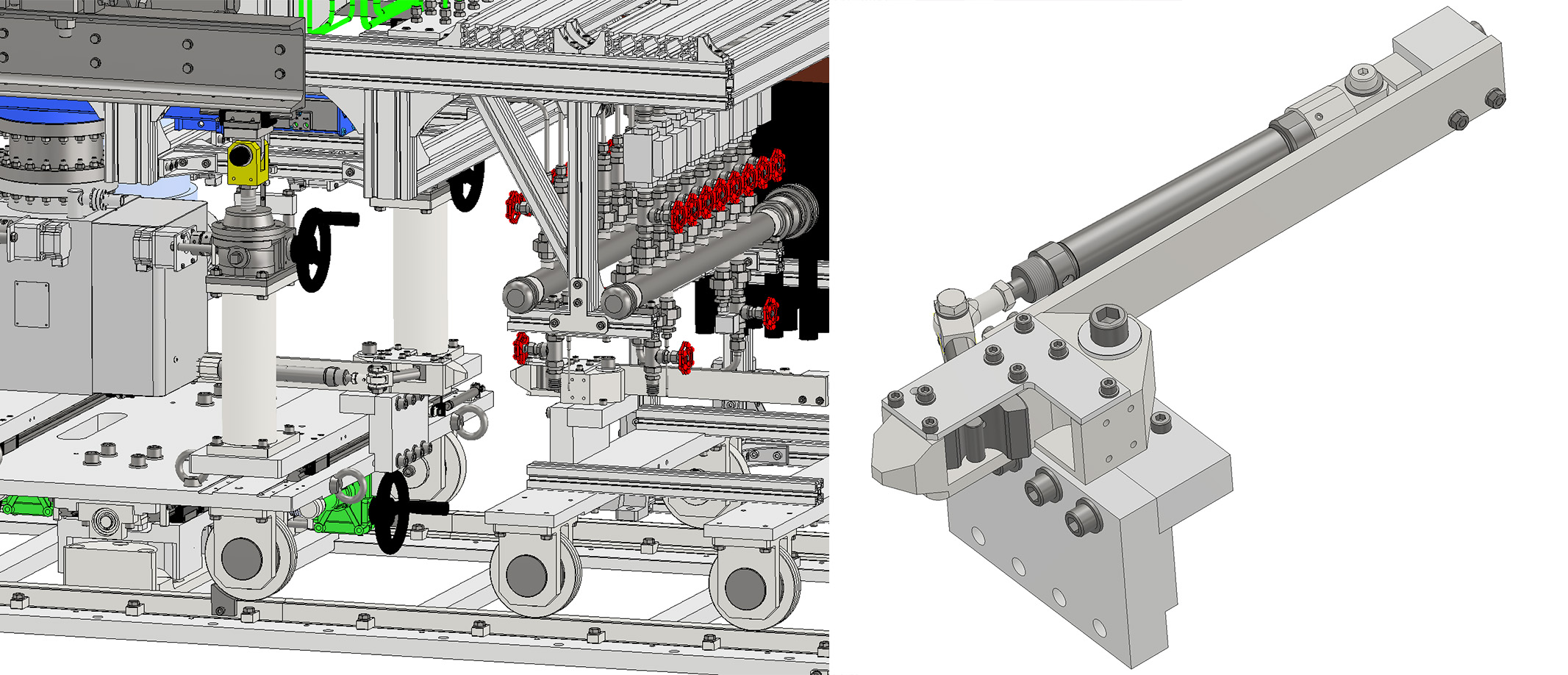}
    \caption{Unit connection mechanism.}
    \label{Fig_unit_coupler}
\end{figure}

To transport the units to their designated positions, some travel mechanism is required. However, installing such a mechanism on each unit makes maintenance of the drive components difficult in the radiation environment.
To avoid this, units are transported by towing them using a motorized trolley. 
These trolleys are stored in areas
accessible
by
personnel during operations.
To realize this approach, it is necessary to develop a mechanism for connecting and disconnecting the towing trolley from each unit.
Figure~\ref{Fig_unit_coupler} shows the coupling mechanism, inspired by automatic couplers used on trains.
Each unit can be remotely connected and disconnected from the motorized trolley.

A pillow seal type flange, which is widely used in J-PARC and other facilities was developed for beamline vacuum connections.
Due to space constraints, the pillow seal flange was combined with a beam collimator.
The first accelerating tube and the target chamber is connected by this flange.

\begin{figure}[b]
    \centering
    \includegraphics[width=0.8\textwidth]{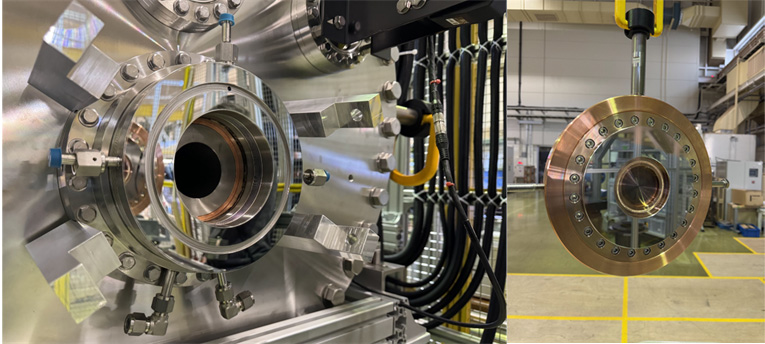}
    \caption{Pillow seal flange (left) and dummy blank flange made of NC50(right).}
    \label{Fig_pillow-seal}
\end{figure}

A composite electrolytic polished stainless steel surface is usually used for the mating flange to the pillow seal for better sealing performance. 
However, this time, the mating part is the accelerator cavity itself made of Cu.
Due to bonding technology constraints, a copper alloy (NC50) was adopted for the surface material.
Figure~\ref{Fig_pillow-seal} shows a photograph of the pillow seal flange and dummy blank flange which is the same dimension and made of the same materials as the accelerating cavity installed in this position.
The contact surface on the copper alloy side is hard chrome plated to prevent oxidation and improve surface hardness.
A vacuum test confirmed a leak rate of $10^{-11}\,\mathrm{Pa\cdot m^3/s}$ or less and an ultimate pressure was equivalent to that achieved with a standard blank flange ($1 \times 10^{-5}\,$Pa or less).

Remote connection mechanisms
are
required for cooling water, compressed air, vacuum, waveguides, solenoid cables, and other electrical wiring.
There are many useful connectors and mechanisms which are used in various facilities.
Since constraints will vary depending on the actual installation environment, details are not being pursued at this time.

\subparagraph{Others}

\begin{figure}
    \centering
    \includegraphics[width=0.70\textwidth]{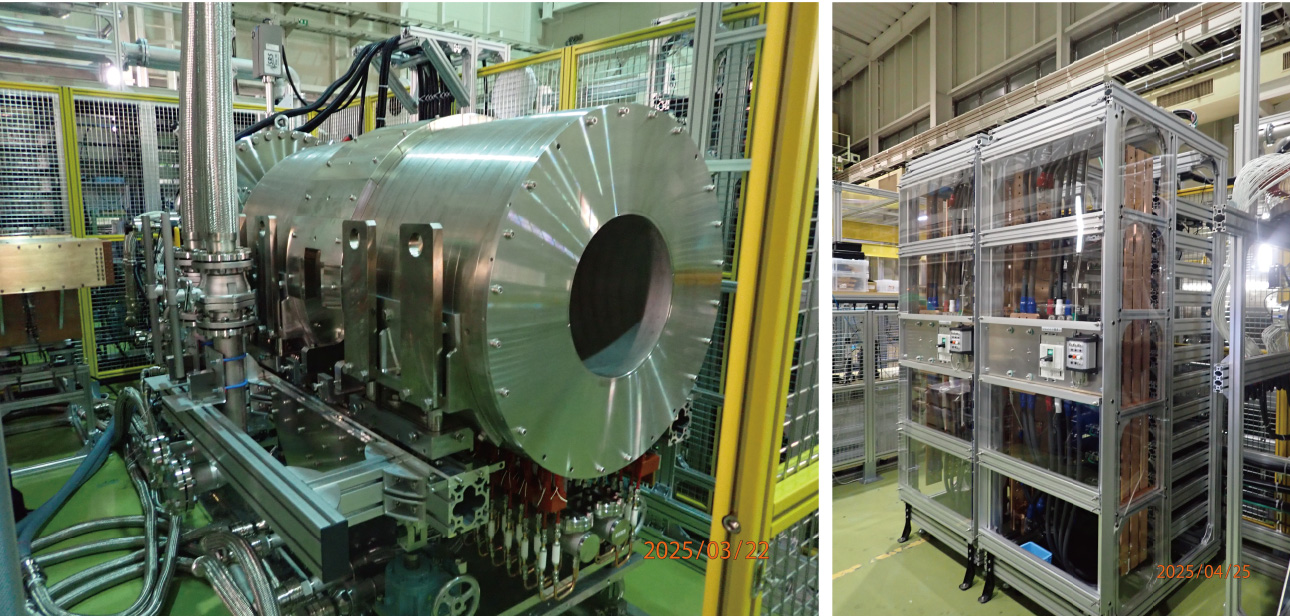}
    \caption{Solenoid (left) and power supply for solenoid (right).}
    \label{others}
\end{figure}

Figure~\ref{others} (left) shows the solenoid.
As shown in Figure~\ref{Fig_cross-section}, the solenoids are installed both sides of a  acceleration cavity and are mounted on a single support frame, forming one unit. 
For the full capture section, 40 such units (approximately 60$\,$m) are installed in series. 
The basic structure of the solenoids follows that of SuperKEKB, but their inner diameter is doubled and outer diameter increased by 1.5 times to match the accelerator cavity size. 
Layout of the cooling water piping have been improved due to space constraints and easy assembly.
Design parameters are summarized in the Table \ref{spec-solenoid}

\begin{table}[hbt]
    \centering
    \caption{Design specification of the solenoid}
    \begin{tabular}{lcc}
        \toprule
            & unit & value \\
        \midrule
            hollow conductor size& $\mathrm{mm}$ &  $\Box$14/$\phi$8 \\
            turns & $\mathrm{}$ &  432 (36$\times$12) \\
            outer diameter & $\mathrm{mm}$ & 900 \\
            inner diameter & $\mathrm{mm}$& 400 \\
            length & $\mathrm{mm}$& 579 \\
            current & $\mathrm{A}$ &  800 \\
            voltage & $\mathrm{V}$ &  86 \\
            power & $\mathrm{kW}$ & 68.8 \\
            magnetic field & $\mathrm{T}$ & 0.55 \\
            cooling water flow & $\mathrm{L/min}$ & 74 \\
            pressure drop & $\mathrm{MPa}$ & 0.2 \\
            temperature rise of cooling water & $\mathrm{K}$ & 14\\
        \bottomrule
    \label{spec-solenoid}
    \end{tabular}
\end{table}

Figure~\ref{others} (right) shows the DC power supply for the solenoid.
To reduce costs, 
modular power supplies (TDK Lamda, HWS3000-24) 
are used.
The modules will be connected in a
3-series, 10-parallel configuration.
Currently, only 9 modules have been delivered, operating in a 3-series, 3-parallel configuration. 
The remaining modules will be delivered by August 2025 and full power test is expected in September.
Each of the two solenoids is connected to an independent power unit, enabling individual control.

For the interlock and control system, NI's CompactRIO was adopted.
The software was mainly developed using LabVIEW and EPICS.
Archiver Appliance was adopted for data archiving and more than 2,500 PVs are archived currently.

\paragraph{Timeline}
All the components of the prototype except for the
accelerating structure and 
the
pulsed power supply for the FC have been 
delivered and installed.
The missing two components are under development 
which
will be completed 
by the end of JFY2026. 
Evaluation of
components, especially the rotating target is in progress.
Complete 
assembly and full power test of the FC are currently
scheduled for JFY2027.

\newpage
\subsection{Nano-beam Area}
\subsubsection{Damping Ring Design (WPP-12)}
\paragraph{Description}
The baseline design of the ILC damping ring with a normal conducting magnet is content of work package WPP-12 \cite{{Pre-lab},{ITN-WP}}. 
Although 
the idea of using permanent magnets 
for
power savings for the ILC
is considered, 
the establishment of a design using the fundamental normal-conducting magnets 
is
the most critical factor in executing the ILC DR engineering design. The present design of the damping ring in ILC TDR\cite{ILC-TDR} is a simple design using a hard-edge magnet model with zero spacing between magnets. The ILC DR is designed to have a very large dynamic aperture to maximize the positron capture yield. The specification of the ILC damping ring is listed in Table~\ref{tab:WPP15-ATFdampingringSpecification}. However, it is pointed out that the dynamic aperture of the circular accelerator decreases when the fringe field of the magnet is considered. Since the spatial distribution of positrons differs depending on the positron capture method, the dynamic aperture is a factor that affects the positron capture yield in each method.
It is important 
that it be
considered 
quantitatively before 
selection of the positron capture method.
Since the establishment of the basic DR design also has a significant impact on the other area systems of the ILC, 
the aim is to establish the design by the end of 2027.
\begin{table}[h]
\caption{Specification of the ILC damping ring.}
\vspace{-4mm}
\begin{center}
\begin{tabular}{cc|c}
    \hline
    && Specification \\
    \hline
    Normalized emittance & 
    $\gamma\varepsilon_x$/$\gamma\varepsilon_y$ at N = 2$\times$$10^{10}$ &
    4.0~$\mu$m / 20~nm  \\
    Dynamic aperture (action variable) & 
    $\gamma$ ($A_x$ + $A_y$) &
    0.07~m  \\
    Longitudinal acceptance &
    $\Delta$$\delta$$\times$$\Delta$z &
    $\pm$0.75~$\%$$\times$$\pm$ 33~mm \\
    \hline
\end{tabular}
\end{center}
\label{tab:WPP15-ATFdampingringSpecification}
\end{table}

\begin{figure}[htbp]
	\centerline{
		\includegraphics[width=13cm] {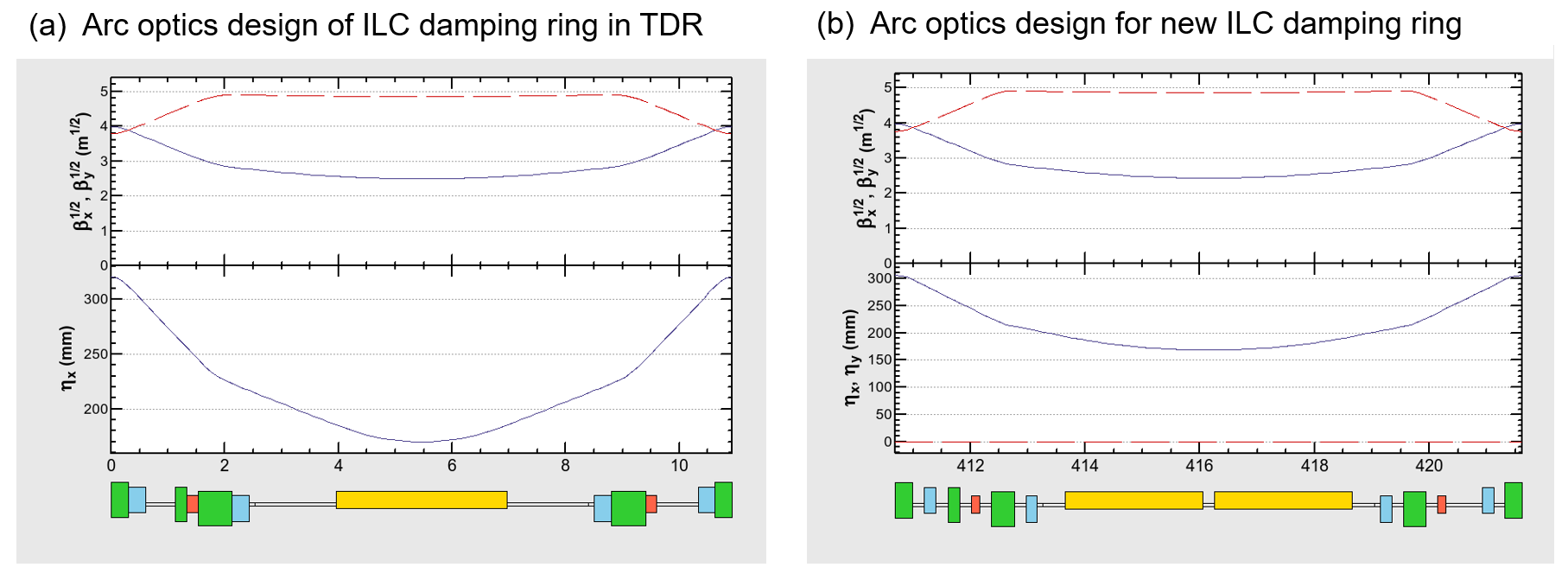}}
	\caption{Arc cell optics of ILC damping ring for TDR (a) and the modified optics (b)\cite{DR_kubo}.}
	\label{fig:WPP12-ILCDRarcCell}
\end{figure}

\paragraph{Status}
Effort 
is focussed on transforming the ILC TDR conceptual optics design into a practical design. The TDR optics deck was modified to accommodate feasible magnet dimensions. After the optics modification, it was found that replacing the single 3m-long bending magnet in the arc cells with two 2.4m-long magnets could reduce the DR horizontal emittance, including intra-beam scattering, from 6.3$\mu$m to 4.1 $\mu$m. Furthermore, it was confirmed that the dynamic aperture remains unchanged with these optics’ modifications, establishing a realistic base design for the ILC DR beam optics. The beam optics of the arc cells for the
TDR and the modified optics 
are shown in Figure~\ref{fig:WPP12-ILCDRarcCell} and the evaluated emittances and the damping rings are listed in Table~\ref{tab:WPP12-ILCDRparameters}.  

\begin{table}
    \caption{Emittance of the ILC damping ring both for TDR optics and the modified optics. The emittance evaluation with intra-beam scattering was calculated by assuming the bunch population of N=2$\times$$10^{10}$\cite{DR_kubo}.}
    \vspace{-7mm}
    \begin{center}
    \begin{tabular}{cc|cc}
        \hline
         && TDR & modified optics \\
         \hline
         Emittance w/o IBS & $\gamma\varepsilon_x$/$\gamma\varepsilon_y$ & 5.74~$\mu$m/20~nm  &  3.27~$\mu$m/20~nm \\
         Emittance with IBS & $\gamma\varepsilon_x$/$\gamma\varepsilon_y$ & 6.27~$\mu$m/20~nm  &  4.08~$\mu$m/20~nm \\
         Damping ring & $\nu_x$/$\nu_y$/$\nu_z$ &
         23.9~ms/23.9~ms/11.9~ms  &
         25.5~ms/25.5~ms/12.8~ms  \\
         \hline
    \end{tabular}
    \end{center}
    \label{tab:WPP12-ILCDRparameters}
\end{table}

However, the bending magnet is currently used to evaluate dynamic aperture by using magnet models with the hard-edge magnets, which have no fringe field. Therefore, in collaboration with KEK, ANSTO, and Korea University, we plan to begin evaluating dynamic apertures using magnet models that include the effects of fringe fields. A kick-off meeting for this international collaborative research 
was
held in October 2025.

The goal for WPP-12 is the damping ring design to achieve the ILC requirements; low emittance for both the horizontal and vertical directions of 4~$\mu$m/20~nm, including intra-beam scattering, while simultaneously realizing a wide dynamic aperture of 0.07~mrad. The design of the ILC damping ring and the evaluations of the realistic model will be performed by the end of 2027.

\subsubsection{Injection \& Extraction System of Damping Ring (WPP-14)}
\paragraph{Description}
WPP-14 is related to the fast injection and extraction system of the damping rings. The beam extraction with a fast kicker system using semiconductor pulsed power supplies with nano-second pulse length  was demonstrated in principle at KEK's ATF about ten years ago. Since semiconductor technology continues to evolve, we will develop a fast kicker power supply using the latest semiconductor technology. 

Injection/extraction stripline kickers for the ILC damping ring have many similarities with the storage ring injection striplines for the UK Diamond-II upgrade project - see Table~\ref{Tab-Phil}. 
Prototype striplines for Diamond-II are under development, with installation and testing planned in the existing Diamond transfer line and storage ring.
Commercial development of a SiC pulser for Diamond-II with the UK company Kentech Ltd. has begun.
Parallel development of a pulser suitable for ILC has been discussed with the same company. Given the synergies between the Diamond-II and ILC specifications, a prototype pulser for ILC could be developed quickly and at relatively low cost. This would also potentially allow a direct test of the prototype ILC pulser in situ with the Diamond-II prototype stripline kicker.

\begin{table}[htb]
\begin{small}
\begin{center}
\begin{tabular}{|l||r|r|r|r|}
    \hline
    & \multicolumn{2}{|c|}{ILC} & \multicolumn{2}{|c|}{DIAMOND-II} \\\hline
Operating mode & Baseline & Hi Luminosity & Aperture-sharing & Kick-and-cancel  \\
& & & injection & injection \\
Pulse structure & 1312 burst & 2625 burst & Single kick & Double kick \\
Rep rate & 5 Hz & 5 Hz & 5 Hz & 5 Hz \\
Pulse duration (FW) & $< 6$ ns & $< 3$ ns & $< 3$ ns & $< 15$ ns \\
Pulse separation & 554 ns & 332 ns & -& $\sim 2$-20 $\rm \mu s$ \\
Voltage	& $\pm$10 kV & $\pm$10 kV & $\pm$20 kV & $\pm$20 kV \\
Technology & DSRD? GaN? & DSRD? GaN? & Avalanche & SiC \\
\hline
\end{tabular}
\end{center}
\end{small}
\caption{Comparisons of between the stripline kickers between ILC and DIAMOND-II}
\label{Tab-Phil}
\end{table}

\paragraph{Status}
An agreement between JAI/Oxford and CERN for the pulser prototype development work to proceed in Oxford has recently been signed. In advance of this, the company has done initial simulations for the preliminary design of a suitable pulser. The design, manufacturing and subsequent testing of a new prototype damping-ring kicker pulser module can now proceed. The module shall fulfil the outline requirements specified and agreed by the ILC Development Team (IDT). Oxford University will take part in the design, construction, testing and analysis of results and simulations, with the following work plan:
\begin{itemize}
    \item Specification of performance goals in consultation with IDT experts.
    \item Procurement exercise, placement of a contract with the identified supplier, and follow-up of technical progress with the supplier via regular meetings.
    \item Testing of the prototype pulser and report on its performance.
\end{itemize}
The aim is to complete the prototype by around October 2026, and to complete its testing and characterisation by September 2027.

\subsubsection{Final Focus (WPP-15)}
\paragraph{Description}
 WPP-15 is the system design of the ILC BDS and the advancing of the beam technology required for it\cite{Pre-lab}. Even though some of the items listed in the original WP-15 can be performed after the ILC Pre-Lab start, it is appropriate for the time-critical 
 work
to 
be
narrowed down 
to 
 the higher priority research topics. Therefore, 
 the following 3 research topics 
 were selected
 as time-critical items for
 WPP-15 \cite{ITN-WP}.
\begin{itemize}
    \item wakefield mitigation
    \item mitigation and correction of higher-order aberration
    \item training for ILC beam tuning (machine-learning etc.)
\end{itemize}
ATF2 beamline is the only existing test accelerator in the world to test the final focus beamline of the linear colliders, and the study of the final focus beamline at ATF is important for ILC. The technical research of the final focus system for the ILC at ATF2 beamline has been proceeded in international cooperation under the ATF international collaboration. The primary goal of the ATF international collaborative research has been to focus the beam to the level required for the ILC and to stabilize the beam position. The ATF2 beamline has achieved a beam size of 41 nm, the world smallest value comparable to the target of 37 nm\cite{{ATF2_OKUGI},{ATF2_ATFreport2020}}. Regarding beam position stabilization, using intra-train feedback technology, the beam position has been stabilized within the resolution of the beam position monitor\cite{{ATF2_FONT01},{ATF2_FONT02}}. The ATF accelerator is constructed on the ground level, making it susceptible to environmental noise much more than ILC. Consequently, keeping stable beam conditions at ATF is significantly more challenging than
expected for
ILC operation.

Therefore, in accordance with the research themes pursued by ITN, ATF research is now focused on concentrating the beam to 37 nm at the ATF focal point and keeping the beam size stable for couple days.
 The latest beam tuning technologies, one of the WPP-15 research topics, are essential to achieve the ATF main goal. 
 Beam diagnostic devices
 havew been developed 
 for various beam tests at the ATF2 beamline to achieve the ATF main goal. Furthermore, achieving the ATF main goal necessitates improving electron beam quality through wakefield mitigation and multipole error mitigation along the ATF2 beamline. 
 The
 WPP-15 research topics
to be developed are:
wakefield mitigation; multipole error mitigation; and,  ILC beam tuning training through the ATF main goal of stably operating the ATF ultra-small beam via the ATF international collaboration.
\paragraph{Status}

\begin{figure}[htb]
	\centerline{
		\includegraphics[width=13cm] {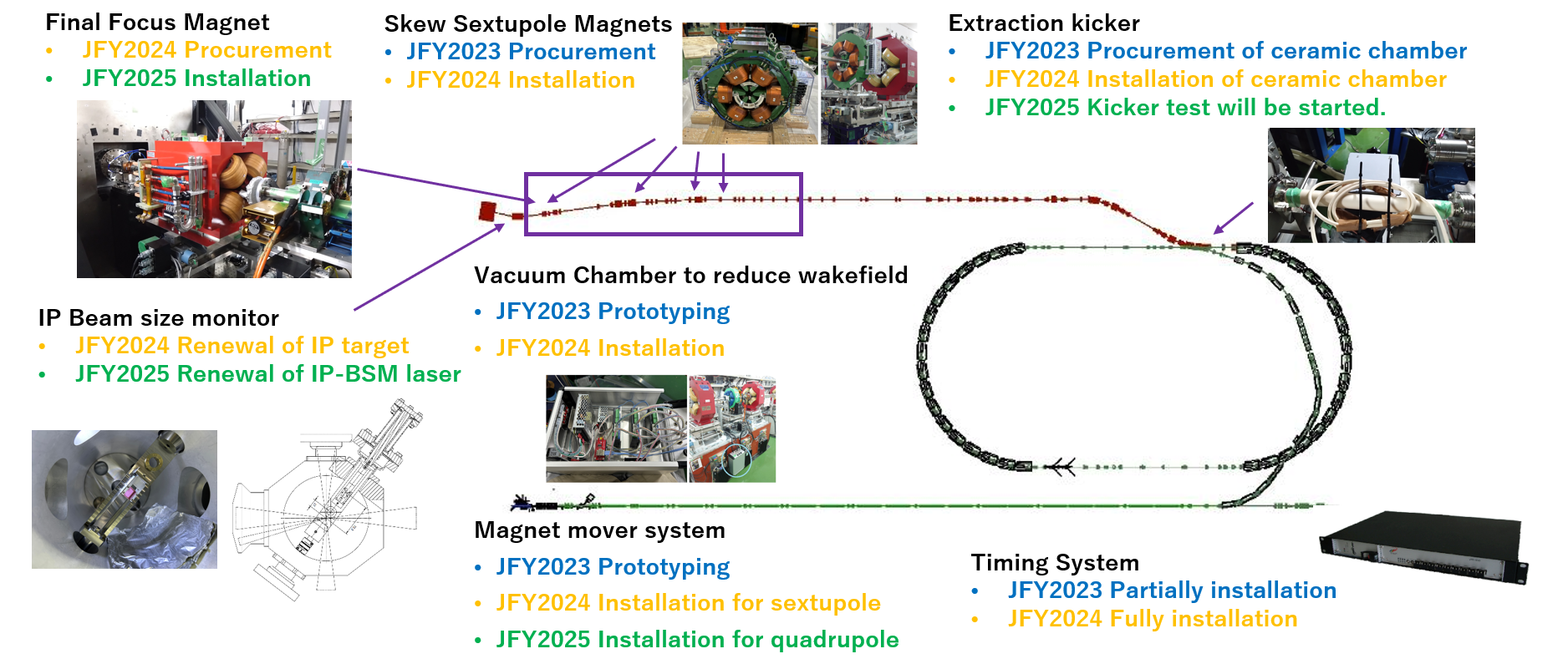}}
	\caption{Present status of the ATF hardware upgrade for the ITN nano-beam study.}
	\label{fig:WPP15-ATF2hardwareUpgrade}
\end{figure}

\begin{figure}[h!]
	\centerline{
		\includegraphics[width=13cm] {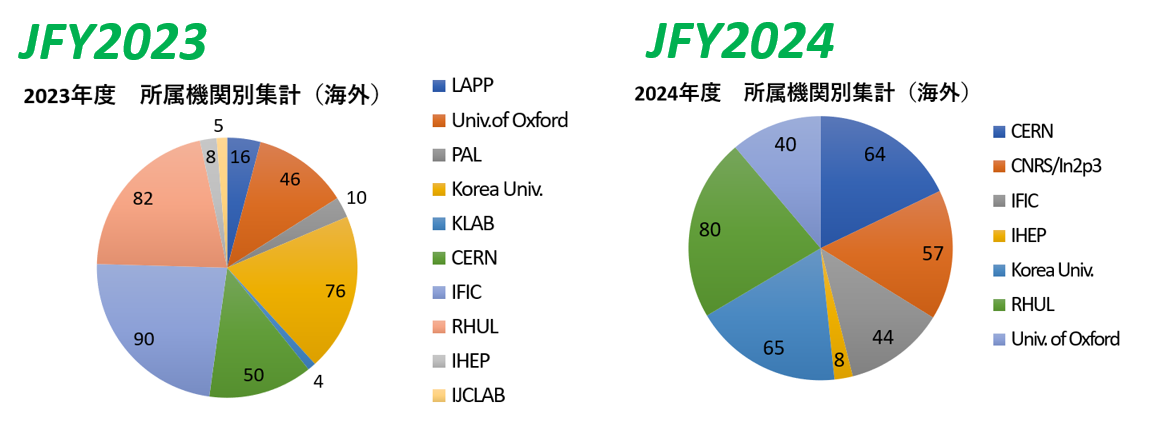}}
	\caption{Distribution of research institution for researchers from outside KEK who came to conduct research at the ATF in 2023 and 2024.}
	\label{fig:WPP15-ATFstatistics}
\end{figure}

International 
development research for WPP-15 
commenced 
under
the ATF
international collaboration.
The first half of the project focussed on hardware upgrades aimed at replacing aging components and enhancing the performance of the ATF accelerator. The second half of the project will conduct full-scale beam tests utilizing the upgraded facilities. JFY2025 is in the final stage of the first half program and the hardware upgrade is proceeding as scheduled. The present status of hardware upgrades is shown in the Figure~\ref{fig:WPP15-ATF2hardwareUpgrade}. The beam studies are also 
progressing
through ATF international collaboration, and the statistics for the researcher involved in ATF research outside KEK since JFY2023 are shown in the Table~\ref{tab:WPP15-ATFstatistics} and Fig~\ref{fig:WPP15-ATFstatistics}. Detailed progress in individual aspects is as follows.
\begin{table}[htb]
    \caption{Statistics by research institution for researchers from outside KEK who came to conduct research at the ATF in 2023 and 2024.}
    \begin{center}
    \begin{tabular}{c|cc}
        \hline
         & JFY2023 & JFY2024 \\
         \hline
         Operation week & 14  &  20 \\
         \hline
         Countries      &  7  &  5  \\
         Institutions   & 10  &  7  \\
         Person-day     & 387 & 358 \\
         \hline
    \end{tabular}
    \end{center}
    \label{tab:WPP15-ATFstatistics}
\end{table}

\begin{figure}[h]
	\centering
		\includegraphics[width=0.85\textwidth]{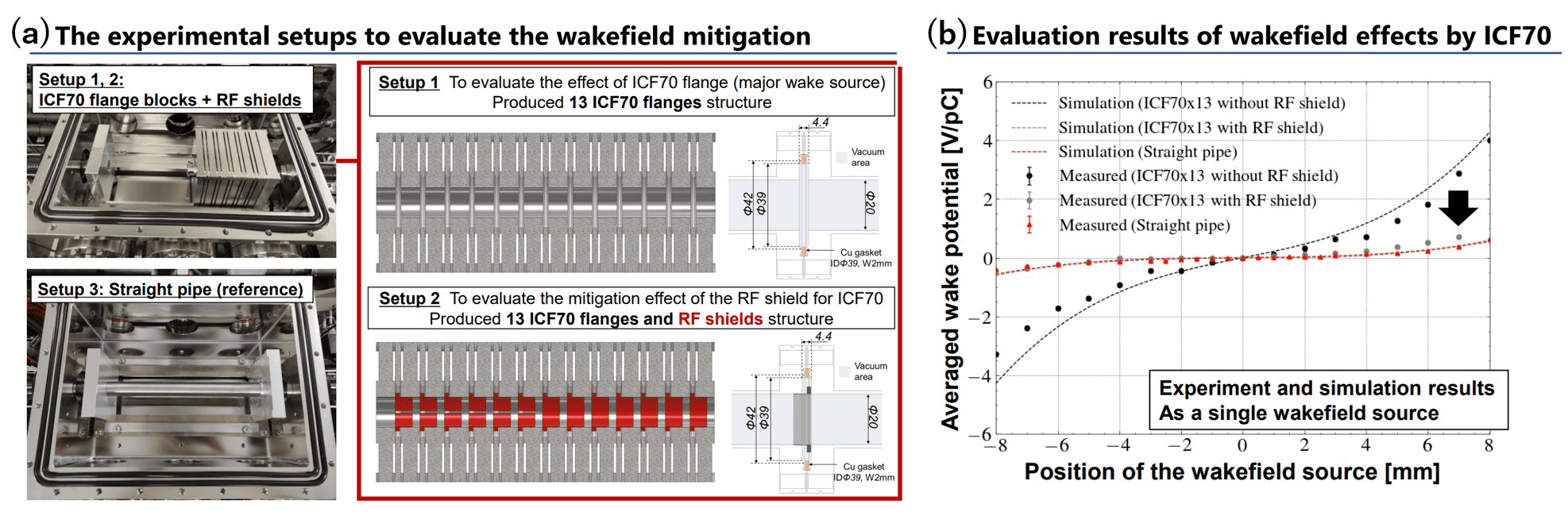}
	\caption{Wakefield effect for ICF70 flange was evaluated at ATF2 beamline. (a) Setup of the wakefield measurement for ICF70 flange. (b) Wakefield measured results for ICF70 flange with/without RF mask.\cite{ATF2_Abe}}
	\label{fig:WPP12-ATF2wakeMeasurement}
\end{figure}

\subparagraph{Wakefield mitigation}
Since the electron beam is focused to the nanometer scale at the focus point of the ATF2 beamline, the effects of beam kicks due to wakefield affect the IP beam size significantly. 
On the ATF2 beamline, which operate ultra-small beams, even minor step changes like the ICF flange gap can cause effects even though it is a beam transport line. 
Various wakefield models that account for step changes in the vacuum chamber throughout the entire beamline 
are being studied 
by ATF international collaboration\cite{{ATF2_ KoryskoThesis},{ATF2_AbeThesis}}. By comparing this wakefield realistic model with actual measurements, 
advances are being made in the 
understanding of the effects of wakefield on nanometer-scale ultra-small beams. During 2023 operations, KEK installed a device on the ATF2 beamline capable of simultaneously moving multiple structures mimicking the shape of ICF flanges. This allowed us to investigate the impact of ICF flanges on the wakefield effect, confirming that ICF flanges do affect the beam (Figure~\ref{fig:WPP12-ATF2wakeMeasurement}).

\begin{figure}[h]
	\centerline{
		\includegraphics[width=13cm] {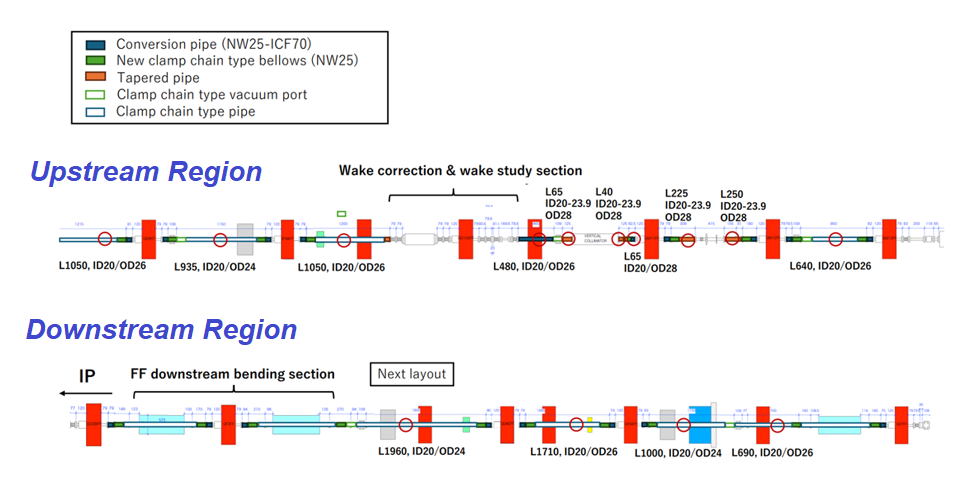}}
	\caption{Schematic Figure~of the vacuum chamber upgrade for ATF2 beamline.}
	\label{fig:WPP12-ATF2vacuumChamber}
\end{figure}

In response, in 2024, the vacuum chamber was rebuilt to minimize step differences on the inner surface of the vacuum chamber, particularly in the sensitive areas for wakefield in ATF2 beamline (Figure~\ref{fig:WPP12-ATF2vacuumChamber}).

\begin{figure}[htbp]
	\centerline{
		\includegraphics[width=13cm] {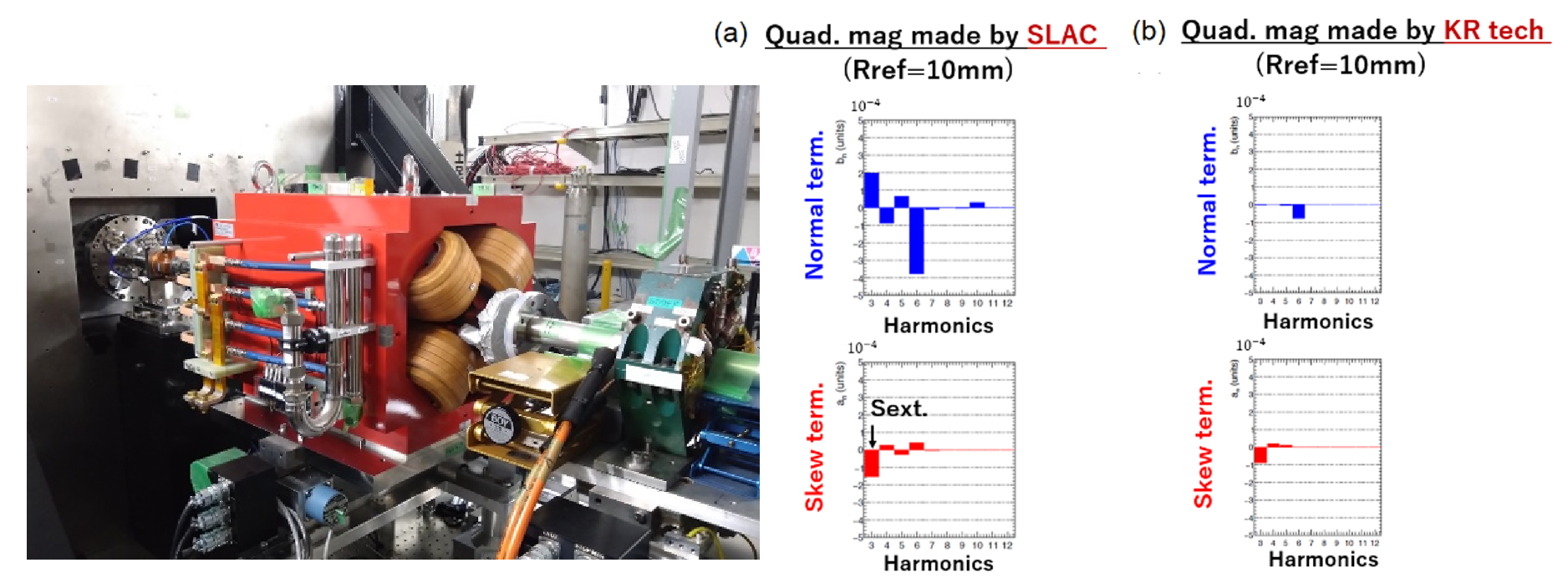}}
	\caption{Renewed ATF final focus magnet and the magnetic field measurement for the original ATF final focus magnet (a) and the renewed ATF2 final focus magnet (b).}
	\label{fig:WPP12-FinalMagnet}
\end{figure}

\subparagraph{Mitigation and correction of higher-order aberration}
Since the electron beam is focused to the nanometer scale at ATF2 IP, the control of the fine multipole field errors of magnets on the beam is highly significant. To suppress the effect of multipole field errors on the nanometer-scale beam, it is crucial to use magnets with small multipole field errors. However, the final focus doublet for the ATF2 beamline was modified and reused from a magnet previously employed in another accelerator, with the cooperation of the ATF international collaboration participating laboratories, during the ATF2 construction phase. It was not originally designed and manufactured as the optimal final focus magnet of ATF2. Therefore, one of the final doublet was newly fabricated in 2022 and 
installed
in 2023. In 2024, the remaining final focusing magnet was manufactured and inserted into the beamline in 2025.
The results of the multipole field measurements both for the original and the renewed ATF2 final focus magnets are shown in Figure~\ref{fig:WPP12-FinalMagnet}. The multipole field of the ATF2 final focus magnet is improved by replacing the magnets.
Furthermore, the magnet mover system used in the ATF2 beamline is also aging significantly, and particularly for the control circuits, there were no replacement parts 
for the eventuality of failures.
Therefore, in 2023, the movers for the sextupole magnets, which have stringent position control accuracy requirements, were updated. Starting in 2024, 
progressive updating has been undertaken of the movers for the quadrupole electromagnets, which also have demanding position control accuracy requirements. 
It is planned
to have the beam test 
to
investigate the impact of the improvement of the magnets and magnet movers, in collaboration with CERN and others within the ATF international collaboration.

\subparagraph{Training for ILC beam tuning (machine-learning etc.)}
Research on beam tuning techniques incorporating machine learning at the ATF accelerator is also progressing through the ATF international collaboration. Since 2022, ATF has been optimizing beam injection efficiency in the damping ring, optimizing the IP beam size tuning knob at the ATF2 focus point, and performing emittance tuning in the damping ring\cite{ATF2_Kurata}. Although these tunings demonstrated the usefulness of machine learning for beam tuning in the ATF accelerator, they were limited to finding optimal values for existing tuning knobs. The next step requires incorporating simulation-based reinforcement learning\cite{{ATF2_ML01},{ATF2_ML02}}. Currently, the ATF collaboration is developing a flight simulator for the ATF accelerator in international cooperation with KEK, CERN, and IJCLab. This aims to efficiently transfer beam tuning methods derived from simulation to the actual accelerator. This is expected to enable efficient reinforcement learning via simulation, even in countries outside KEK participating in the ATF international collaboration. Through such efforts, 
it is aimed
to establish a new optimization process for beam size tuning at the ATF focus point and reduce optimization time. The beam tuning under more challenging beam tuning conditions by utilizing ultra-low beta optics in collaboration with CERN is also planned in order to advance the beam tuning technology\cite{ATF2_UltraLow}.

Furthermore, advancing beam stabilization techniques through feedback is crucial to achieve stable beam operation. Feedback technology within the train at the ATF accelerator had been advanced by the University of Oxford, but progress stalled for a period due to the impact of COVID-19. This feedback research was resumed after ITN started and is currently being vigorously pursued.
\subparagraph{Technological development required to advance WPP-15}
The ATF international collaboration is advancing efforts to stabilize the beam through the stabilization of the timing system and to enhance the beam diagnostics system to advance WPP-15. To accommodate the increased number of researchers following the launch of ITN, ATF is prioritizing upgrades starting with equipment that requires significant time for fault resolution. The timing system 
has highest 
priority. Therefore, since 2021, with the cooperation of the SuperKEKB group, we have been migrating the timing system from the legacy CAMAC-based system to an event-based timing system. To date, all major subsystems and beam diagnostic equipment have been updated to the new timing system. Currently, in collaboration with the IJClab, we are vigorously investigating the stability of synchronization between the RF electron gun laser pulse and the RF clock, as well as its impact on electron beam parameters both directly beneath the electron gun and along the ATF beam transport line. These tools involved the use of CERN White Rabbit protocol embedded in low jitter IDOROGEN boards developed by IJCLab\cite{ATF2_Timing}.

Furthermore, the ATF International Collaboration is actively developing beam diagnostic devices necessary for beam tuning at ATF and beam diagnostics devices required for the ILC accelerator. Regarding beam diagnostics at ATF, the IP beam size monitor is being upgraded. The IP target was replaced in 2025, and the laser used in the IP beam size monitor is planned for replacement in 2026. These upgrades are expected to enable stable and accurate beam size measurements at the IP. IFIC and Korea University are developing beam position monitors for the ILC main linac\cite{{ATF2_Jang},{ATF2_Laura01},{ATF2_Laura02}} and CERN is developing the Cherenkov diffraction radiation monitor as non-destructive beam size monitor\cite{{ATF2_ChDR01},{ATF2_ChDR02}} using the ATF accelerator.

\subparagraph{Plan for the Completion of WPP-15}
WPP-15 goal is to achieve stable operation of 37~nm beam for a couple days at the ATF IP by March 2028. 
The hardware preparations 
underway
will advance 
that target date 
for the WPP-15 to 
by March 2026, including procuring the magnet with small multipole field error and the laser of IP beam size monitor and updating the vacuum chamber with minimum step to reduce the wakefield. In parallel, the ATF accelerator will focus on creating an accelerator operation platform that integrates simulations using the latest beam tuning techniques with accelerator operation through ATF international collaboration, preparing for full-scale tuning after hardware upgrades starting in 2026. The ATF international collaboration efforts for WPP-15 over the next two years were discussed at the ATF project meeting held at the end of November 2025.

The IP-BSM laser, essential for IP beam size measurement for ATF accelerator, is scheduled for procurement in January 2026. Beam focusing will proceed with the goal of achieving 37nm beam focus in the ATF by spring 2027, utilizing the latest beam tuning techniques while advancing the commissioning of this IP beam size monitor. At the ATF project meeting, it was confirmed that the latest beam tuning techniques will continue to be advanced using the Flight Simulator, which reproduces the actual ATF accelerator by simulation and which applies the tuning procedure developed by the simulation to ATF accelerator, primarily led by CERN within the ATF international collaboration. Furthermore, the contributions of each research institution involved in this effort were also confirmed.

Regarding beam stabilization,  in addition to stabilizing the second beam using intra-train feedback with FONT, the ATF project meeting confirmed that the consideration of the feedforward of position and angle of the extraction beam based on ground vibration and damping ring orbit information using the FONT system will be started with relevant institutions to explore the feasibility.
Furthermore, it was confirmed that big data acquisition will proceed by simultaneously collecting long-term beam and hardware information from the ATF accelerator. This data will be utilized for machine learning to investigate beam instability factors and corresponding countermeasures. Through these efforts, development will advance with the goal of stably maintaining the focused beam at the ATF accelerator for several days by spring 2028.

\subsubsection{Evaluation of Magnet Vibration (WPP-16)}
\paragraph{Description}
The purpose of WPP-16 is to design the final doublet (FD) of the ILC and to evaluate the associated QD0 cryostat vibration\cite{{Pre-lab},{ITN-WP}}. In the TDR baseline, the 1.9 K superfluid helium supply for QD0 and the interface to external magnet power leads are via the Service Cryostat\cite{ILC-TDR}. The Service Cryostat connects to QD0 via a long He-II cryogenic line that must pass through a labyrinth in the end Pacman radiation shielding to avoid having a direct path for beamline radiation to the 
presumed occupied experimental detector hall. The vertical beam fluctuation to QD0 must be stable to the order of 50 nm, to stay within the capture range of the intra-train collision feedback. 
It will 
greatly affect
not only the ILC accelerator design, but also the design of the detectors and their interfaces, so we need to conclude the effect of the FD vibration before ILC Pre-Lab starts. 
\paragraph{Status}
Since BNL has been taking the lead in the investigation of FD at the ILC, it was expected that BNL will be the main laboratory of this vibration experiment. Furthermore, the equipment required for this research was prepared at BNL during the TDR phase, and it is essential to continue using the equipment currently located at BNL. However, since BNL participation in the ITN is not yet approved, contrary to the importance of the plan, WPP-16 itself has not progressed.

On the other hand, the vibration requirements for the QD0 magnets used in SuperKEKB have become comparable to those for the FD magnets to be used in the ILC. The SuperKEKB magnets are superconducting magnets cooled by 4.2~K helium and do not use superfluid helium. Through meticulous design, SuperKEKB has reduced the vibration of the cryomodule of  this non-superfluid 4.2~K helium-cooled superconducting final focus magnets to 50~nm\cite{FD_Yamaoka}.

At present stage, the FD system using superfluid helium described in the TDR is considered the primary candidate for the ILC. However, 
as the 
the vibration measurement by BNL 
may not
be performed before the ILC Pre-lab is established, it is considered necessary to design a new FD system based on the cryostat design of SuperKEKB.
\subsubsection{Beam Dump Design (WPP-17)}
\paragraph{Description}
The main beam dump absorbs the electron or positron beam after collision at the end of each beamline. Because the beam power of full power operation after the 1~TeV upgrade will be rated at 14~MW, a pressurized water dump that is capable of 17~MW, including a 20~\% safety margin, had been designed based on the 2.2~MW water dump at SLAC. 
The design work was conducted under the GDE in the 2000s, and the results were published \cite{MainDump} and summarized in the TDR.

The concept of this MW beam dump is as follows:
The beam, consisting of 1,312 or 2,652 bunches, will be injected into the main dump by rasterizing its injection point, which becomes circular about 12~cm in diameter, to prevent a local heat accumulation in the dump. Similarly, heated water along the beam axis in the dump needs to be swept out as quickly as possible. This can be achieved by creating a vortex water flow in the dump vessel. Additionally, the water should be pressurized to about 1~MPa to prevent a boiling due to maximum heat deposition on the beam injection axis.

\begin{figure}[htbp]
	\centerline{
	\includegraphics[width=13cm] {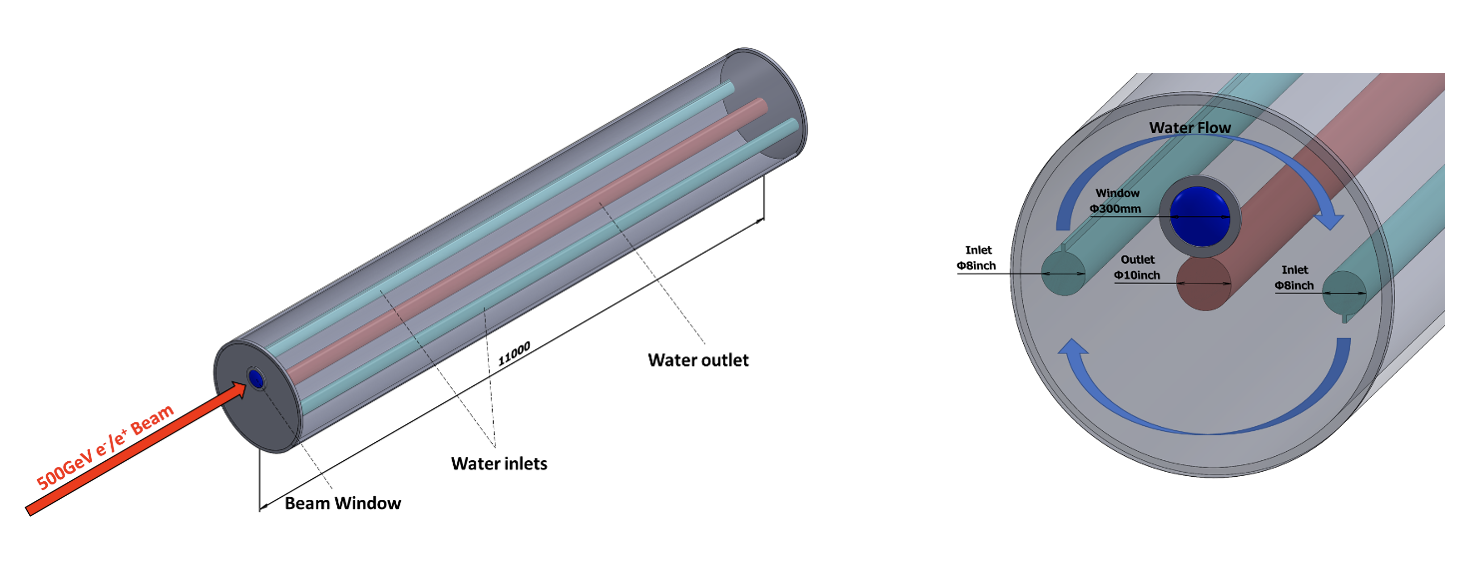}}
	\caption{Conceptual structure of the main beam dump in TDR}
	\label{fig:WPP17-tdr}
\end{figure}

The beam dump vessel was designed with a circular cross section of 1.8~m in diameter to create a vortex flow. The beam window is mounted off-center on the dump to receive a effective vortex flow performance. Two water supply pipes with ejection slits are mounted on the left and right sides, and one water return pipe is mounted at the center of the vessel.  The water vessel is 11~m long to provide an efficient radiation length for the beam. Figure~\ref{fig:WPP17-tdr} show the conceptual structure in TDR.
The water flow rate at the ejection slits in the design is 2.2~m/s and it results in a total amount of water supply to the dump 210~l/s.

The activation of the beam dump and its surrounding shields have been evaluated by the FLUKA simulation code. It revealed that the radiation level during expected maintenance work would reach 100~mSv/h if the ILC operated at maximum beam power for one year.
Access to the dump room must be limited. The beam dump system requires a robust design. For example, the beam window will need to be maintained every few years or less. The dump vessel, which includes a structure for vortex water flow, must also be robustly constructed to prevent mechanical failure.

The work package items for the main beam dump are selected below to prepare the robust design for the engineering design documentation.

\begin{itemize}
\item Develop the design of vortex water flow system in the dump vessel and verify it with a small-scale prototyping.				
\item Develop the design of beam-window remote-exchange system and verify the key functions with a small-scale prototyping of the components.				
\item Design overall water flow system				
\item Design the countermeasure for failures / safety system
\end{itemize}

First, develop a robust design for the vortex-generating structure within the dump vessel. Based on simulations, advance the design, build a mini-model of the key section, and verify vortex generation. Then, incorporate seismic resistance and leak failure countermeasures into the updated design to finalize the circulating water system.
For the remotely-operated beam window replacement mechanism design, examine the window mounting structure first. Then, prototype verification will be conducted on the key mechanism to refine the design. Finally, design the alignment mechanism for the replacement device to complete the overall system design.

These developments will be executed over five years. The first three years will focus on refining the basic design, and the final two years will move to prototype testing and finalizing the overall design.

\paragraph{Status}
KEK and CERN will carry out the beam dump work package. KEK will develop the design and conduct prototype testing. CERN will participate in discussions and provide assistance.

In order to mitigate the risk of mechanical failure of the beam dump body, a revision of the vortex water flow structure design is underway. The original design incorporates pipes with water ejection slits; however, the water flow through a slit requires a velocity of 2.2~m/s, and the total water output of a single pipe exceeds 100~l/s. This has the potential to increase the risk of failure due to the occurrence of mechanical vibrations in the pipes and the thinning of the material, particularly in the vicinity of the slits, as a result of the elevated water flow speed. The present investigation focuses on the examination of several designs that have been developed to ensure the maintenance of flow speed along the beam absorption area, 0.6~m/s for the TDR design. 

\begin{figure}[htbp]
	\centerline{
		\includegraphics[width=13cm] {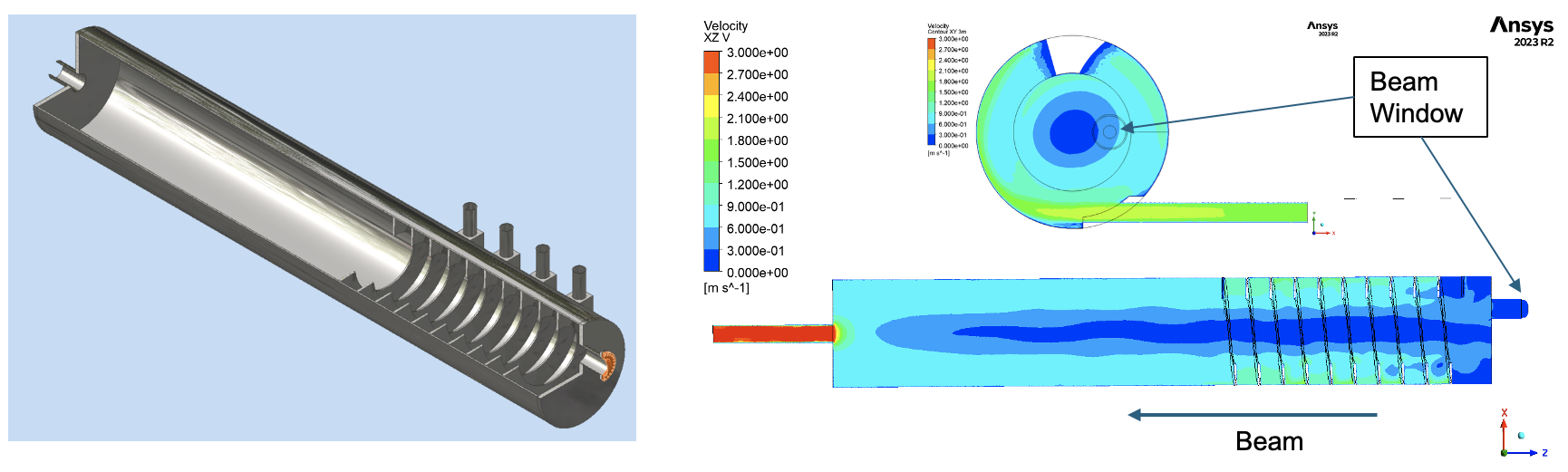}}
	\caption{Update of the main beam dump design}
	\label{fig:WPP17-vortex}
\end{figure}

One potential solution to avoid a piped structure within the dump vessel is
shown in Figure~\ref{fig:WPP17-vortex}. The spiral fins guide water and make a vortex flow down to the end of vessel. 
According to the ANSYS simulation, this structure provides a flow rate 1.5 times higher than the original design. 
It will provide a way to relax the load of the entire system. Further studies are in progress.

Since the prototyping test of vortex flow in a real size dump vessel is presently not realistic,
we are planning to 
make a
1/5 scale model which is currently being designed with industry. The realization of this model is scheduled for 2026. 
Comparison of the results of water flow, as determined by simulation and model, will serve to validate the design and provide a basis for the identification of additional improvements.

The design work for the beam window remote exchange system is proceeding in parallel.
The beam window will be circular and greater than 30~cm in width and a thickness of a few millimeters. It will be made of titanium alloy Ti-6Al-4V, which has been utilized in the proton beam target for MW. The damage of the window is studied by simulation and evaluated to have no severe damage for a few years operation, because of the relatively low energy deposit on the thin window by electron beam.

\begin{figure}[htbp]
	\centerline{
	\includegraphics[width=13cm] {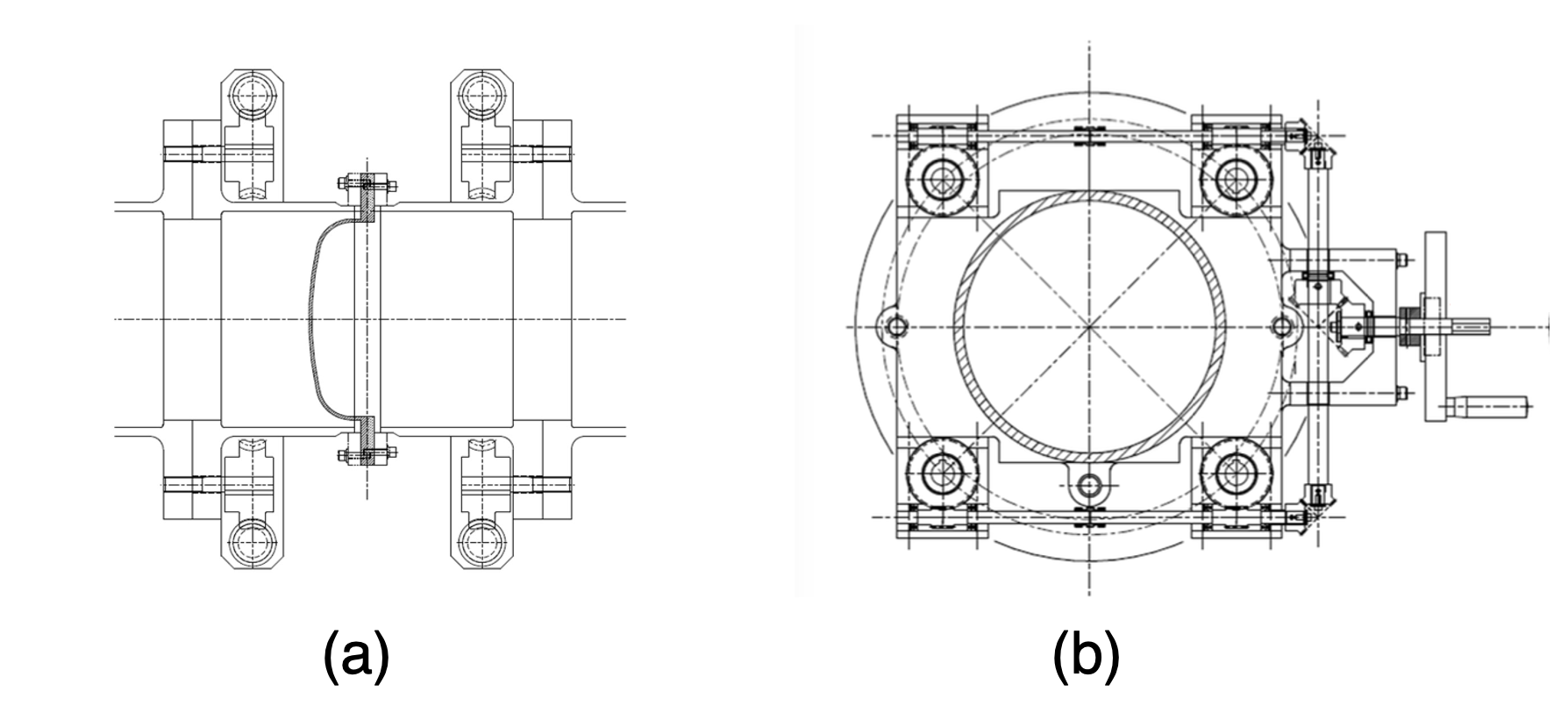}}
	\caption{Design of the beam window chamber}
	\label{fig:WPP17-window}
\end{figure}

Remote handling of the thin window during maintenance is deemed challenging. 
Furthermore, to seal both the vacuum and water sides, it is assumed that a window will be incorporated into a vacuum pipe unit.
Figure~\ref{fig:WPP17-window} (a) shows an example of this window chamber. This unit can be connected to the dump vessel through a flange located on one end.
Several structures have been investigated for this connection, including a clamp type and a conventional bolting configuration. 
Conventional clamping and connection systems seem feasible and are being pursued for the fastening mechanism.
For example, an idea to fasten all bolts together by a cam-based structure is proposed by industry as shown in Figure~\ref{fig:WPP17-window} (b).

The design study of the flange-fixture mechanism with remote handling will take about a year. We plan to start prototyping key components in the second half of 2026. The final design, including the safety measures, will be finalised in 2027.

\newpage
\section{Conclusions}
Responsible institutes have been assigned for the execution of all ITN work packages, with the exception of the crab cavity work package, for which the identification of a responsible institute is still required. 
Although the start-up phase took considerably longer than originally anticipated, the work is now making steady progress in accordance with the agreed plan. At present, it is anticipated that the ongoing work packages will be completed by April 2027, when the current MEXT funding for advanced accelerator R\&D is scheduled to come to an end.

While the primary goal of the ITN 
is to carry out an engineering design study for the ILC, the work packages have strong relevance to other particle physics accelerator projects, including the FCC at CERN, as well as to synchrotron light source facilities. It should also be noted that R\&D activities aimed at long-term future linear colliders, such as the development of high-gradient accelerating cavities and energy recovery linac technology, although not within the current scope of the ITN, represent a natural extension of the work being carried out by the ITN.
The ITN is thus a very important program for future advanced accelerator technology.



\section{ILC Technology Network}\label{ITN-members}
{\bf This document was prepared by }\\
Khaled Alharbi,
Philip Burrows, 
Enrico Cenni,
Yoshinori Enomoto,
Angeles Faus-Golfe,
Klaus Floettmann,
Manuel Formela,
Joe Grames, 
Niclas Hamann,
Tim Lengler,
Benno List, 
Gregor Loisch,
Dieter Lott,
Peter McIntosh,
Shinichiro Michizono (Editor),
Laura Monaco, 
Gudrid Moortgat-Pick, 
Tatsuya Nakada (Editor), 
Toshiyuki Okugi, 
Samanwaya Patra,
Sabine Riemann,
Hiroshi Sakai, 
Peter Sievers,
Nobuhiro Terunuma,
Carmen Tenholt,
Malte Trautwein,
Grigory Yakopov,
Akira Yamamoto, 
Yasuchika Yamamoto,
Kaoru Yokoya, \\[3mm]

\noindent
{\bf Members of the ITN are}
\begin{itemize}
\item ANSTO\\ 
Rohan Dowd

\item CEA\\
Stephane Berry,
Enrico Cenni,
Fabien Eozenou,
Gregoire Jullien 

\item CERN\\
Steffen Doebert, 
Conrad Callieri, 
Rogelio Tomas Garcia, 
Lewis Kennedy, 
Robert Kieffer, 
Kacper Lasocha, 
Andrea Latina, 
Wiktoria Malek, 
Matthias Mentink,
Franck Peauger, 
Karl-Martin Schirm, 
Steinar Stapnes, 
Orson  Vermare

\item KEK \\
Yuki Abe, 
Mitsuo Akemoto, 
Sora Arai, 
Dai Arakawa, 
Sakae Araki, 
Hayato Araki, 
Yasushi Arimoto, 
Alexander Aryshev, 
Rishabh Bajpai, 
Takeshi Dohmae, 
Masato Egi, 
Yoshinori Enomoto, 
Masafumi Fukuda, 
Takeyoshi Goto, 
Takafumi Hara, 
Kazufumi Hara, 
Masahiko Hiraki, 
Teruya Honma, 
Hayato Ito, 
Eiji Kako, 
Hiroaki Katagiri, 
Ryo Katayama, 
Shiori Kessoku, 
Kiyoshi Kubo, 
Takayuki Kubo, 
Ashish Kumar, 
Masakazu Kurata, 
Shigeru Kuroda, 
Shuji Matsumoto, 
Toshihiro Matsumoto, 
Shinichiro Michizono, 
Takako Miura, 
Yu Morikawa, 
Hirotaka Nakai, 
Hiromitsu Nakajima, 
Eiji Nakamura, 
Kota Nakanishi, 
Toshiyuki Okugi, 
Mathieu Omet, 
Konstantin Popov, 
Takayuki Saeki, 
Hiroshi Sakai, 
Motoki Sato, 
Safwan Shanab, 
Hirotaka Shimizu, 
Nobuhiro Terunuma, 
Ryuichi Ueki, 
Kensei Umemori, 
Eric Viklund, 
Yuichi Watanabe, 
Tomohiro Yamada, 
Akira Yamamoto, 
Yasuchika Yamamoto, 
Kaoru Yokoya 

\item IFIC \\
Juan Carlos, 
Javier Olivares Herrador, 
Eduardo Martínez López, 
Nuria  Fuster Martinez, 
Laura Pedraza Motavia, 
Juan Fernandez Ortega 

\item IJCLAB \\
Daniel Charlet, 
Cedric Esnault, 
Zomer Fabian, 
Angeles Faus Golfe, 
Anna Korsun, 
Aurelien Martens

\item INFN Milano\\
Michele Bertucci,
Elisa Del Core,
Laura Monaco,
Daniele Sertore

\item Korea University \\
Hojun Jeong, 
Eun-San Kim, 
Byeong Rok Ko, 
Soohyung Lee, 

\item University of Hamburg \\
Khaled Alharbi,
Manuel Formela,
Niclas Hamann,
Tim Lengler,
Gudrid Moortgat-Pick,
Samanwaya Patra,
Malte Trautwein

\item University of Oxford \\
Douglas Bett, 
Philip Burrows,
Pierre Korysko

\item STFC \\
Peter McIntosh
\end{itemize}

\subsection*{Acknowledgements}
ITN activities of KEK are supported by MEXT for development of key element technologies to improve the performance of future accelerators program, under Grant Number JPMXP1423812204. The work of the ITN is supported by the European Union’s Horizon Europe Marie Sklodowska-Curie Staff Exchanges programme under grant agreement no. 101086276 (EAJADE). Spanish activities are supported by the Ministry of Science, Innovation and Universities, Generalitat Valenciana, and European Union’s NextGeneration EU program. 
Appendix 26 of the Agreement on Collaborative Work (ICA-JP-0103) between and KEK and CERN provides a framework for the ITN participation of European laboratories with CERN as a hub-laboratory. 
We thank G. Taylor for his careful reading of this manuscript with numerous comments and suggestions. 

\end{document}